\documentclass[preprint2]{aastex}

\usepackage{comment}
\usepackage{mathtools}
\usepackage{natbib}
\usepackage{floatrow}
\usepackage{caption}
\usepackage{float}
\restylefloat{table}
\usepackage{placeins}

\newcommand{\Lsun}{$L_{\odot}$}
\newcommand{\Msun}{$M_{\odot}$}
\newcommand{\Rsun}{$R_{\odot}$}
\newcommand{\Mdot}{$\dot{M}$}
\newcommand{\Av}{A$_V$}
\newcommand{\msunyr}{$\rm{M_{\sun} \, yr^{-1}}$}
\newcommand{\Teff}{$T_{\rm eff}$}
\newcommand{\mic}{$\mu$m}

\begin{document}

\title{\textit{Herschel} Observations of Protoplanetary Disks in Lynds 1641}

\author{Sierra L. Grant\altaffilmark{1}, Catherine C. Espaillat\altaffilmark{1}, S. Thomas Megeath\altaffilmark{2}, Nuria Calvet\altaffilmark{3}, William J. Fischer\altaffilmark{4}, Christopher J. Miller\altaffilmark{3}, Kyoung Hee Kim\altaffilmark{5}, Amelia M. Stutz\altaffilmark{6,7}, \'{A}lvaro Ribas\altaffilmark{1}, Connor E. Robinson\altaffilmark{1}}

\shortauthors{Grant et al.}

\altaffiltext{1}{Department of Astronomy, Boston University, 725 Commonwealth Avenue, Boston, MA 02215, USA}
\altaffiltext{2}{Ritter Astrophysical Research Center, Department of Physics and Astronomy, University of Toledo, Toledo, OH 43606, USA}
\altaffiltext{3}{Department of Astronomy, University of Michigan, Ann Arbor, MI 48109, USA}
\altaffiltext{4}{Space Telescope Science Institute, Baltimore, MD 21218, USA}
\altaffiltext{5}{Department of Earth Science Education, Kongju National University, 56 Gongjudaehak-ro, Gongju-si, Chungcheongnam-do 32588, Republic of Korea}
\altaffiltext{6}{Departmento de Astronom\'{i}a, Universidad de Concepci\'{o}n, Casilla 160-C, Concepci\'{o}n, Chile}
\altaffiltext{7}{Max-Planck-Institute for Astronomy, K\"onigstuhl 17, 69117 Heidelberg, Germany}

\shortauthors{Grant et al.}

\begin{abstract}
We analyze \emph{Herschel Space Observatory} observations of 104 young stellar objects with protoplanetary disks in the $\sim$1.5 Myr star-forming region Lynds 1641 (L1641) within the Orion A Molecular Cloud. We present spectral energy distributions from the optical to the far-infrared including new photometry from the \emph{Herschel} Photodetector Array Camera and Spectrometer (PACS) at 70 $\mu$m. Our sample, taken as part of the \textit{Herschel} Orion Protostar Survey, contains 24 transitional disks, eight of which we identify for the first time in this work. We analyze the full disks with irradiated accretion disk models to infer dust settling properties. Using forward modeling to reproduce the observed $n_{K_{S}-[70]}$ index for the full disk sample, we find the observed disk indices are consistent with models that have depletion of dust in the upper layers of the disk relative to the midplane, indicating significant dust settling. We perform the same analysis on full disks in Taurus with \emph{Herschel} data and find that Taurus is slightly more evolved, although both samples show signs of dust settling. These results add to the growing literature that significant dust evolution can occur in disks by $\sim$1.5 Myr.
\end{abstract}

\keywords{Infrared: stars --- protoplanetary disks --- accretion disks --- circumstellar matter}

\section{Introduction} %Sect 1
Protoplanetary disks are composed of gas and dust and are formed in the collapse of a slowly rotating, dense core in the star's natal, molecular cloud \citep{terebey84}.  The details of how these disks evolve from initially well-mixed distributions of gas and dust to systems composed mostly of rocky planets and gas giants are not well understood. As the disk evolves, the dust grains will settle to the midplane where they will grow and the gas will dissipate (e.g., \citealt{goldreich&ward73, weidenschilling80, d'alessio06, testi14}). Many complex processes occur between the initial formation of the disk and the dissipation of gas, including photoevaporation from the central source, grain growth and settling to the midplane, disk instabilities, sculpting due to companions, and planet formation (e.g., \citealt{marsh&mahoney92,clarke01,dullemond&dominik05,lubow&d'angelo06,chiang&mc07,alexander14,rosotti16, pinilla17, armitage18, hendler18, vanderMarel18}).

We can use spectral energy distributions (SEDs) of young objects and their circumstellar disks to analyze these systems and constrain their properties. The circumstellar material is irradiated by the central star and re-emits radiation primarily at infrared and millimeter wavelengths (e.g., \citealt{k&h87, calvet92, chiang&goldreich97, calvet&gullbring98, d'alessio99, d'alessio01, d'alessio06, mcclure13a,ingleby13}). In particular, spectral indices of the disk emission can be used to study some of its properties; these indices are defined as \begin{equation}n_{\lambda_{1}-\lambda_{2}} = \frac{\log(\lambda_{2}F_{\lambda_{2}})-\log(\lambda_{1}F_{\lambda_{1}})}{\log(\lambda_{2})-\log(\lambda_{1})},\label{eq: indice}\end{equation} which are essentially slopes between $\lambda_{1}$ and $\lambda_{2}$. These indices can distinguish evolutionary stages, especially when used in the NIR to mid-infrared (MIR; e.g., \citealt{adams87,lada87,andre&montmerle94,calvet94, greene94,evans09,dunham14,kryukova14,furlan16}). In this paper, we adopt the criteria of \citet{furlan16} and \citet{kryukova14}, where Class 0/I protostars have $n_{[4.5]-[24]}>0.3$, flat-spectrum objects have $-0.3<n_{[4.5]-[24]}<0.3$, and Class II objects (i.e., a pre-main sequence star surrounded by a disk) have $-1.6<n_{[4.5]-[24]}<-0.3$. We can also use spectral indices to infer disk properties, such as gaps, dust settling, and truncation (e.g., \citealt{mcclure10,manoj11,furlan11,mauco16}).

The large size and relatively close proximity of the Orion Molecular Cloud complex makes it an ideal region to study active star formation. These clouds, which span roughly 90 pc in length \citep{megeath12}, are at a distance of $\sim$400 pc \citep{kounkel17}. \citet{kounkel17} find the Orion Nebula Cluster lies at 388 $\pm$ 5 pc and the southern portion of Lynds 1641 (L1641) is located at 428 $\pm$ 10 pc. The Orion Molecular Cloud complex houses regions of both high-mass and low-mass star formation in a relatively dense area. The complex is composed of two clouds, A and B. Cloud A contains the Orion Nebula Cluster to the north, with low-mass and high-mass star formation in its dense center, and L1641 to the south with a large population of low-mass stars (e.g., \citealt{megeath16}). L1641 has been estimated to have over 1600 Class II and Class III (i.e., pre-main sequence stars without disks) objects \citep{hsu12, pillitteri13} with an age of 1--3 Myr (e.g., \citealt{galfalk&olofsson08, fang09, fang13, hsu12}). L1641 contains a population of young stars comparable in size to the Orion Nebula Cluster, but due to its lower spatial density and lack of O stars, it has a much lower infrared background and a lower degree of source confusion. This makes it ideal for sampling spectral energy distributions in the infrared with the modest angular resolution of the \textit{Spitzer Space Telescope} and \textit{Herschel Space Observatory}. L1641 is undergoing more active star formation than Taurus and other nearby star-forming regions and provides us with a large sample of targets that span the first phases of evolution while having roughly the same age and initial chemical composition (e.g., \citealt{megeath16}).

We present \emph{Herschel Space Observatory}\footnote{\textit{Herschel} is an ESA space observatory with science instruments provided by European-led Principal Investigator consortia and with important participation from NASA.} \citep{pilbratt10} photometry taken with the Photodetector Array Camera and Spectrometer (PACS; \citealt{poglitsch10}) of 104 Class II objects in   L1641. These protoplanetary disks were observed as part of the \textit{Herschel} Orion Protostar Survey (HOPS), an open-time key program (e.g., \citealt{manoj13, stutz13,furlan16, fischer17}; B. Ali et al. 2018, submitted).  We include ancillary data from the literature to construct dereddened 0.55--70 \mic \ SEDs for 98 targets. We use a forward modeling process to compare the \citet{d'alessio98, d'alessio99,d'alessio01,d'alessio06} self-consistent, irradiated, accretion disk models to our full disk sample indices. We also perform this forward modeling analysis on a sample of Class II objects in Taurus with PACS data from \citet{howard13} to put our sample into context with a more well-studied region. 

In Section 2, we present the \emph{Herschel} observations and the reduction procedures used to obtain photometry. In Section 3, we analyze the disk properties of the sample, including a description of the stellar sample, presentation of SEDs of individual objects and the median SED for the sample (as well as the full disks and transitional disks separately), and disk classification. In Section 4, we compare our full disk sample with irradiated disk models. We discuss our findings in Section 5 and summarize this work in Section 6. 

\vspace{-1.1mm}

\section{Observations}\label{sect: obs} %Sect 2
\begin{figure*}[h!]
\epsscale{0.65}
\plotone{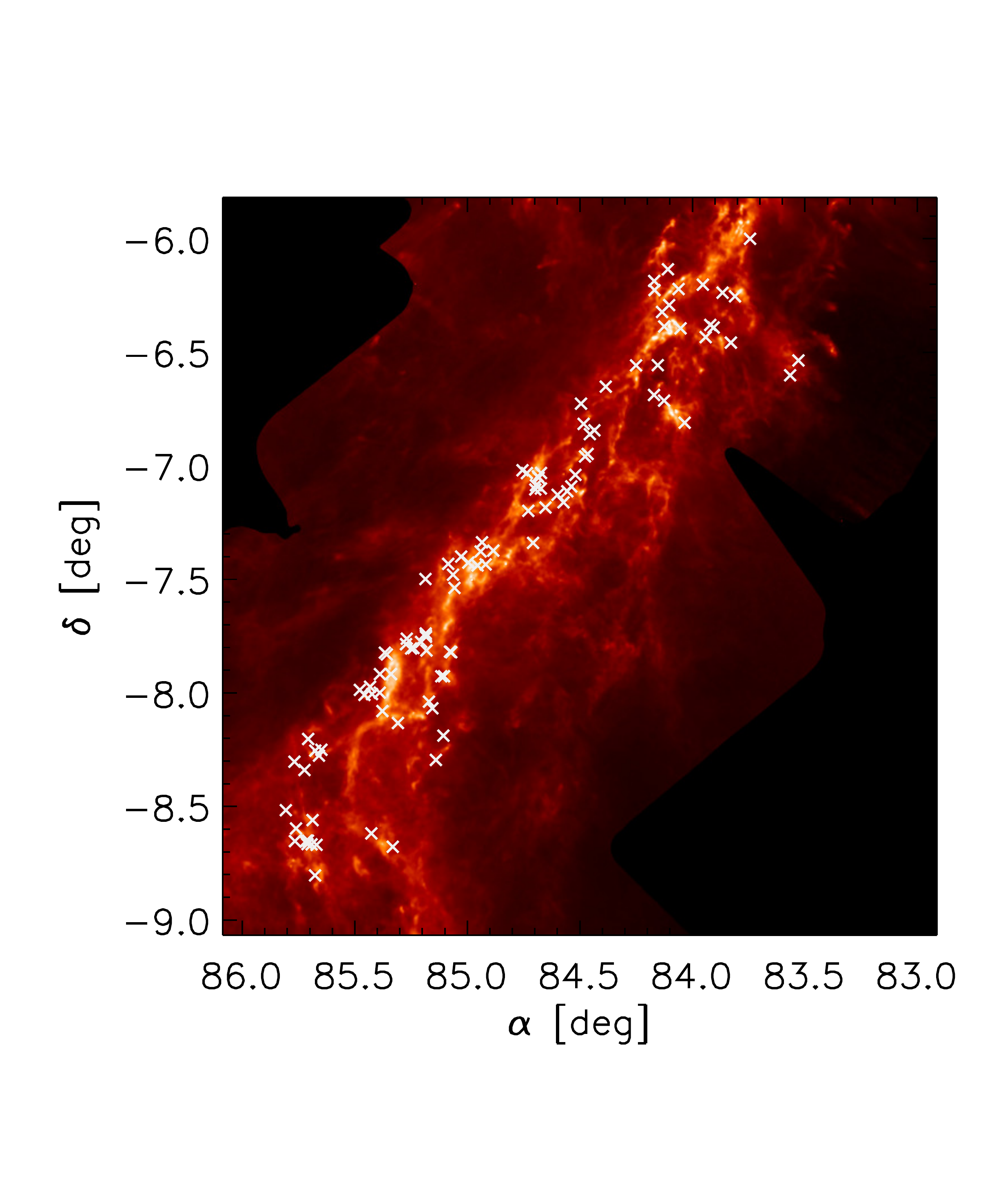}
\centering
\caption{L1641 column density, N(H), map shown on a log scale. The 104 objects in the sample analyzed here are marked with white $\times$-symbols. Figure adapted from \citet{stutz&kainulainen15, stutz&gould16}.}\label{fig: l1641}
\end{figure*}

We compiled an initial sample of young stellar objects (YSOs) on the basis that they were classified as Class II objects in \citet{megeath12, megeath16} and were located in L1641 (we require that our sample be located below --6$\arcdeg$ declination, see Figure~\ref{fig: l1641}). For this sample, we extracted photometry from the \textit{Herschel} Orion Protostar Survey (HOPS) open-time key program in 2010 and 2011 (e.g., \citealt{manoj13, stutz13,furlan16, fischer17}; B. Ali et al. 2018, submitted). We note that \citet{megeath12} used \textit{Spitzer} Infrared Array Camera (IRAC, \citealt{fazio04}) and Multiband Imaging Photometer for SIRTF (MIPS, \citealt{rieke04}) observations. There are an additional 11 sources that were reclassified from Class II objects to Class 0/I/flat-spectrum sources using the \textit{Herschel} observations and they are presented in \citet{furlan16}; we do not include them here. Removing these objects results in a reduced sample of 169 objects out of the 180 Class II L1641 objects in \citet{megeath12} for which we extracted photometry with the HOPS maps. (We discuss further reducing our analysis sample to 104 objects in Section~\ref{subsect: stellar sample}.)
    
The HOPS maps are square maps of 5 or 8 arcmin on a side that were optimized to detect \textit{Spitzer}-identified protostars with expected 70 \mic\ flux densities greater than 42 mJy. Each field was scanned and then cross-scanned in the orthogonal direction to reduce noise. The presence of Class II sources was serendipitous. 

The maps used for point-source photometry were reduced with a high-pass filtering method that reduces the contribution of smoothly varying extended structure while preserving the integrity of point sources. This process is described in detail by B. Ali et al. 2018, submitted. We obtained photometry for each source in a circular aperture of 9.6 arcsec in radius with background subtraction of the signal measured in an annulus extending from 9.6 to 19.2 arcsec. These aperture parameters are not much larger than the 70 \mic\ angular resolution of PACS (5.2 arcsec) in order to reduce the influence of the bright, spatially varying nebular emission in Orion. The resulting values were divided by 0.7331 to account for flux in the wings of the point-spread function that is not included in the aperture. The uncertainty for each measurement is dominated by a 5\% floor that was included to account for the global calibration uncertainty. The data were processed with version 9 of the \textit{Herschel} Interactive Processing Environment, using the FM7 version of the PACS calibration. The \textit{Herschel} fluxes are listed in Table~\ref{tab:observations} and Table~\ref{tab: flagged observations}. Table~\ref{tab:observations} contains objects included in our analysis and  Table~\ref{tab: flagged observations} contains objects that were removed from the sample as described in Section~\ref{subsect: stellar sample}.

\section{Observed Properties}
    \subsection{Stellar Sample}\label{subsect: stellar sample}

    L1641 is a relatively well-studied region, and the works of \citet{fang09,fang13}, \citet{hsu12}, \citet{carattiogaratti12}, and \citet{ kim13, kim16} provide stellar properties for a subset of its YSO population.
    \citet{galfalk&olofsson08} derive an age of $\sim$1 Myr for region. \citet{fang13} found similar median ages, with 1.5 Myr for the ``clustered'' YSOs and 1.6 Myr for the ``isolated'' YSOs. We adopt 1.5 Myr for our analysis.\footnote{Our sources were observed in fields chosen to target protostars, so we expect that our sources are associated with dense gas filaments/clumps and are similar in age to the ``clustered'' age found by \citet{fang13}. Additionally, the derived age depends on the chosen pre-main-sequence track. \citet{fang13} used the \citet{dotter08} isochrones while \citet{hsu12} find an age of 3 Myr using the \citet{siess00} and \citet{baraffe98} isochrones. We note that age differences in the literature will not significantly impact our results.} 
    
    From the 169 Class II objects discussed in Section~\ref{sect: obs}, we remove objects that meet any of the following criteria: 1) their \textit{Herschel} maps show nearby close sources or nebulosity; 49 are flagged for this reason. 2) They have visual extinctions, \Av, equal to or greater than 15; this applies to 18 systems. 3) They have colors uncharacteristic of classical T Tauri stars (CTTSs); this applies to one object. Including the targets with nearby sources or nebulosity would make the 70 \mic\ fluxes for those objects more uncertain due to contamination, thus making our analysis using the 70 \mic\ data unreliable. Objects with \Av$\geq$15 will have higher uncertainties and we remove them to avoid introducing biases.  The one source that does not have colors characteristic of a CTTS is an A star and should not be compared directly to the rest of the sample. We present the \textit{Herschel} photometry and stellar properties for these flagged objects in the Appendix, but these objects are not included in the rest of the analysis. The 104 protoplanetary systems analyzed in this work are marked in the L1641 column density map shown in Figure~\ref{fig: l1641} \citep{stutz&kainulainen15, stutz&gould16}.
    
\begin{figure*}[]
\epsscale{0.73}
\plotone{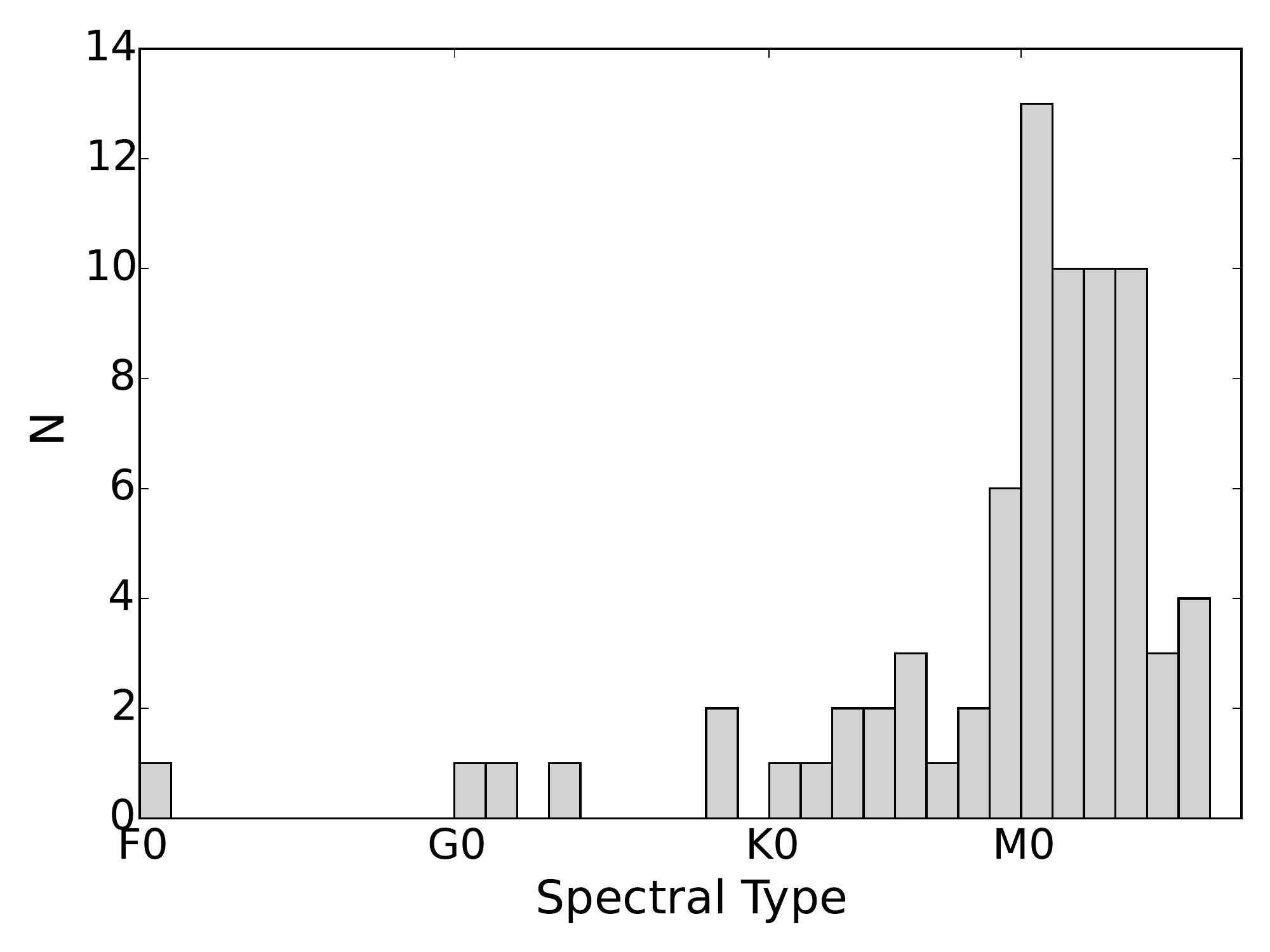}
%\centering
\caption{Histogram of spectral types available in the literature for our L1641 sample. Bin size is one subclass. The median spectral type is M1. }\label{fig:spt hist}
  %% From SpT_hist.py
\end{figure*}

    For our sample of 104 systems, we obtained literature values of spectral types for 75 objects, \Av\ values for 98 objects, luminosities (L) for 77 objects, and mass accretion rates (\Mdot) for 61 objects. These values and their references are listed in Table~\ref{tab: stellar properties}. More than 75\% of the spectral types in this sample come from \citet{hsu12} using the \citet{hernandez04} spectral-typing process, with typical uncertainties of less than one subclass. The median spectral type for the sample is M1 (Figure~\ref{fig:spt hist}). Additionally, for the flagged sample we have spectral types for 31, \Av\ values for 59, luminosities for 40, and mass accretion rates for 24 objects. These values are available in Table~\ref{tab: stellar properties flagged} in the Appendix.

\begin{figure*}[]
\epsscale{0.73}
\plotone{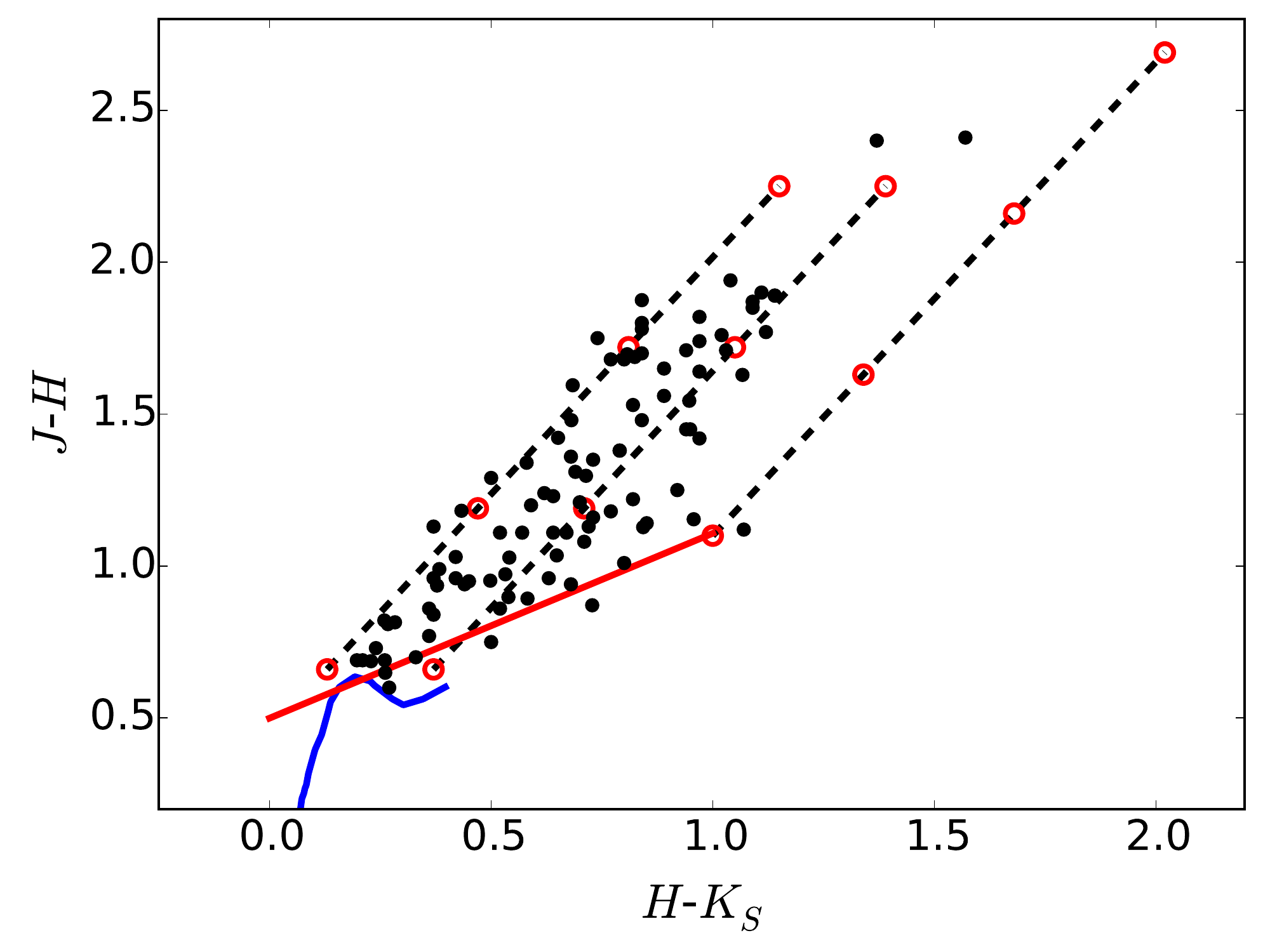}
%\centering
\caption{\textit{J}-\textit{H} vs. \textit{H}-\textit{K$_{S}$} color-–color diagram. The solid lines are the intrinsic colors of stars on the CTTS locus (red, \citealt{meyer97}) and dwarf branch (blue, \citealt{bessell&brett88}). The dashed lines are extinction vectors and the open red points show the extinction in increments of \Av=5.  }\label{fig: jhk}
  %% From JHK_new.py
\end{figure*}

\begin{figure*}[]
\epsscale{0.73}
\plotone{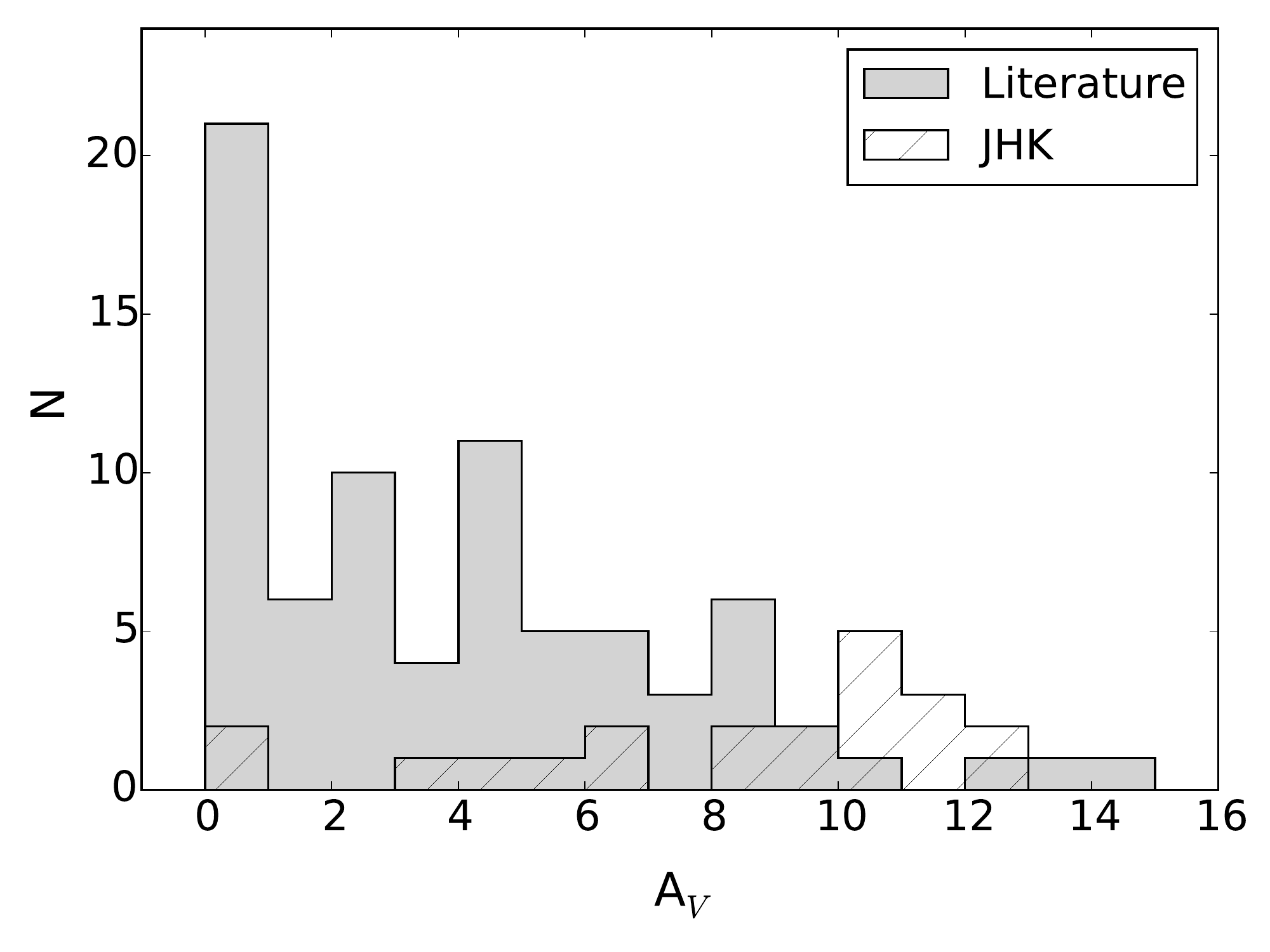}
%\centering
\caption{Histogram of visual extinctions in our sample. Objects with values of \Av\ from the literature are shown in gray.  The hatched region represents those objects whose \Av\ values were calculated in this work from the \textit{J}--\textit{H} and \textit{H}--\textit{K$_S$} colors and the CTTS locus.}\label{fig: Av hist}
  %% From Av_new.py
\end{figure*}

    Of the 98 \Av\ determinations, 77 were obtained from the literature (see the Appendix for the flagged sample). The rest of the \Av\ determinations were calculated in this work using the observed \textit{J}, \textit{H}, and \textit{K$_{S}$} photometry bands from the Two Micron All-Sky Survey (2MASS, \citealt{skrutskie06}), the \citet{mathis90} reddening law (with an $R_{V}=3.1$; the extinction curve in the \textit{JHK}-bands is similar for the \citealt{mcclure09} law and we adopt $A_H/A_J = 0.624$ and $A_K/A_J = 0.382$), and by calculating the intrinsic color from the CTTS locus from \citet{meyer97},
    \begin{equation}
    (J-H)_{\rm CTTS} = 0.58 \pm 0.11 (H-K_{S})_{\rm CTTS} +
    0.52 \pm 0.06,
    \end{equation} 
    after converting the locus to the 2MASS photometric system using conversions presented by \citet{carpenter01}. For objects with \citet{megeath12} ID numbers 612 and 792, the \Av\ calculated in this manner leads to slightly negative values that are nonphysical, and thus here we adopt \Av\ values of 0.0 (Table~\ref{tab: stellar properties}). Figure~\ref{fig: jhk} shows the \textit{J}-\textit{H} vs. \textit{H}-\textit{K$_{S}$} color--color diagram for the sample with the CTTS locus shown as a solid red line. The dwarf branch from \citet{bessell&brett88} and the CTTS locus shown in Figure~\ref{fig: jhk} have been converted to the 2MASS photometric system as described above. The distribution of the visual extinction values gathered from the literature and calculated in this work from the CTTS locus for the sample is shown in Figure~\ref{fig: Av hist}. The \Av\ distribution indicates that the literature values tend to be lower than those calculated from the CTTS locus, likely because the literature values (taken mostly from \citealt{fang09,fang13} and \citealt{kim16}) were calculated using photospheric colors based on available spectral types, which biases the distribution towards low extinction values where spectral typing is more accurate.

    The values for L and \Mdot\ are collected from the literature \citep{fang09, fang13,kim13,kim16}. \citet{fang09, fang13} use the H$\alpha$, H$\beta$, and HeI line luminosities, as well as the full width of H$\alpha$ at 10\% to derive \Mdot. Accretion rates from \citet{fang09} and \citet{fang13} agree within 50--250\%. \citet{kim13,kim16} use the Pa$\gamma$, Pa$\beta$, and Br$\gamma$ line luminosities.  Accretion rates from \citet{kim16} range from being $\sim$3\% to $\sim$7,000\% of the values in \citet{fang13} for the same object. These differences in mass accretion rates could be due to variability and/or differences in observational proxies (i.e., the lines used to calculate the mass accretion rates). We adopt the most recently published values.
    
    We note that  using Infrared Telescope Facility/SpeX \citep{rayner03} observations \citet{kim16} determined objects with \citet{megeath12} ID numbers 250 and 980 to be binaries. \citet{kounkel16}, using the \textit{Hubble Space Telescope} and the NICMOS and WFC3 cameras, searched for binaries with separations between 100 and 1,000 AU; 25 out of 169 of the objects in our Class II sample are included in this survey. Of these 25, eight are found to be binaries; these are objects 315, 369, 421, 523, 526, 561, 579, and 950. Of the known binaries, only two (561 and 579) remain in the sample after object flagging. With roughly 15\% of the sample studied for binaries (and mostly at large separations), we cannot rule out that other objects in our sample are binaries.

    \subsection{Spectral Energy Distributions}\label{sect: seds}
    
    We constructed the SEDs of the sources using data from 2MASS at the \textit{J}-, \textit{H}-, and \textit{K$_{S}$}- bands, \emph{Spitzer} IRAC at bands 3.6, 4.5, 5.8, and 8 $\mu$m, and \emph{Spitzer} MIPS at 24 \mic, along with the \emph{Herschel} PACS 70 $\mu$m photometry presented in this work. Additionally, 40 of the objects have \textit{V}- and \textit{I}-band photometry from \citet{hsu12} and 64 have \emph{Spitzer} Infrared Spectrograph (IRS; \citealt{houck04}) spectra from \citet{kim13,kim16} with an additional 18 available in the Combined Atlas of Sources with Spitzer IRS Spectra archive \citep{lebouteiller11}. In Figure~\ref{fig: seds}, we present the SEDs of the 98 sources that have \Av\ determinations.  In Figure~\ref{fig: flagged seds}, we show the SEDs of the 58 sources that have \Av\ determinations, but have been flagged as described in Section~\ref{subsect: stellar sample}. Objects without \Av\ values lack 2MASS data, and thus we cannot calculate \Av\ from the CTTS locus. The values of \Av\ that are listed in Table~\ref{tab: stellar properties} were used to correct the photometry, and spectroscopy if available, for extinction. Objects with $A_{Ks}<0.3$ are dereddened using the \citet{mathis90} extinction law with $R_V=3.1$, and those with $A_{Ks}\geq0.3$ are dereddened using the \citet{mcclure09} extinction curves. Each object is listed with it’s corresponding identification number from \citet{megeath12}; spectral type, if available; and \Av.

	The dereddened SEDs are plotted with the photospheric fluxes corresponding to the adopted spectral type \citep{k&h95}, scaled at the \textit{J}-band where the peak of the photospheric emission of K- and M-type stars occurs, although excesses from accretion, hot gas, and hot dust may still contribute at this wavelength \citep{fischer11,mcclure13a}. For objects with spectral types later than M6.0 (i.e., M7.0 or M7.5 objects), M6.0 photospheres are shown. For objects with spectral type K8.0, K7.0 photospheres are shown. This is due to the lack of M7.0 and K8.0 spectral types in \citet{k&h95}. The SEDs are also plotted with the median dereddened SED of Taurus in the K5.0--M2.0 spectral type range \citep{furlan06}. We use the Taurus median as a proxy for a typical protoplanetary disk. The Taurus median is scaled at the \textit{H}-band, where the flux is still dominated by stellar emission in T~Tauri stars. Here we have extended the median out to the 70 \mic\ PACS wavelength using data from the \emph{Herschel} Open Time Key Project's Gas in Protoplanetary Systems, presented in \citet{howard13}. Additionally, we plot the Taurus quartiles (i.e., the range within which 50\% of the objects fall) for reference.

\subsection{Disk Classification}\label{sect: disk classification}
	To characterize the properties of the disks in our sample, we first determine if our objects are full disks (FDs) or transitional disks (TDs). FDs have optically thick material extending from the dust sublimation radius near the star to the outer disk.  TDs contain holes or gaps in their dust distribution that can be inferred from a dip in the NIR and MIR flux in the SED (see \citealt{espaillat14} for a review).
	
	A considerable number of ways to identify TDs are outlined in the literature. These include the use of photometric colors and spectral indices, essentially slopes between photometry or spectral points (e.g., \citealt{furlan06,furlan09a,furlan11, gutermuth08, fang09,fang13, cieza10, merin10, muzerolle10, kim13,kim16, koenig12,ribas13}). The photometric bands for the colors and indices should be chosen such that they effectively separate the TDs from the FD population, while minimizing biases from emission features, such as silicate features. We implement some of the color and index criteria used in the literature on our sample; we did not use classification systems that relied on IRS spectra since we do not have these data for our entire sample. Here we list the criteria implemented on our sample,
	
	\begin{enumerate}
	   
		\item \citet{fang09,fang13} use $K_{S}$-[5.8] vs. [8]-[24] with boundaries for TDs defined by [8]-[24] $\geq$ 2.5 and $K_{S}$-[5.8] $\leq$ 0.56 + ([8]-[24]) $\times$ 0.15.
	
	    \item \citet{merin10} use [3.6]-[8] vs. [8]-[24] with two regions separating systems with only photospheric fluxes and systems with some excess flux above the photosphere in the IRAC bands: Region A: 0.0 $<$ [3.6]-[8] $<$ 1.1 and 3.2 $<$ [8]-[24] $<$ 5.3; Region B: 1.1 $<$ [3.6]-[8] $<$ 1.8 and 3.2 $<$ [8]-[24] $<$ 5.3. \citet{kim13}, using their Orion A sample, finds that Region A corresponds to classical TDs, defined as disks with Inner Disk Excess Fractions (IDEF) $<0.25$ (see Equation 8 of \citealt{kim13}) and Region B to weak-excess TDs, $0.25\leq$ IDEF $<0.5$, and pre-transitional disks, IDEF $\geq0.5$. 
	
	    \item \citet{cieza10} use [3.6]-[24] $>1.5$ and [3.6]-[4.5] $<0.25$ as the broad criteria for TDs in their Ophiuchus sample. 
	    
	    \item \citet{muzerolle10} use slopes $\alpha_{[3.6]-[5.8]}$ $<-1.8$ and $-1.5\leq$ $\alpha_{[8]-[24]}$ $\leq0.0$ for weak excess sources (sources with some MIR excess that is above the photosphere but below what we would expect for optically thick material; see \citealt{muzerolle10} for full discussion) and $\alpha_{[8]-[24]}>0.0$ for classical TDs. 
	 
	\end{enumerate}

\begin{figure*}[]
\epsscale{1}
\plotone{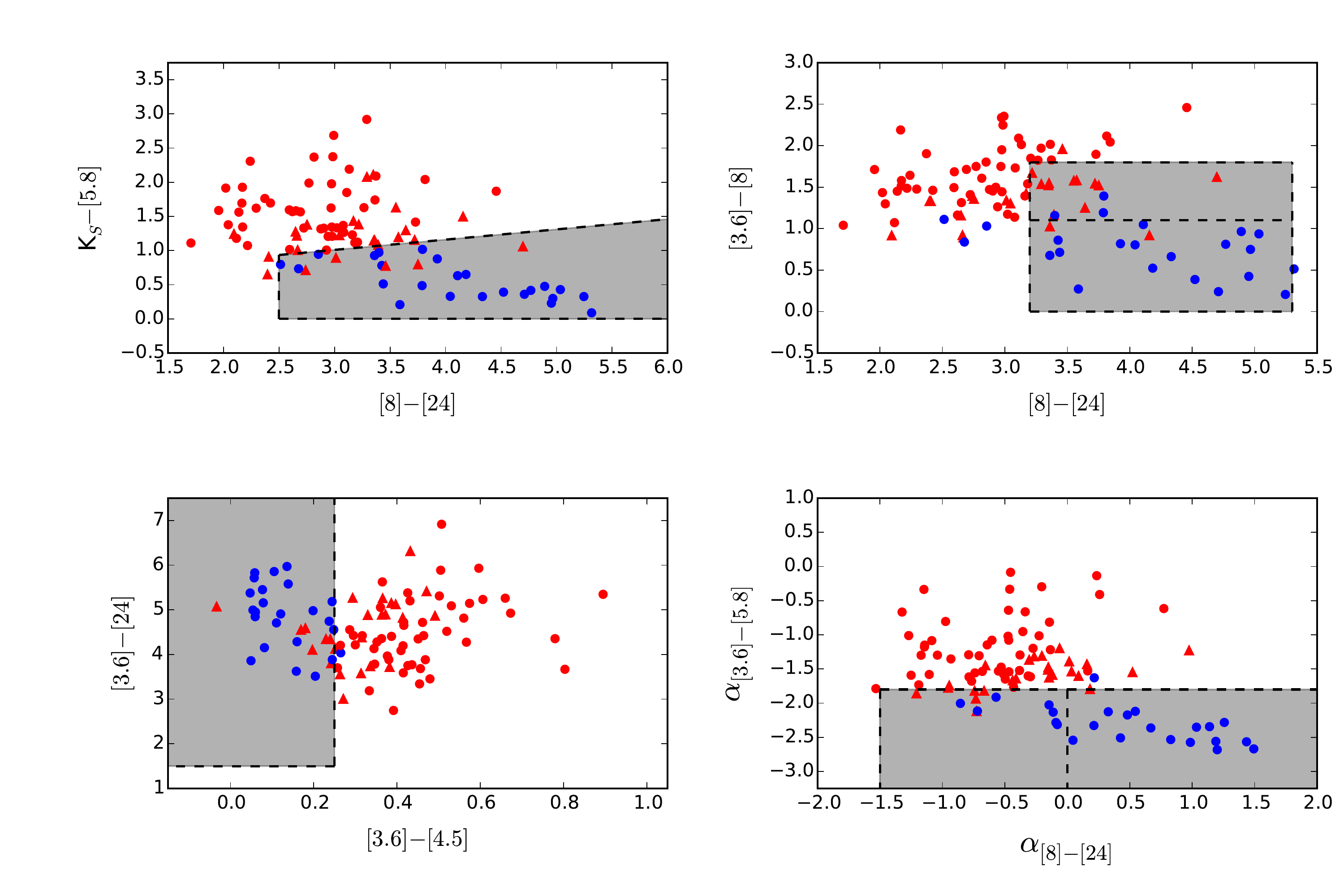}
\centering
\caption{Transitional disk selection criteria (gray shaded area) from \citet[top left]{fang09,fang13}, \citet[top right; Region A is bottom box, Region B is top box]{merin10}, \citet[bottom left]{cieza10}, and \citet[bottom right; weak-excess sources in left box, classical TDs in right box]{muzerolle10} applied to our L1641 sample (points).  Colors and indices utilize 2MASS and \textit{Spitzer} photometry. In this work we identify a protoplanetary disk as transitional (blue circle) if it meets three of the four selection criteria. Red triangles are used to mark candidate transitional disks, objects that meet one or two of the four selection criteria.  Full disks are denoted with red circles.
}\label{fig: TD criteria comparison}
  %% From Indices_MegaTable.py
\end{figure*}

    The criteria discussed here are shown in Figure~\ref{fig: TD criteria comparison}. To classify disks as transitional in our sample, we require that they meet three out of the four criteria listed above (we consider objects to be candidate TDs if they meet one or two, but not three, of the criteria).  We identify 24 objects in our sample as TDs and an additional 25 as candidates; the TDs and candidate TDs are listed in Table~\ref{tab: TDs}. This leads to $\sim$23\% TDs among the sample with available \Av, eight of which are newly identified. The eight newly identified TDs have \citet{megeath12} ID numbers of 227, 250, 278, 387, 403, 619, 689, and 811. A TD fraction of $\sim$23\% is higher than the $\sim$10\% at 1.5 Myr inferred by \citet{muzerolle10} using \textit{Spitzer} photometry. The discrepancy could be due to broader selection criteria than \citet{muzerolle10}, a bias that the TDs tend to be bright at PACS 70 \mic\ because of illuminated edge of the disk (discussed more in Section~\ref{subsect: index comparison with models}), or to uncertainties in the age of L1641. We note that a TD identification criterion using Wide-field Infrared Survey Explorer (WISE; \citealt{wright10}) 12 \mic\ photometry and \textit{Herschel} PACS 70 \mic\ photometry has been used by \citet{ribas13}, \citet{bustamante15}, and \citet{rebollido15}. These groups used the spectral index $\alpha_{[12]-[70]}>0.0$ to identify TDs. A similar method can be used for this sample, using the IRAC 8 \mic\ data instead of the WISE photometry. Using $\alpha_{[8]-[70]}>0.0$ as an identifier for this sample leads to 21 TDs. Compared to the identifications reported in this work, 16 overlap with the TD sample of 24 (67\% overlap), 3 with the candidate TDs, and 2 with the full disks.

   \subsection{L1641 Median SED}\label{sect: median SED}
\begin{figure*}[h]
\epsscale{0.7}
\plotone{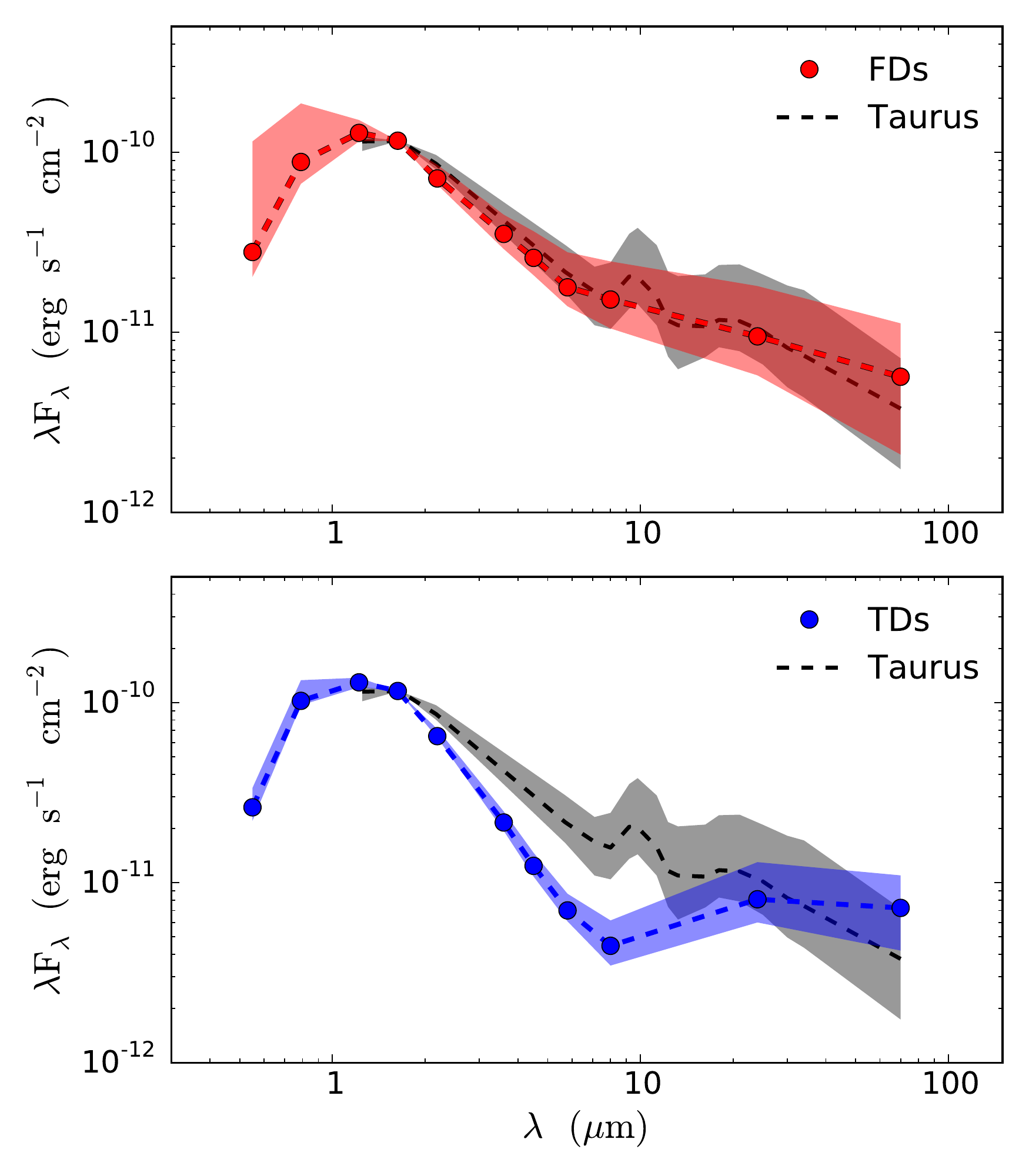}
%\centering
\caption{Top: The median SED for full disks (FDs) in the L1641 sample with \Av\ determinations (red points and dashed line) compared to the Taurus median from \citet{furlan06} and extended here to 70 \mic\ using data from \citet[black dashed line]{howard13}. Quartiles are shown as the shaded regions.
Bottom: The median SED for the transitional disks (TDs) in the L1641 sample (blue points and dashed line) compared to the Taurus median (black dashed line). Quartiles are shown as the shaded regions.}\label{fig: median SED}
  %% From MedianSEDs.py
\end{figure*}

   We present the median SEDs for the entire L1641 sample (Table~\ref{tab: median SED all}), as well as the FD median (Table~\ref{tab: median SED FD}) and TD median (Table~\ref{tab: median SED TD}), along with their quartiles. In Figure~\ref{fig: median SED} we show the median of the FD and TD samples in L1641 compared to the Taurus median. We use photometry from the \textit{V}-band to 70 \mic \ to construct the median. We determine the dereddened fluxes using the \Av \ values listed in Table~\ref{tab: stellar properties}, as described in Section~\ref{sect: seds}. In order to reduce spread in the SEDs due to differences in stellar parameters and unknown distances, we normalize each object's SED to the median \textit{H}-band flux for the sample, as is done for the Taurus median computed by \citet{furlan06}, before computing the median at each photometry point. Flagged objects are not included in the median SED.
   
   The median SED for the whole sample (FDs and TDs) is dominated by the FD flux as they are more numerous. The L1641 FD median follows the Taurus median quite well. When the TDs are separated from the FDs, a clear distinction arises between the two, namely, that the TD median is much lower than the FD median at NIR and MIR wavelengths, while it is slightly higher at 70 \mic. A similar result was found for TDs with \textit{Herschel} data in Chamaeleon by \citet{ribas13}.

\section{Comparison with Models}\label{sect: comparison with models}

    \subsection{Description of Models}\label{subsect: description of models} 
    
    We use the D'Alessio irradiated accretion disk (DIAD) models  \citep{d'alessio98,d'alessio99,d'alessio01,d'alessio06} to study to our protoplanetary disk sample, allowing us to identify trends in disk properties. These models calculate the structure and emission of accretion disks irradiated by the central star. The models include the effects of dust settling, by having two grain populations, each with a size distribution $n(a) \propto a^{-3.5}$ between $a_{\rm min}$ and $a_{\rm max}$. The population in the upper layers has $a_{\rm max}=0.25$ \mic, while the disk midplane population has $a_{\rm max}=1$ mm \citep{d'alessio06}. For both populations, $a_{\rm min} = 0.005$ \mic. We assume a dust mixture of silicates and graphite with silicate mass abundances relative to gas of 0.004 and graphite fractional abundances of 0.0025, with opacities calculated with Mie theory using optical constants from \citet{dorschner95} for olivine silicates and from \citet{draine&lee84} for graphite. The input model parameters that we keep fixed are the disk radius, the height of the directly irradiated inner disk edge or the ``wall", the viscosity parameter ($\alpha$) which is parameterized using the methods of \citet{shakura&sunyaev73}, and the stellar parameters for the central star (mass, radius, and effective temperature). We vary the following input parameters: the mass accretion rate (\Mdot), the inclination angle of the disk to the line of sight ($i$), and the dust-to-gas mass ratio in the upper layers of the disk in units of the total dust-to-gas mass ratio ($\epsilon$, with a value of 1 corresponding to an unsettled disk, while a low value of 0.0001 corresponds to a factor 10$^4$ depletion of small grains in the upper layers of the disk atmosphere). We discuss the input values in Section~\ref{subsect: index comparison with models}. 

    \subsection{Index Comparisons with Models}\label{subsect: index comparison with models}
    
    We calculate theoretical SEDs from our grid of models, convolving the theoretical fluxes with the filter response at each band to create synthetic colors which we use to calculate the indices for each model. We select the $n_{K_{S}-[70]}$ index (see Equation~\ref{eq: indice}) as an indicator of the total flux in the PACS 70 \mic \ band relative to the star. We adopted \textit{K$_{S}$}-band magnitudes as our anchoring point instead of a shorter wavelength filter because the \textit{K$_{S}$}-band is less affected by reddening and a large fraction of our sample is heavily
    reddened (Fig. \ref{fig: Av hist}). We performed a two-sample Kolmogorov-Smirnov (K--S) test between the $n_{K_{S}-[70]}$ and $n_{J-[70]}$ distributions for the total samples, the FDs, and the TDs separately and find that there is no significant difference between using $J$ and $K_{S}$ as our anchoring point.

    In this work, we focus on the effects of varying mass accretion rates, inclinations, and dust settling and how these parameters affect our index comparison with observations in L1641. The mass of the disk depends on the surface density and radius.  The surface density depends on \Mdot/$\alpha$. Here, when we fix $\alpha$ and the radius and vary \Mdot, we are essentially varying the mass of the disk \citep{d'alessio98}.

    Inclination can have significant effects on model SEDs \citep{d'alessio06}, especially in the edge-on case. When the disk is edge-on, the star is extinguished by the disk. We note that we do not expect any of the disks in this sample to be edge-on, since the extinction of the central star in the optical and near-infrared wavelengths would have prevented these disks from being identified as Class II objects. Besides the edge-on case, the closer to edge-on the disk is, the more of the hot inner wall that will be observed, making the NIR flux higher than a disk that is closer to face-on. The DIAD models assume a completely vertical inner wall at the dust sublimation radius for FDs; in the future, changes may be made to include curved walls \citep{mcclure13b}. 
    
    Dust settling is a marker of disk evolution. As small dust grains settle from the upper layers of the disk atmosphere to the disk midplane, the flux from MIR to FIR wavelengths decreases \citep{d'alessio06}. The NIR is dominated by the wall, which is unaffected by disk settling. Thus, NIR fluxes anchor our indices such that we can probe the degree of dust settling with the FIR.
    
\begin{figure*}[h!]
\epsscale{1}
\plotone{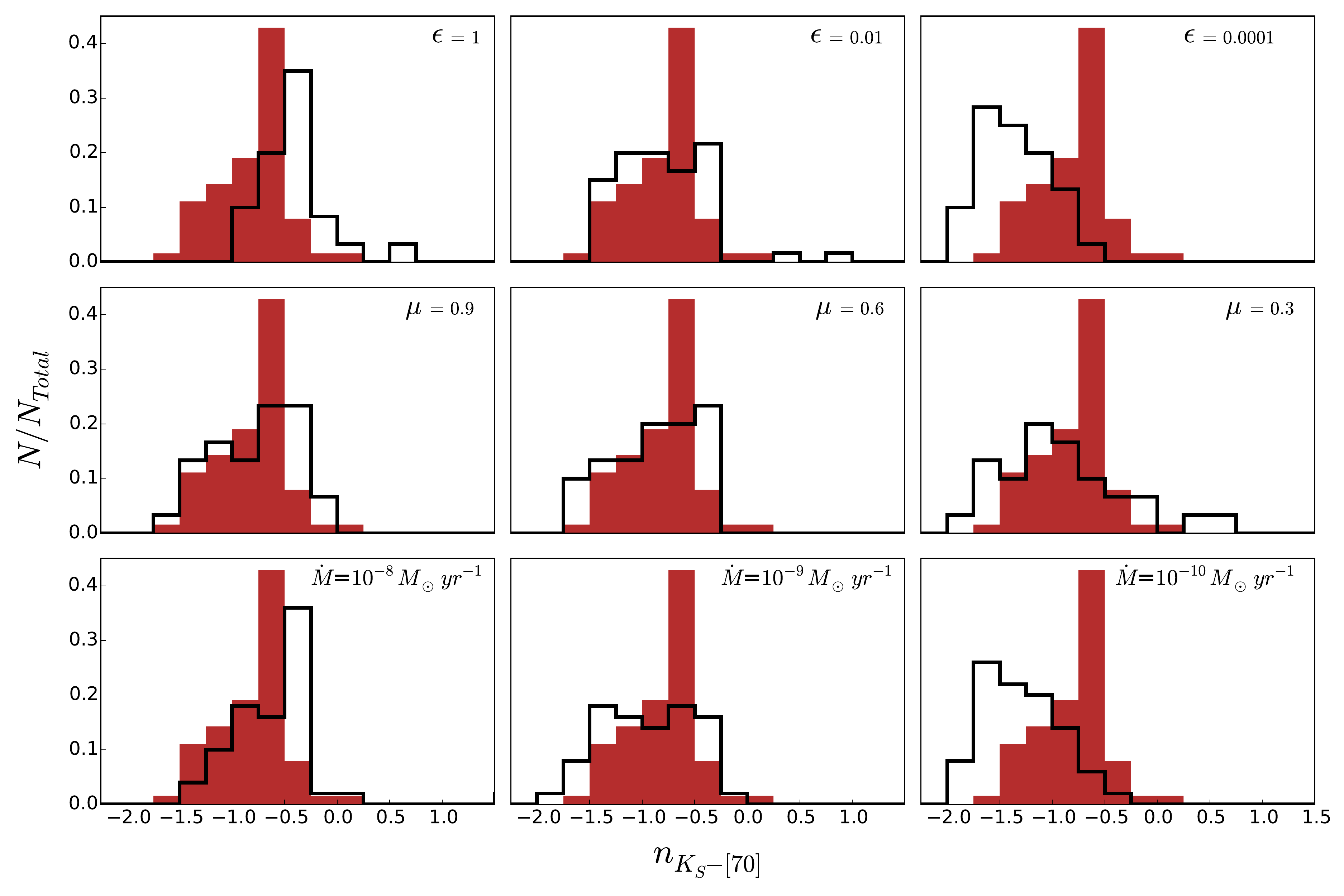}
\centering
\caption{Comparison of the observed $n_{K_{S}-[70]}$ of L1641 FDs (red) to synthetic $n_{K_{S}-[70]}$ from the disk model grid (black outline) described in Section~\ref{subsect: description of models}. Each panel shows the $n_{K_{S}-[70]}$ of models corresponding to the dust settling, inclination, or accretion rate noted in the upper right corner, letting the other parameters vary over the chosen grid range.  Each distribution has been normalized by the  total number in that sample.
}\label{fig: indice data and model}
  %% From ModelColorsandIndices.py
\end{figure*}

    We adopt the median spectral type of the sample, M1, to describe the stellar properties that we use as input to the model. %median spt of just FDs is M0.5.
    Although the sample contains a wide range of spectral types, the use of a spectral index anchored at a stellar-dominated wavelength mitigates the effects caused by the choice of one spectral type to represent the sample. We adopt an age of 1 Myr and a mass of $\sim$0.4 \Msun\ and a radius of $\sim$2 \Rsun\ for our M1 star, consistent with the \citet{siess00} isochrones. We use a \Teff \ of $\sim$3600 K based on stellar temperatures given in \citet{k&h95}. We vary (1) the mass accretion rate as $\log_{10}$(\Mdot) from --10 to --7.5 \msunyr, sampled in steps of 0.5, (2) the inclination of the disk parameterized by $\mu=\cos(i)$ from 0.1 to 0.9, in steps of 0.1, and at 0.99 (nearly edge-on at $\mu=0.1$, $\sim$84$^{\circ}$, and very nearly face-on at $\mu=0.99$, $\sim$8$^{\circ}$), and (3) the dust settling, $\log_{10}(\epsilon)$, from --4 to 0, in steps of 1.0. This results in a grid of 300 models. We use a constant value of $\alpha$ set to 0.01 and an outer radius for the disk at 300 AU. In Figure~\ref{fig: indice data and model}, we show that the synthetic model index distribution covers the observed distribution.
    
\begin{figure*}[h!]
\epsscale{0.48}
\plotone{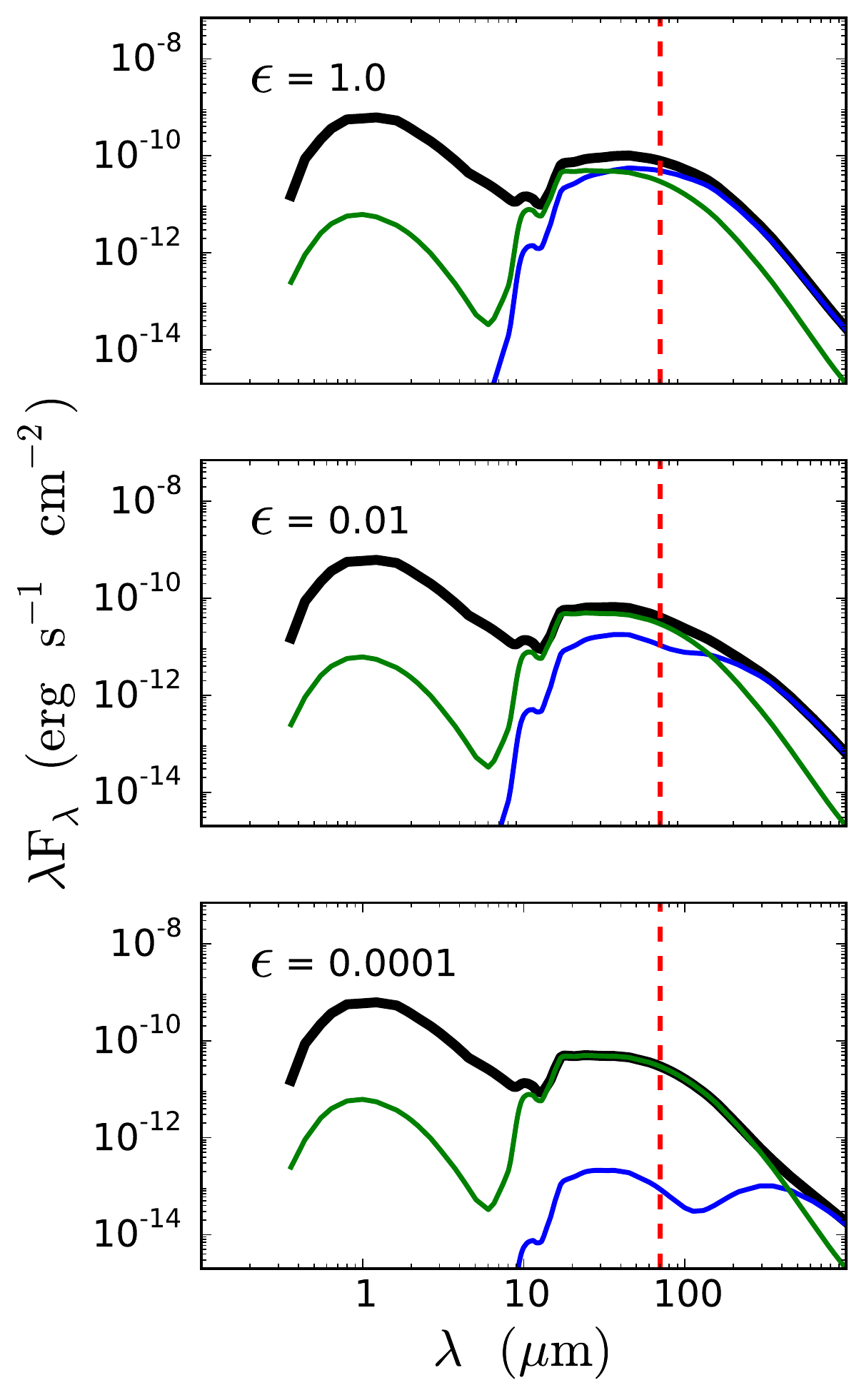}
\centering
\caption{Transitional disk models with varying degrees of settling. The total model (black) includes the photosphere (not shown) and the total disk emission which is composed of emission from the wall (green) and the outer disk (blue). This model has a wall temperature of 150 K and a gap size of $\sim$13 AU. Even for moderate settling, the wall emission dominates the flux at 70 \mic\ (red dashed line).}\label{fig: model TD sed}
  %% From TDModels_T150.py
\end{figure*}

    We do not include forward modeling analysis of the TDs in this work. Depending on the size of the gap in the disk, the 70 \mic\ flux could be dominated by the wall emission, as opposed to the disk emission. Figure~\ref{fig: model TD sed} shows an example TD model with a wall temperature of 150 K and a gap size of $\sim$13 AU showing that for only the most unsettled disks does the disk emission dominate over the wall emission at 70 \mic.  As we aim to derive information about the disk properties, we cannot use 70 \mic\ fluxes unless the gap size is known. Future work will be done to use the DIAD models to model the TDs in detail. 
    
% \vspace{-4.85mm}
% \vspace{-4mm}

    \subsection{Forward Modeling}\label{sect: stat analysis} 
\begin{figure*}[]
\epsscale{1}
\plotone{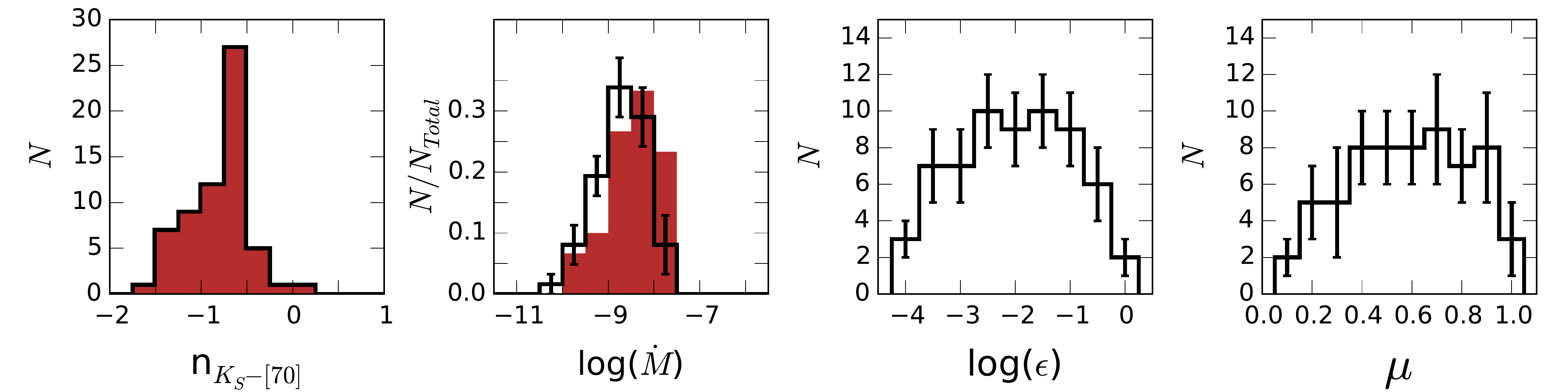}
\centering
\caption{Statistical analysis results for the L1641 FD sample. We show the distributions for observed (red) and predicted (outline) distributions of $n_{K_{S}-[70]}$, mass accretion rate, dust settling, and inclination  ($\mu = \cos(i)$). Error bars are standard deviations of the forward-modeling realizations.}
\label{fig: l1641 posteriors_all}
  %% From Plot_StatDistributions_line.py in FDs folder
\end{figure*}

    We use forward modeling to reproduce the observed FD L1641 $n_{K_{S}-[70]}$ distribution using the DIAD models. We carried out 100 realizations (above this, there is minimal change), and in each, we randomly sampled the cumulative distributions of the grid of $\log_{10}$(\Mdot), $\cos(i)$, and $\log_{10}(\epsilon)$ and calculated the $n_{K_{S}-[70]}$ index for that combination of model parameters. We repeated this until we found the distributions of these parameters that reproduced the observed distribution of $n_{K_{S}-[70]}$. We adopted a uniform distribution for parameters $\mu$ (assuming that disks are randomly oriented) and $\epsilon$, i.e., we sample uniformly from an even distribution. To determine a cumulative distribution for the mass accretion rate, we used the \Mdot\ values from the literature for the sample, as listed in Table~\ref{tab: stellar properties}. We only have mass accretion rates for a subset of the sample and these are likely biased towards higher values because these are more easily measured. Therefore, we do not expect to reproduce the observed distribution. We restrict the analysis sample to have mass accretion rates between $10^{-7.5}$ and $10^{-10}$ \msunyr. The highest mass accretion rates ($>$10$^{-7.5}$ \msunyr) are largely associated with high visual extinction values and we remove these to avoid uncertain measurements. We also remove low mass accretion rates ($<$10$^{-10}$ \msunyr) as these may have chromospheric contamination \citep{ingleby11}.
    The adopted distribution of mass accretion rates is shown in the second column of Figure~\ref{fig: l1641 posteriors_all}. We note that we use the 63 FD objects that have $n_{K_{S}-[70]}$ values (i.e., are not missing \textit{K$_{S}$}-band measurements or extinctions), have mass accretion rates between $10^{-7.5}$ and $10^{-10}$ \msunyr, and have not been flagged (for contamination by close sources and/or nebulosity, high \Av, or colors not representative of CTTS objects), as discussed in Section~\ref{sect: obs}.  
    
    The resulting mean of the distributions that reproduce the observed $n_{K_{S}-[70]}$ distribution are shown in Figure~\ref{fig: l1641 posteriors_all}, where the error bars indicate the standard deviation between the results of each realization. The results in Figure~\ref{fig: l1641 posteriors_all} indicate that the observed \Mdot\ distribution is biased towards high values and that we may be missing lower accretion rate values for the sample, as expected. The distribution for $\mu$ is consistent with a uniform distribution, except at very high inclinations. This is to be expected, since it is difficult to identify edge-on disks. The $\epsilon$ distribution is relatively uniform except at the extreme unsettled and settled ends of the parameter space. We discuss this result and put it into context with other regions in Section~\ref{sect: discussion}.
    
\begin{figure*}[h]
\epsscale{0.73}
\plotone{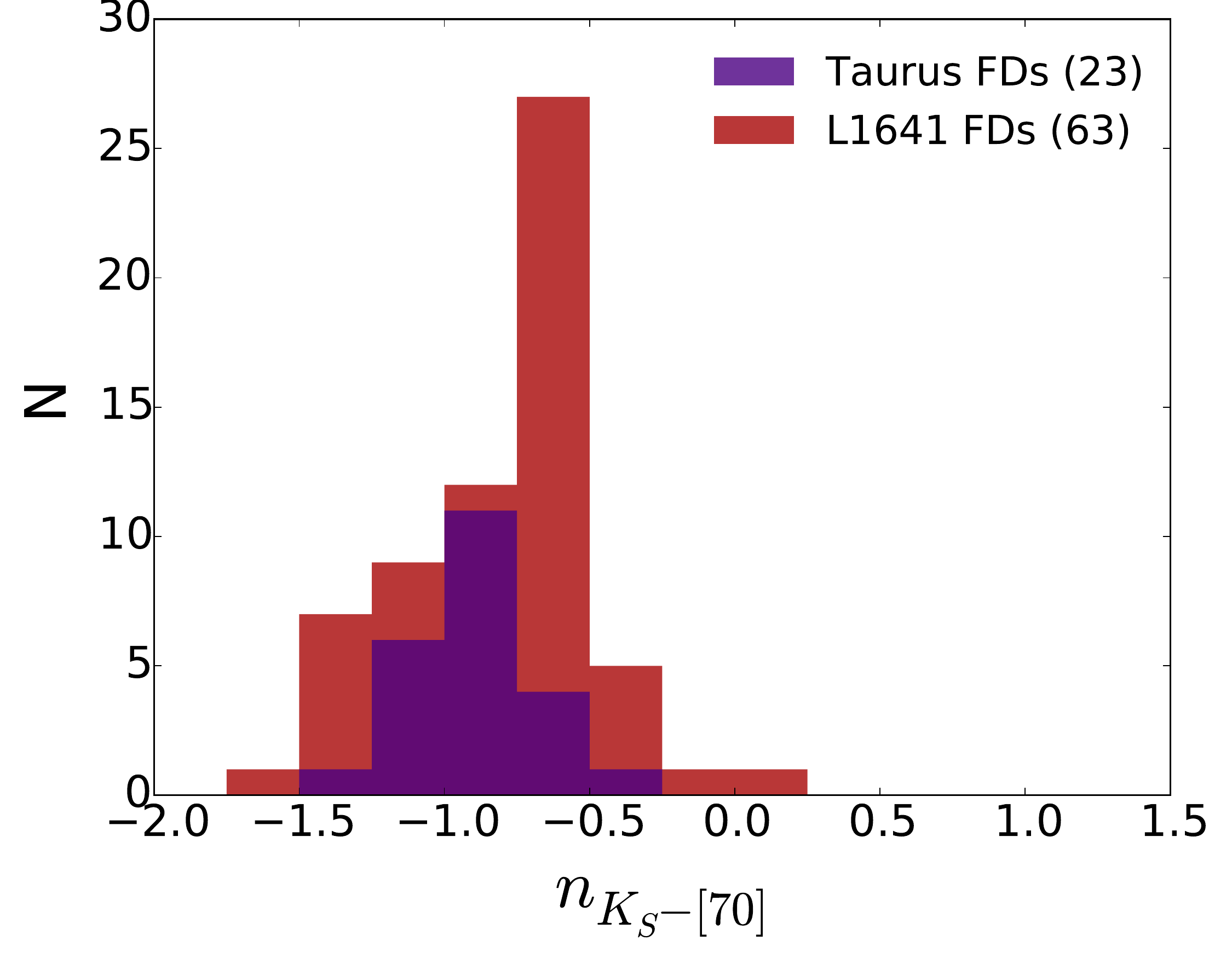}
\centering
\caption{{The distribution of the $n_{K_{S}-[70]}$ index for L1641 FDs (red) and Taurus FDs (purple).}}\label{fig: Taur nK_70 comparison} %no flagged objects in L1641
  %% From Taur_comparison.py
\end{figure*}

\begin{figure*}[]
\epsscale{1}
\plotone{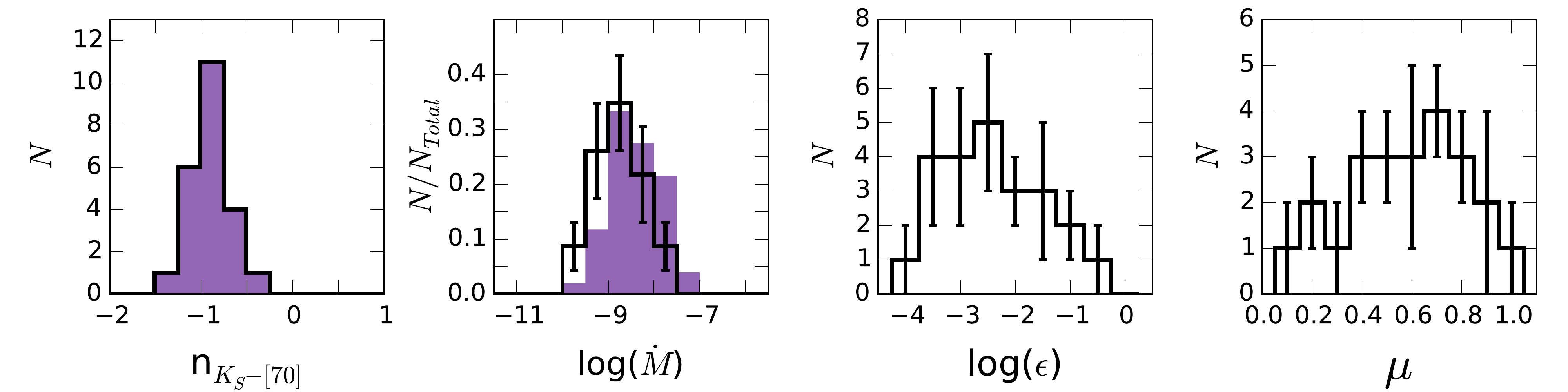}
\centering
\caption{Statistical analysis results for the Taurus FD sample. We show the distributions for observed (purple) and predicted (outline) distributions of $n_{K_{S}-[70]}$, mass accretion rate, dust settling, and inclination  ($\mu=\cos(i)$). Error bars are standard deviations of the forward-modeling realizations.}
\label{fig: taurus posteriors_all}
  %% From PlotStatDistributions.py
\end{figure*}

    We also performed this analysis on a Taurus sample in order to compare the two regions. The Taurus $n_{K_{S}-[70]}$ indices are calculated for 23 Class II objects with \textit{K$_{S}$}-band magnitudes from \citet{k&h95} and PACS 70 \mic\ fluxes from \citet{howard13}. We only use objects that were not classified as transitional disks in \citet{howard13} and have \Av\ values in \citet{k&h95}. The objects are dereddened in the same way as described for the L1641 sample in Section~\ref{sect: seds}. We compare the FD $n_{K_{S}-[70]}$ distributions for the two samples in Figure~\ref{fig: Taur nK_70 comparison}, finding that L1641 peaks at a slightly higher index. Mass accretion rates for 51 systems in Taurus are taken from \citet{hartmann98} and \citet{white&ghez01} to produce the cumulative distribution that is sampled to reproduce the distribution of $n_{K_{S}-[70]}$. The sample for which we have mass accretion rates for and for which we have $n_{K_{S}-[70]}$ overlaps partially.  We restrict the objects in the Taurus sample to have $-10.0\leq log_{10}(\dot{M})\leq-7.5$ as we did for L1641. We use the same model grid and parameter space for both L1641 and Taurus since the index distributions cover the same range. These results are shown in Figure \ref{fig: taurus posteriors_all} and are discussed in Section \ref{sect: discussion}.

\section{Discussion}\label{sect: discussion}  

    Here we discuss the degree to which the dust is processed or settled in our L1641 sample and compare this sample to other regions to place it into a larger context. The L1641 FDs show a wide range of dust settling. We find a trend in the dust settling toward a depletion factor of 100--1,000 ($\log_{10}(\epsilon)\sim-2$ -- $-3$; Figure~\ref{fig: l1641 posteriors_all}). For disks with $\log_{10}(\epsilon)=-1$, the gas-to-dust ratio in the atmosphere of the disk is a factor of ten larger than the $100:1$ usually assumed for the ISM, with larger grains having settled toward the midplane. The observed disk indices are consistent with models that have low values of $\epsilon$ and thus may be settled.
    
    Age spreads and uncertainties make it difficult to directly link age and dust settling for a given region; however, we can compare global trends for L1641 to those found for the Taurus, Chamaeleon I, and Ophiuchus star-forming regions. Ophiuchus is slightly younger than our sample at $<$1 Myr \citep{luhman&rieke99}, Taurus is similarly aged at $\sim1-2$ Myr \citep{k&h95,hartmann01,luhman10,andrews13}, and Chamaeleon I is slightly older with a median age of 2 Myr \citep{luhman04}, although all regions have some age spread. \citet{furlan06,furlan09a}, also using the DIAD models, find that Taurus, Ophiuchus, and Chamaeleon I all have dust depletion factors of 100 to 1,000 ($\log_{10}(\epsilon)\sim-2$ to $-$3), indicating that dust evolution and settling are well underway by even the first $\sim$1 Myr. Similar conclusions were reached by \citet{lewis&lada16}, who found that the median SEDs from 2--8 \mic\ of all Class II objects in Orion A from \citet{megeath12}, are consistent with a flat disk, indicative that a significant number of these disks has undergone a high degree of dust settling.
    
    We performed a two-sample K--S test on the FD $n_{K_{S}-[70]}$ distributions for the L1641 and the Taurus samples and found that these distributions are statistically different (p-value $\sim$ 0.02). The higher index values of L1641 could be caused by slightly less disk settling compared to Taurus or additional contamination from molecular cloud structures, or a combination of the two. The contamination may come from remnant material associated with the star or structures in the line of sight. We performed our statistical analysis on the Taurus sample described in Section~\ref{sect: stat analysis} (Figure~\ref{fig: taurus posteriors_all}). The mean of the realizations shows that the Taurus FD $n_{K_{S}-[70]}$ distribution is only reproduced by a non-zero degree of dust settling. The Taurus $\log_{10}(\epsilon)$ distribution peaks at $\sim-3$, corresponding to a dust depletion factor of $\sim$1,000. The distribution trends toward lower $\epsilon$ values than L1641, indicating that Taurus is slightly more evolved than the similarly aged L1641. However, both distributions indicate that these young regions already show advanced degrees of dust processing and settling. 

    Theoretical predictions show that small dust grains (with large surface-to-mass ratios) will experience a strong drag force as they move through the gas, causing them to settle to the midplane in a few thousand years while large, compact grains can settle even faster (e.g., \citealt{dullemond&dominik04b}). The process of settling is complicated by the fact that turbulence can cause a mixing of dust grains in disks, counteracting settling; however, turbulence is not effective high in the disk atmosphere where the gas density is sufficiently low \citep{dullemond&dominik04b}. If the gas velocities in the upper layers are low, then dust settling can occur more efficiently (e.g., \citealt{ciesla07,flahery15,flaherty17}). Thus the dust depletion in the FDs of L1641 may be indicative of low turbulence in the upper disk atmosphere. Dust settling increases the dust density in the disk midplane, enhancing further dust growth, indicating that even the FDs in L1641 may be in the process of forming planets (e.g., \citealt{goldreich&ward73, takeuchi&lin02, dullemond&dominik04b, testi14}).

\section{Summary and Conclusions}

    We present \emph{Herschel Space Observatory} 70 \mic\ PACS observations of 104 Class II objects in the Lynds 1641 region of the Orion Molecular Cloud A. We construct the SEDs of the 98 objects that have \Av\ determinations and perform a statistical analysis by comparing our results to the D'Alessio irradiated disk models \citep{d'alessio98, d'alessio99, d'alessio01,d'alessio06} to obtain global properties of the full disk sample. Additionally, we compare our Lynds 1641 sample directly to a sample in the Taurus star-forming region. From this analysis, we reach the following conclusions:
    
    \begin{enumerate} 
	
	\item We identify 24 TDs ($\sim$23\%) in our sample, including eight newly classified objects. There are 74 FDs in the sample.

	\item We calculate the median SED for L1641 from \textit{V}-band to 70 \mic. The FD median follows the scaled Taurus median well, while the TD median shows a much lower flux in the NIR and MIR and a slightly higher 70 \mic\ flux than the scaled Taurus median. 
	
	\item We use forward modeling in conjunction with the D'Alessio models to find model parameter values that reproduce the observed FD $n_{K_{S}-[70]}$ distribution in L1641 and Taurus. This analysis indicates that Taurus is slightly more evolved than our L1641 sample, which is already showing signs of dust processing with dust depletion factors of $\sim$100--1,000.
    
    \end{enumerate}
    
    This work provides a large population of YSOs with FIR \emph{Herschel} observations in a single star-forming region at a common distance. The \emph{Herschel} data presented here offer further insight into this region, and more work can be done to study trends as a function of environment, evolutionary state, stellar mass and disk mass, and more. (Sub-)millimeter data will provide additional information that can be used with more detailed modeling to study these systems in further detail.

\acknowledgments {
We thank the referee for comments that improved the paper. SLG, CCE, and AR acknowledge support from the National Science Foundation under CAREER Grant Number AST-1455042 and the Sloan Foundation. STM and WJF were supported by NASA through awards issued by JPL/Caltech. AMS acknowledges funding from Fondecyt regular (project code 1180350), the ``Concurso Proyectos Internacionales de Investigaci\'{o}n, Convocatoria 2015” (project code PII20150171), and the BASAL Centro de Astrof\'{i}sica y Tecnolog\'ias Afines (CATA) PFB-06/2007. This work is based on observations made with the \textit{Herschel Space Observatory}, a European Space Agency Cornerstone Mission with significant participation by NASA. The \textit{Herschel} spacecraft was designed, built, tested, and launched under a contract to ESA managed by the \textit{Herschel/Planck} Project team by an industrial consortium under the overall responsibility of the prime contractor Thales Alenia Space (Cannes), and including Astrium (Friedrichshafen) responsible for the payload module and for system testing at spacecraft level, Thales Alenia Space (Turin) responsible for the service module, and Astrium (Toulouse) responsible for the telescope, with in excess of a hundred subcontractors. PACS has been developed by a consortium of institutes led by MPE (Germany) and including UVIE (Austria); KU Leuven, CSL, IMEC (Belgium); CEA, LAM (France); MPIA (Germany); INAF-IFSI/OAA/OAP/OAT, LENS, SISSA (Italy); IAC (Spain). This development has been supported by the funding agencies BMVIT (Austria), ESA-PRODEX (Belgium), CEA/CNES (France), DLR (Germany), ASI/INAF (Italy), and CICYT/MCYT (Spain). This work is also based [in part] on observations made with the \textit{Spitzer Space Telescope}, which is operated by the Jet Propulsion Laboratory, California Institute of Technology under a contract with NASA. This publication makes use of data products from the Two Micron All Sky Survey, which is a joint project of the University of Massachusetts and the Infrared Processing and Analysis Center/California Institute of Technology, funded by the National Aeronautics and Space Administration and the National Science Foundation. This paper utilizes the \citet{d'alessio98,d'alessio99,d'alessio01,d'alessio05,d'alessio06} Irradiated Accretion Disk (DIAD) code. We wish to recognize the work of Paola D'Alessio, who passed away in 2013. Her legacy and pioneering work live on through her substantial contributions to the field. }

\appendix
\section{Appendix}
Here we present information about the protoplanetary systems that were flagged as described in Section~\ref{subsect: stellar sample}. In Tables~\ref{tab: flagged observations} and \ref{tab: stellar properties flagged} we present the \textit{Herschel} PACS fluxes and the stellar properties, respectively. For the flagged objects, we have spectral types for 31, luminosities for 40, \Av\ values for 59, and mass accretion rates for 24. Of the 59 \Av\ values, 40 are from the literature and the additional 19 are calculated in this work as described in Section~\ref{subsect: stellar sample}. In Figure~\ref{fig: flagged seds} we present the SEDs of the flagged systems. SEDs are computed as discussed in Section~\ref{sect: seds}.

\FloatBarrier
\begin{deluxetable}{ccccc}
\centering
\tablewidth{0pt}
\tablecaption{\textit{Herschel} Fluxes}
\tablehead{
\colhead{M12 Num} & \colhead{RA(J2000)} & \colhead{Dec(J2000)} & \colhead{F$_{70}$}  \\ \colhead{}& \colhead{}& \colhead{}&  \colhead{(Jy)} 
}
\startdata
198&      05:42:42.4&      -08:48:13.8&      0.403$\pm$0.020   \\ 
217&      05:41:19.6&      -08:40:38.7&      0.086$\pm$0.005   \\ 
223&      05:42:40.8&      -08:40:08.6&      0.041$\pm$0.003   \\ 
225&      05:42:46.1&      -08:40:00.8&      0.088$\pm$0.005   \\ 
227&      05:42:50.5&      -08:39:57.5&      0.265$\pm$0.014   \\ 
228&      05:42:52.5&      -08:39:16.6&      0.121$\pm$0.006   \\ 
231&      05:43:03.9&      -08:39:09.2&      0.063$\pm$0.004   \\ 
232&      05:42:50.0&      -08:39:02.8&      0.0247$\pm$0.0025   \\ 
233&      05:42:51.4&      -08:39:01.9&      0.042$\pm$0.003   \\ 
250&      05:41:42.4&      -08:37:07.2&      0.277$\pm$0.014   \\ 
256&      05:43:02.6&      -08:35:48.8&      0.089$\pm$0.005   \\ 
263&      05:42:45.0&      -08:33:36.3&      0.087$\pm$0.005   \\ 
269&      05:43:13.5&      -08:31:00.5&      0.189$\pm$0.010   \\ 
278&      05:42:53.6&      -08:20:22.6&      0.062$\pm$0.004   \\ 
282&      05:43:04.4&      -08:18:10.8&      0.069$\pm$0.004   \\ 
284&      05:40:33.7&      -08:17:43.4&      0.184$\pm$0.010   \\ 
291&      05:42:38.2&      -08:16:35.4&      0.161$\pm$0.008   \\ 
294&      05:42:42.1&      -08:15:15.1&      0.050$\pm$0.003   \\ 
296&      05:42:35.6&      -08:15:01.8&      0.059$\pm$0.004   \\ 
307&      05:42:49.8&      -08:12:10.3&      0.090$\pm$0.005   \\ 
313&      05:40:25.7&      -08:11:16.8&      0.0350$\pm$0.0025   \\ 
342&      05:41:14.0&      -08:07:57.4&      0.106$\pm$0.006   \\ 
378&      05:41:30.6&      -08:04:47.9&      0.85$\pm$0.04   \\ 
383&      05:40:37.3&      -08:04:03.0&      2.21$\pm$0.11   \\ 
387&      05:40:41.0&      -08:02:18.6&      0.067$\pm$0.004   \\ 
399&      05:41:49.7&      -08:00:32.1&      0.89$\pm$0.04   \\ 
400&      05:41:41.7&      -08:00:18.4&      0.083$\pm$0.004   \\ 
402&      05:41:33.4&      -07:59:56.2&      0.296$\pm$0.015   \\ 
403&      05:41:54.6&      -07:59:12.4&      0.053$\pm$0.004   \\ 
411&      05:41:43.7&      -07:58:22.4&      0.145$\pm$0.008   \\ 
428&      05:40:27.8&      -07:55:36.3&      0.115$\pm$0.006   \\ 
429&      05:40:24.9&      -07:55:35.4&      0.053$\pm$0.003   \\ 
434&      05:41:33.2&      -07:55:02.1&      0.107$\pm$0.006   \\ 
435&      05:41:21.4&      -07:55:01.1&      0.050$\pm$0.003   \\ 
463&      05:41:25.9&      -07:49:50.6&      0.202$\pm$0.010   \\ 
466&      05:41:28.0&      -07:49:22.4&      0.048$\pm$0.003   \\ 
468&      05:40:17.1&      -07:49:14.4&      0.280$\pm$0.014   \\ 
471&      05:40:18.5&      -07:49:06.7&      0.043$\pm$0.003   \\ 
474&      05:40:43.6&      -07:48:47.8&      0.063$\pm$0.004   \\ 
476&      05:40:59.9&      -07:48:16.0&      0.0179$\pm$0.0017   \\ 
477&      05:40:57.5&      -07:48:08.8&      0.236$\pm$0.012   \\ 
483&      05:41:05.5&      -07:47:07.6&      0.219$\pm$0.011   \\ 
485&      05:40:49.3&      -07:46:32.5&      0.154$\pm$0.008   \\ 
487&      05:41:04.6&      -07:45:40.1&      0.231$\pm$0.012   \\ 
488&      05:40:44.1&      -07:45:09.7&      0.0379$\pm$0.0029   \\ 
491&      05:40:44.2&      -07:44:43.5&      0.060$\pm$0.004   \\ 
494&      05:40:44.4&      -07:44:16.7&      0.0337$\pm$0.0029   \\ 
512&      05:40:13.8&      -07:32:16.1&      1.03$\pm$0.05   \\ 
525&      05:40:44.7&      -07:29:54.5&      0.418$\pm$0.021   \\ 
530&      05:40:15.3&      -07:28:46.8&      0.227$\pm$0.012   \\ 
546&      05:39:49.1&      -07:26:17.2&      0.0153$\pm$0.0026   \\ 
553&      05:39:40.5&      -07:26:03.5&      0.021$\pm$0.003   \\ 
556&      05:40:20.4&      -07:25:54.0&      0.107$\pm$0.006   \\ 
561&      05:39:58.9&      -07:25:33.5&      0.65$\pm$0.03   \\ 
574&      05:40:06.5&      -07:23:58.7&      0.405$\pm$0.020   \\ 
579&      05:39:45.8&      -07:22:37.2&      0.0274$\pm$0.0022   \\ 
582&      05:39:32.3&      -07:22:24.2&      0.050$\pm$0.003   \\ 
597&      05:38:50.0&      -07:20:18.5&      0.62$\pm$0.03   \\ 
598&      05:39:44.3&      -07:20:10.5&      0.020$\pm$0.003   \\ 
619&      05:38:55.0&      -07:11:53.7&      0.069$\pm$0.004   \\ 
626&      05:38:36.6&      -07:11:00.2&      0.248$\pm$0.013   \\ 
633&      05:38:17.4&      -07:09:39.5&      0.349$\pm$0.018   \\ 
637&      05:38:23.9&      -07:07:38.9&      0.076$\pm$0.005   \\ 
641&      05:38:13.4&      -07:06:43.4&      0.061$\pm$0.004   \\ 
644&      05:38:47.7&      -07:06:14.8&      0.029$\pm$0.003   \\ 
645&      05:38:41.5&      -07:05:59.3&      0.156$\pm$0.008   \\ 
653&      05:38:09.1&      -07:05:25.8&      0.061$\pm$0.004   \\ 
654&      05:38:46.8&      -07:05:09.1&      0.059$\pm$0.004   \\ 
663&      05:38:44.0&      -07:03:09.5&      0.0201$\pm$0.0024   \\ 
666&      05:38:44.8&      -07:02:47.0&      0.0338$\pm$0.0027   \\ 
673&      05:38:04.8&      -07:02:21.7&      0.0322$\pm$0.0028   \\ 
677&      05:38:56.5&      -07:01:55.4&      0.057$\pm$0.004   \\ 
680&      05:38:41.5&      -07:01:52.5&      0.076$\pm$0.004   \\ 
689&      05:39:01.2&      -07:01:09.5&      0.047$\pm$0.003   \\ 
729&      05:37:54.5&      -06:57:31.1&      0.057$\pm$0.004   \\ 
734&      05:37:51.7&      -06:56:51.8&      0.062$\pm$0.004   \\ 
751&      05:37:49.3&      -06:51:37.4&      0.0347$\pm$0.0028   \\ 
755&      05:37:44.5&      -06:50:36.8&      0.112$\pm$0.006   \\ 
761&      05:37:56.0&      -06:48:54.9&      0.244$\pm$0.012   \\ 
762&      05:36:08.3&      -06:48:36.3&      0.311$\pm$0.016   \\ 
792&      05:37:58.8&      -06:43:33.7&      0.039$\pm$0.003   \\ 
798&      05:36:30.2&      -06:42:46.3&      0.032$\pm$0.004   \\ 
811&      05:36:41.0&      -06:41:17.8&      0.097$\pm$0.006   \\ 
818&      05:37:32.4&      -06:39:05.1&      0.059$\pm$0.004   \\ 
832&      05:34:15.8&      -06:36:04.5&      0.64$\pm$0.03   \\ 
847&      05:37:00.1&      -06:33:27.4&      0.150$\pm$0.008   \\ 
848&      05:36:36.9&      -06:33:24.2&      0.106$\pm$0.007   \\ 
853&      05:34:06.9&      -06:32:08.0&      0.071$\pm$0.005   \\ 
874&      05:35:18.9&      -06:27:25.3&      0.109$\pm$0.007   \\ 
887&      05:35:45.9&      -06:25:59.1&      0.169$\pm$0.009   \\ 
914&      05:36:12.6&      -06:23:39.4&      0.155$\pm$0.008   \\ 
920&      05:35:37.3&      -06:23:26.7&      0.060$\pm$0.004   \\ 
926&      05:36:30.1&      -06:23:10.1&      0.251$\pm$0.013   \\ 
930&      05:35:41.0&      -06:22:45.4&      0.204$\pm$0.011   \\ 
971&      05:36:32.3&      -06:19:19.9&      0.210$\pm$0.011   \\ 
980&      05:36:24.8&      -06:17:30.4&      0.64$\pm$0.03   \\ 
994&      05:35:14.6&      -06:15:12.5&      0.186$\pm$0.010   \\ 
1001&      05:35:27.9&      -06:14:15.0&      0.0275$\pm$0.0029   \\ 
1006&      05:36:40.4&      -06:13:33.3&      0.078$\pm$0.005   \\ 
1007&      05:36:14.7&      -06:13:16.9&      0.071$\pm$0.004   \\ 
1011&      05:35:48.9&      -06:12:07.7&      0.171$\pm$0.009   \\ 
1020&      05:36:40.8&      -06:11:08.2&      0.106$\pm$0.007   \\ 
1039&      05:36:26.1&      -06:08:03.7&      0.230$\pm$0.012   \\ 
1086&      05:34:58.5&      -06:00:00.5&      0.162$\pm$0.009   \\ 

\enddata
\tablecomments{For each object, we list the corresponding identification number from \citet{megeath12}.}
\label{tab:observations}
\end{deluxetable}
\FloatBarrier

\FloatBarrier
\begin{deluxetable}{ccccccccc}
\tablewidth{0pt}
\tablecaption{Stellar Properties}
\tablehead{
\colhead{M12 Num} & \colhead{SpT} & \colhead{SpT Ref.}  & \colhead{\Av} & \colhead{\Av \ Ref.} & \colhead{L} & \colhead{L Ref.} & \colhead{log\Mdot} & \colhead{\Mdot \ Ref.} \\
\colhead{} & \colhead{} & \colhead{} & \colhead{} & \colhead{} & \colhead{(\Lsun)} & \colhead{} & \colhead{(\msunyr)} & \colhead{}
}
\startdata
  198 & M2.0 & a & 7.71 & b & 2.17 & b & $<$ -8.29 & b  \\
  217 &  &  & 10.82 & CTTS J-H &  &  &  &   \\
  223 & M2.5 & a & 0.91 & b & 0.25 & b & -8.2 & c  \\
  225 & K4.0 & a & 5.46 & b & 4.04 & b & $<$ -8.5 & c  \\
  227 & K0.0 & a & 7.2 & c & 2.501 & c &  &   \\
  228 &  &  & 6.21 & CTTS J-H &  &  &  &   \\
  231 & M1.5 & a & 4.7 & b & 0.95 & b & -8.6 & c  \\
  232 &  &  & 6.21 & CTTS J-H &  &  &  &   \\
  233 &  &  & 6.72 & b & 3.54 & b &  &   \\
  250 & K1.0 & a & 5.43 & b & 11.71 & b & -8.2 & b  \\
  %256 &  &  &  &  &  &  &  &   \\
  263 & M2.5 & a & 5.3 & c & 0.527 & c & $<$ -9.4 & c  \\
  269 & M3.0 & b & 6.92 & b & 0.47 & b & -8.02 & b  \\
  278 & K6.0 & a & 4 & c & 0.896 & c & $<$ -8.6 & c  \\
  282 & M2.5 & a & 2.92 & b & 0.35 & b & -9.2 & b  \\
  284 &  &  & 10.82 & CTTS J-H &  &  &  &   \\
  291 &  &  & 8.25 & CTTS J-H &  &  &  &   \\
  294 & M3.0 & a & 3.4 & b & 0.32 & b & -9.15 & c  \\
  296 & M2.0 & a & 2.14 & b & 0.35 & b & -9.05 & b  \\
  307 & K7.0 & a & 4.1 & c & 0.575 & c & -7.75 & c  \\
  313 & M3.0 & a & 5.3 & CTTS J-H &  &  &  &   \\
  342 & M2.0 & a & 1.41 & b & 0.95 & b & $<$ -8.52 & b  \\
  378 & K7.0 & d & 6.88 & b & 1.95 & b & -6.74 & d  \\
  383 & F & d & 6.6 & b & 59.31 & b &  &   \\
  387 & M0 & d & 7.9 & d & 1.126 & d &  &   \\
  399 & K5.0 & d & 0.03 & b & 5.83 & b & -6.69 & d  \\
  400 & M0.0 & a & 4.5 & d & 0.047 & d &  &   \\
  402 & M1.5 & d & 8 & b & 1.05 & b & $<$ -9.09 & b  \\
  403 & K3.5 & d & 5.07 & b & 1.9 & b & -8.4 & d  \\
  411 & M7.5 & a & 0.59 & b & 0.4 & b & -8.22 & b  \\
  428 &  &  & 8.46 & b & 1.05 & b &  &   \\
  429 & M3.0 & a & 8.2 & d & 1.723 & d & -7.15 & d  \\
  434 & K7.0 & a & 0.9 & b & 0.55 & b & -8.05 & c  \\
  %435 &  &  &  &  &  &  &  &   \\
  463 & K7.0 & a & 4.26 & b & 1.03 & b & $<$ -8.96 & b  \\
  466 & M4.0 & a & 0 & c & 0.014 & c & -10.7 & c  \\
  468 &  &  & 8.55 & CTTS J-H &  &  &  &   \\
  471 &  &  & 9.07 & CTTS J-H &  &  &  &   \\
  %474 &  &  &  &  &  &  &  &   \\
  476 & M3.0 & a & 4.2 & d & 0.187 & d & -8.92 & d  \\
  477 & M1.0 & a & 4.6 & d & 0.566 & d & -8.26 & d  \\
  483 & G1.0 & d & 10.6 & d & 8.105 & d & -6.54 & d  \\
  485 & M0.0 & a & 6.33 & b & 0.52 & b & $<$ -9.55 & b  \\
  487 & M0.5 & a & 2.1 & c & 1.134 & c & $<$ -9.15 & c  \\
  488 &  &  & 10.77 & CTTS J-H &  &  &  &   \\
  491 &  &  & 12.26 & CTTS J-H &  &  &  &   \\
  494 &  &  & 9.85 & b & 0.36 & b &  &   \\
  512 & M5.5 & a & 4.1 & CTTS J-H &  &  &  &   \\
  525 & M1.0 & a & 4.26 & b & 1.13 & b & -6.8 & c  \\
  530 &  &  & 9.94 & CTTS J-H &  &  &  &   \\
  %546 &  &  &  &  &  &  &  &   \\
  %553 &  &  &  &  &  &  &  &   \\
  556 & M1.5 & a & 0.87 & b & 0.33 & b & -8.67 & b  \\
  561 &  &  & 13.1 & b & 2 & b &  &   \\
  574 & M3.0 & a & 5.7 & c & 0.509 & c & -8.3 & c  \\
  579 &  &  & 14.6 & b & 4.17 & b &  &   \\
  582 & K3.0 & a & 0.39 & b & 0.14 & b & -9.92 & b  \\
  597 &  &  & 3.38 & CTTS J-H &  &  &  &   \\
  598 &  &  & 11.95 & CTTS J-H &  &  &  &   \\
  619 &  &  & 12.54 & CTTS J-H &  &  &  &   \\
  626 & M4.0 & a & 4.2 & c & 0.8 & c & -7.75 & c  \\
  633 & M3.0 & a & 4.2 & b & 1.07 & b & $<$ -8.47 & b  \\
  637 & K8.0 & a & 1.5 & c & 0.978 & c & -8.8 & c  \\
  641 & K4.5 & d & 8.89 & b & 2 & b &  &   \\
  644 & M1.0 & d & 10.67 & CTTS J-H &  &  &  &   \\
  %645 &  &  &  &  &  &  &  &   \\
  653 & K2.0 & a & 3.4 & c & 2.045 & c & -8.4 & c  \\
  654 & M0.5 & a & 4.3 & d & 0.243 & d & -8.94 & d  \\
  663 &  &  & 11.26 & CTTS J-H &  &  &  &   \\
  666 &  &  & 11.34 & CTTS J-H &  &  &  &   \\
  673 & M5.0 & a & 2.7 & d & 0.136 & d &  &   \\
  677 &  &  & 10.7 & CTTS J-H &  &  &  &   \\
  680 & M2.0 & a & 2.6 & d & 0.183 & d &  &   \\
  689 & M3.0 & a & 3.5 & c & 0.339 & c & -8.8 & c  \\
  729 & K7.0 & a & 0.9 & d & 0.939 & d &  &   \\
  734 & M2.4 & a & 1.4 & d & 1.324 & d & -7.29 & d  \\
  751 & M3.5 & a & 1.41 & b & 0.31 & b & -9.31 & b  \\
  755 & M1.0 & a & 0.66 & b & 0.45 & b & -7.9 & c  \\
  761 & M0.0 & b & 8.25 & b & 1.04 & b &  &   \\
  762 & M2.5 & a & 2.31 & b & 0.85 & b & -7.79 & d  \\
  792 &  &  & 0.0 & CTTS J-H &  &  &  &   \\
  798 & M1.5 & a & 8.01 & b & 1.41 & b & -8.09 & d  \\
  811 & M5.0 & a & 0.4 & c & 0.197 & c & -9.85 & c  \\
  818 & M3.0 & a & 0 & c & 0.228 & c & -10.1 & c  \\
  832 & G8.0 & e & 0.42 & CTTS J-H &  &  &  &   \\
  847 & G3.0 & a & 2.61 & b & 10.44 & b & -7.89 & b  \\
  848 & K7.5 & a & 3.94 & b & 2.05 & b & -6.35 & b  \\
  853 & M0.0 & a & 0.24 & b & 0.15 & b & $<$ -10.6 & c  \\
  874 & M4.0 & a & 0 & d & 0.187 & d & -8.9 & d  \\
  887 & G0.5 & c & 9.8 & c & 13.259 & c & -6.9 & c  \\
  914 & M5.0 & a & 0.4 & d & 0.232 & d & -9.83 & d  \\
  920 & M2.5 & a & 0.21 & b & 0.3 & b & $<$ -8.82 & b  \\
  926 & K6.5 & a & 0.9 & d & 0.45 & d &  &   \\
  930 & M0.0 & a & 1 & d & 1.839 & d & -7.93 & d  \\
  971 & G8.0 & a & 2.8 & c & 4.469 & c & $<$ -8.15 & c  \\
  980 &  &  & 0.35 & b & 8.82 & b & -7.85 & b  \\
  994 & M1.0 & a & 2.3 & d & 0.569 & d & -8.92 & d  \\
  1001 & K8.0 & a & 2.6 & c & 0.723 & c & -8.35 & c  \\
  1006 & M0.5 & a & 0.41 & b & 0.45 & b & $<$ -9.62 & b  \\
  1007 & M1.5 & a & 0.2 & d & 0.355 & d & -8.18 & d  \\
  1011 & K4.0 & a & 0.9 & c & 1.736 & c & -7.65 & c  \\
  1020 & K2.0 & a & 12.3 & b & 1.6 & b &  &   \\
  1039 & M0.0 & a & 1.78 & b & 1.5 & b & $<$ -8.77 & b  \\
  1086 & M0.0 & a & 0.08 & b & 0.66 & b & -8.51 & b  \\

\enddata
\tablecomments{References: $^a$\citet{hsu12}, $^b$\citet{kim16}, $^c$\citet{fang13}, $^d$\citet{fang09}, $^e$\citet{hsu13}. For sources with an \Av\ reference of “CTTS J-H,” visual extinctions were measured in this work as described in Section~\ref{subsect: stellar sample}.}  
%From StellarPropsTable_noflagged which is made with StellarProperties_full.csv and Topcat
\label{tab: stellar properties}
\end{deluxetable}
\FloatBarrier

% \centering
% \newpage
% \FloatBarrier
\begin{figure*}
\includegraphics[angle=0,scale=0.7]{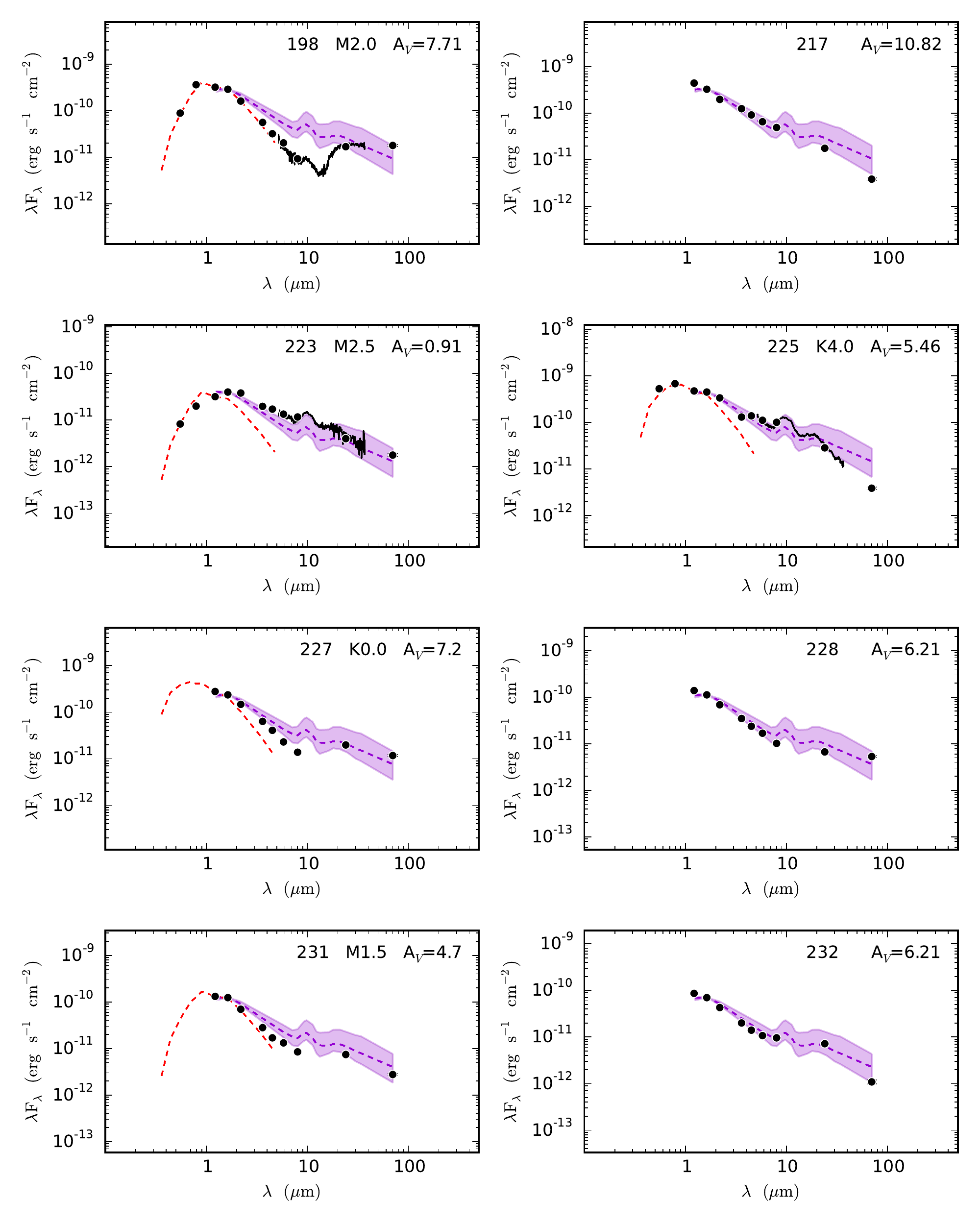}
\caption{SEDs for the non-flagged L1641 sample. Photometry and IRS spectra (where available) shown here are dereddened, as described in Section \ref{sect: seds}. Objects are identified by their \citet{megeath12} number, and we include the adopted spectral type and value of reddening, as given in Table~\ref{tab: stellar properties}. Photospheric fluxes for the adopted spectral type (red dashed lines) are from \citet{k&h95}. The Taurus median, shown as the purple dashed lines, and the Taurus quartiles, shown as the shaded regions, are from \citet{furlan06} and data are from \citet{howard13}.}\label{fig: seds}
\end{figure*}

% \centering
\begin{figure*}%[h]
\ContinuedFloat
\includegraphics[angle=0,scale=0.7]{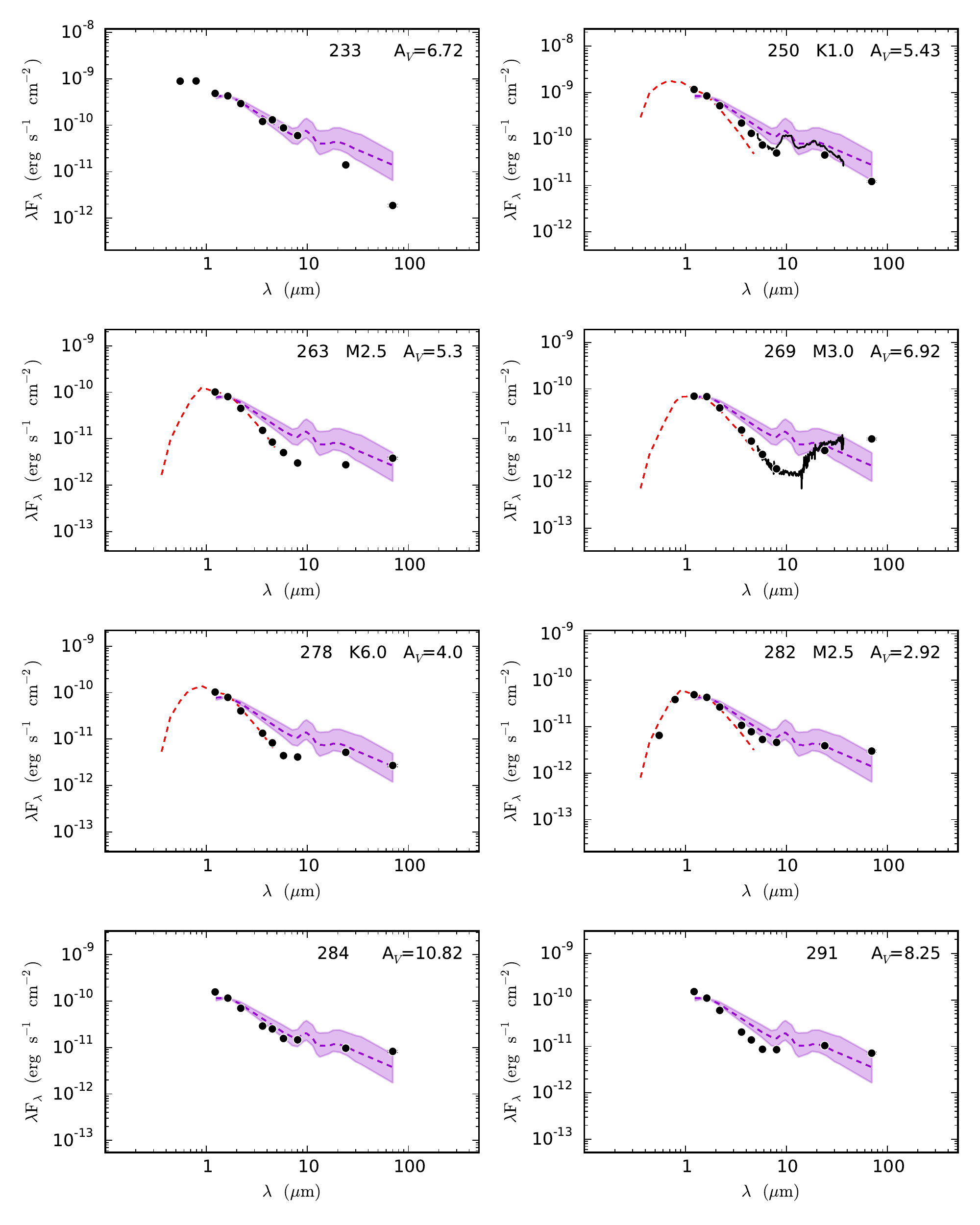}
\caption[]{Continued}
\end{figure*}
% \clearpage

% \centering
% \clearpage
\begin{figure*}
\ContinuedFloat
\includegraphics[angle=0,scale=0.7]{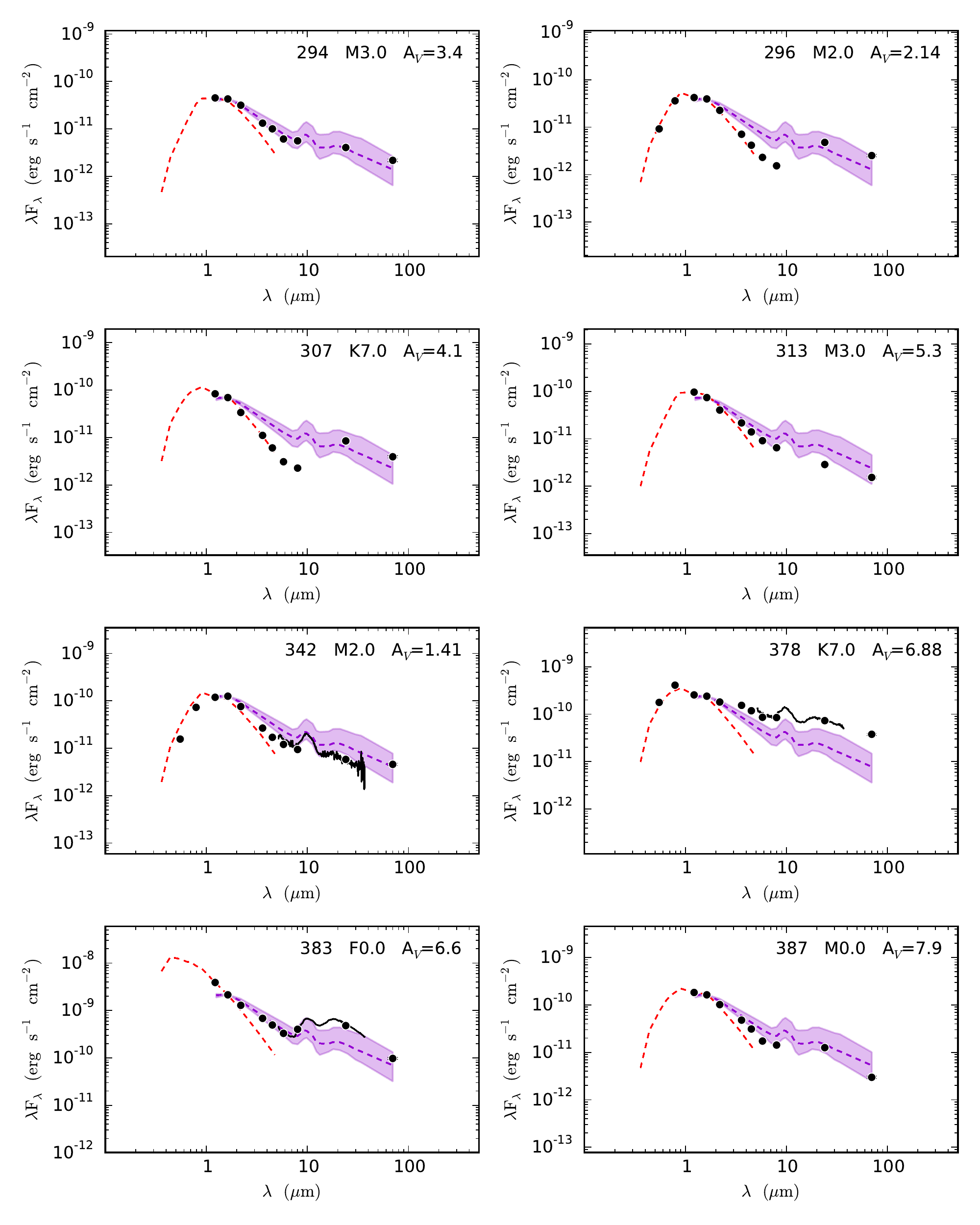}
\caption[]{Continued}
\end{figure*}
% \clearpage

% \centering
% \clearpage
\begin{figure*}
\ContinuedFloat
\includegraphics[angle=0,scale=0.7]{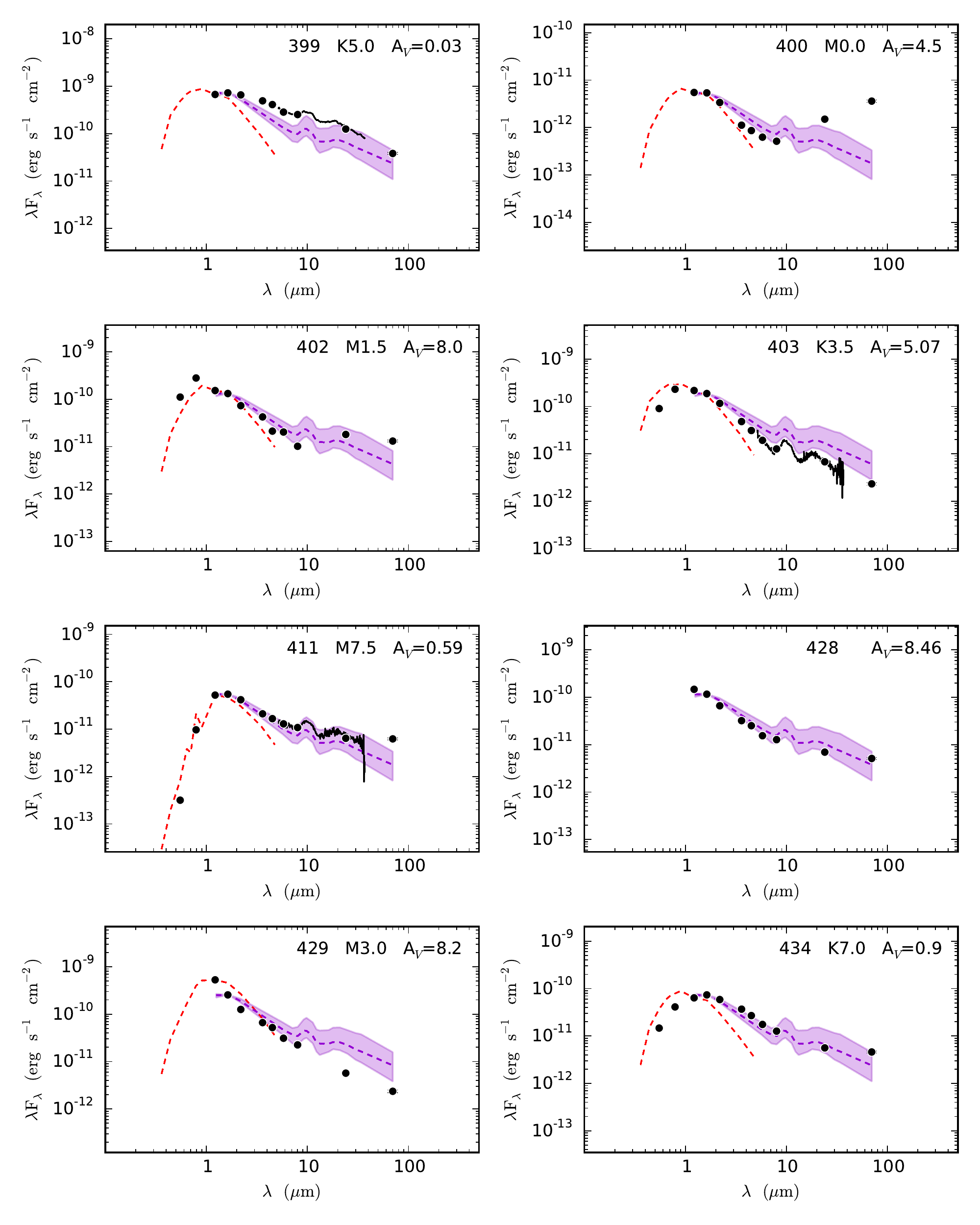}
\caption[]{Continued}
\end{figure*}
% \clearpage

% \centering
% \clearpage
\begin{figure*}
\ContinuedFloat
\includegraphics[angle=0,scale=0.7]{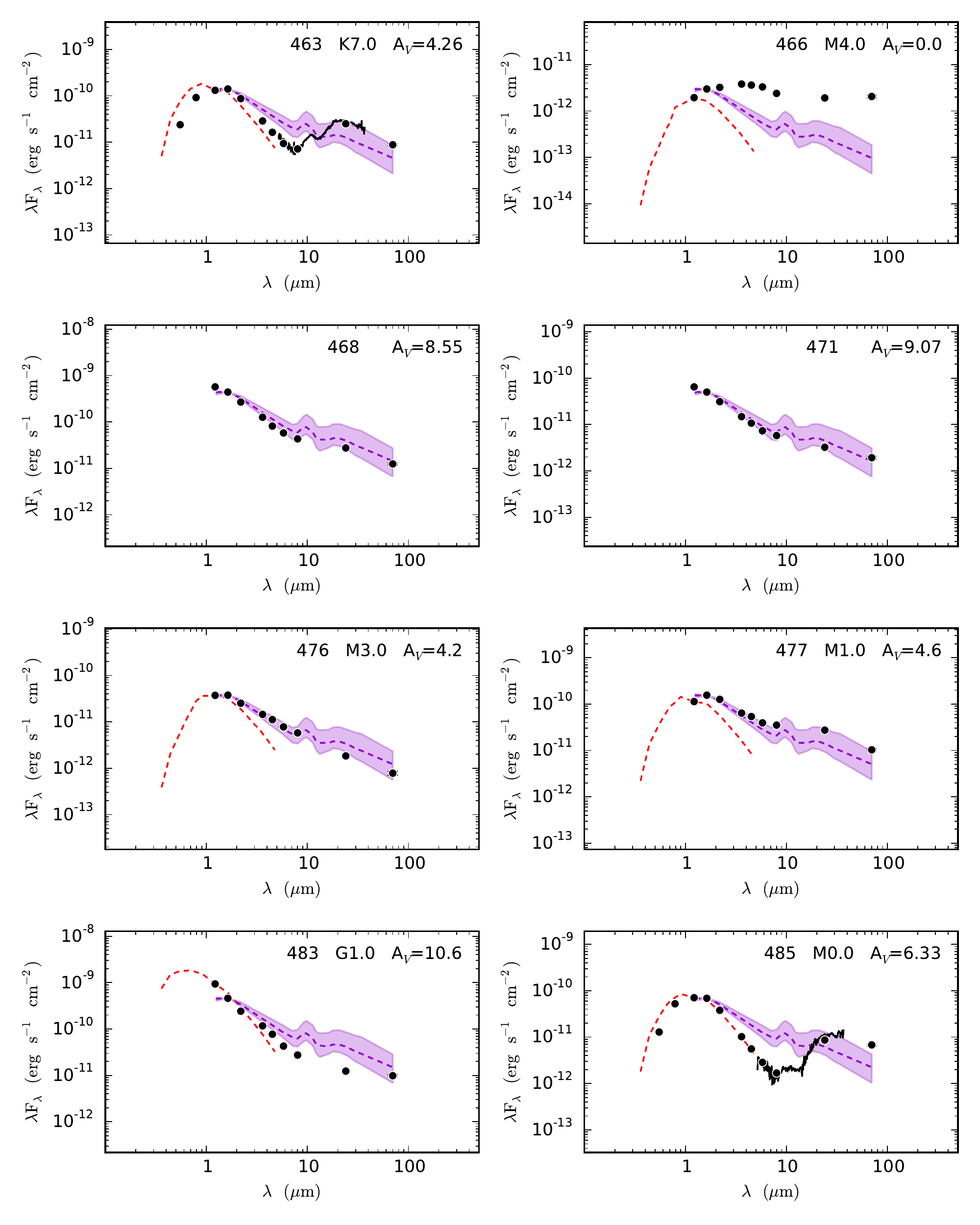}
\caption[]{Continued}
\end{figure*}
% \clearpage

% \centering
% \clearpage
\begin{figure*}
\ContinuedFloat
\includegraphics[angle=0,scale=0.7]{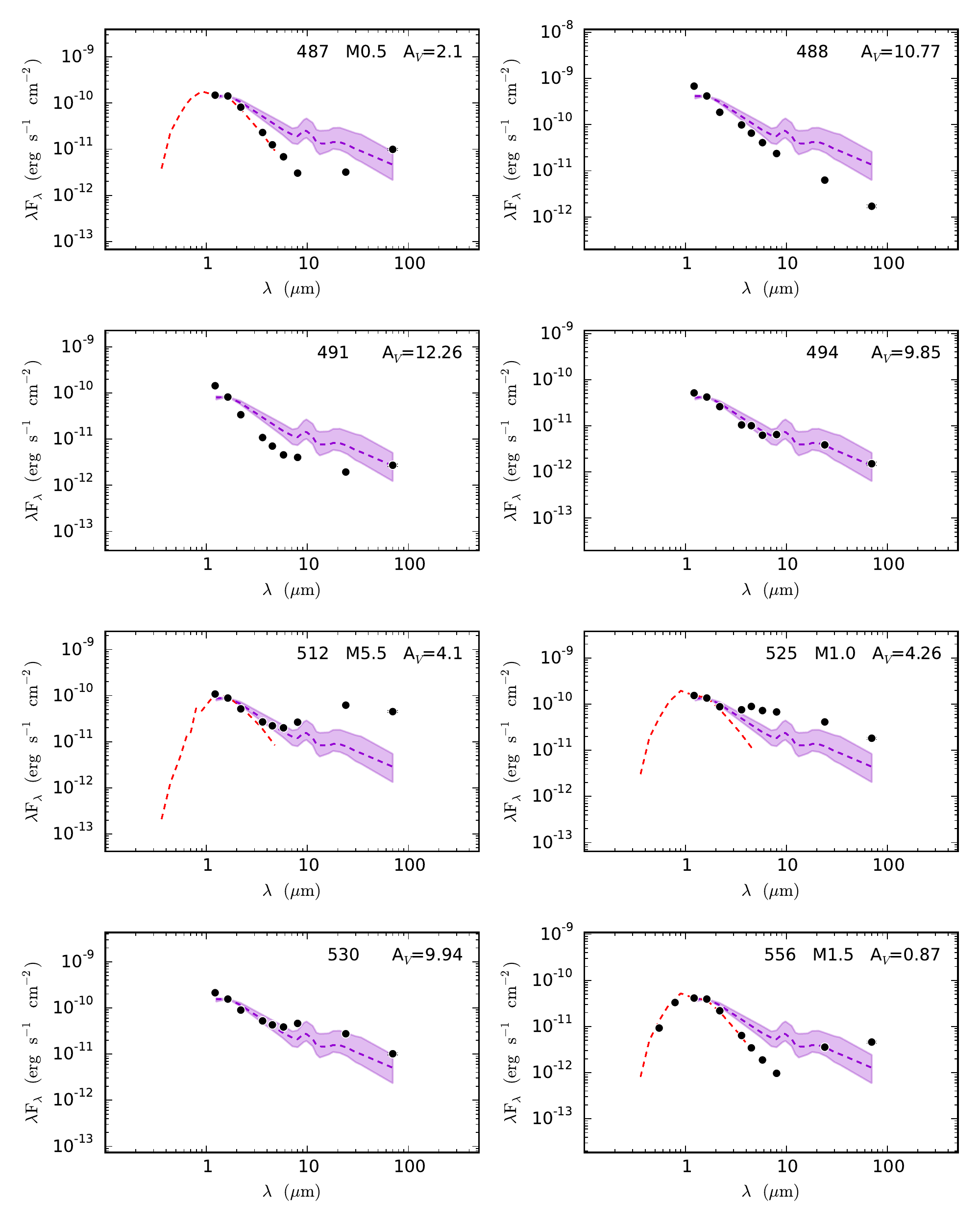}
\caption[]{Continued}
\end{figure*}
% \clearpage

% \centering
% \clearpage
\begin{figure*}
\ContinuedFloat
\includegraphics[angle=0,scale=0.7]{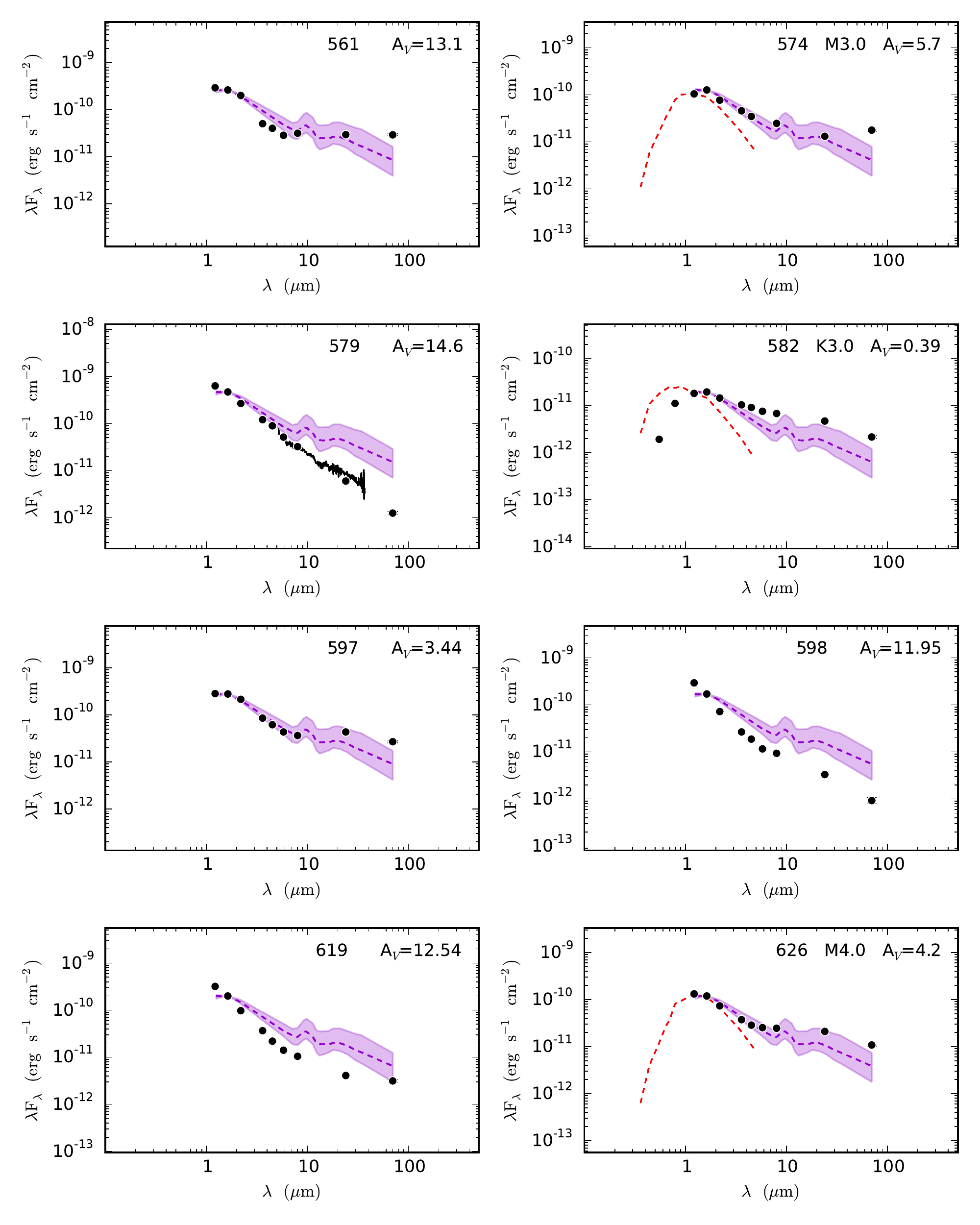}
\caption[]{Continued}
\end{figure*}
% \clearpage

% \centering
% \clearpage
\begin{figure*}
\ContinuedFloat
\includegraphics[angle=0,scale=0.7]{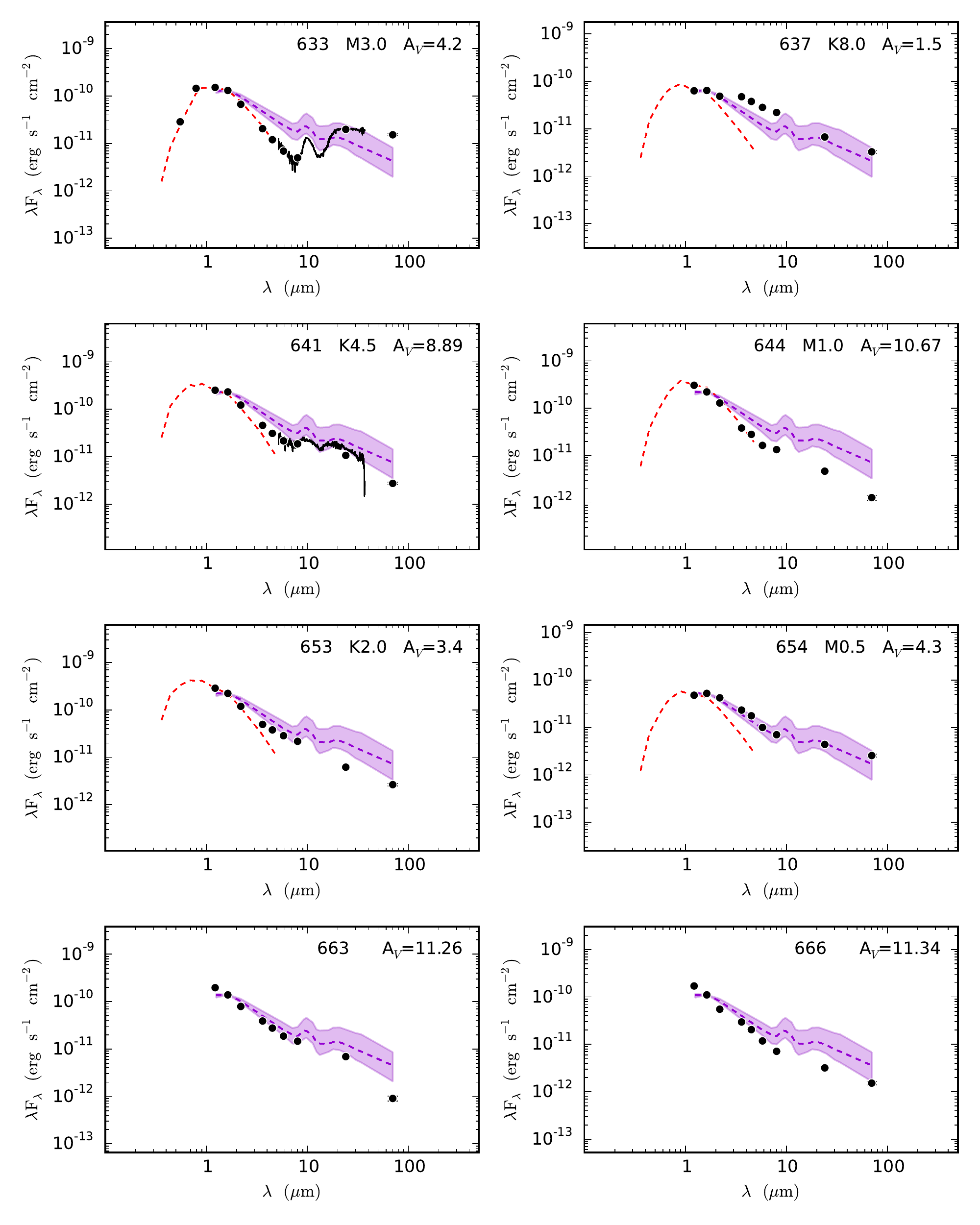}
\caption[]{Continued}
\end{figure*}
% \clearpage

% \centering
% \clearpage
\begin{figure*}
\ContinuedFloat
\includegraphics[angle=0,scale=0.7]{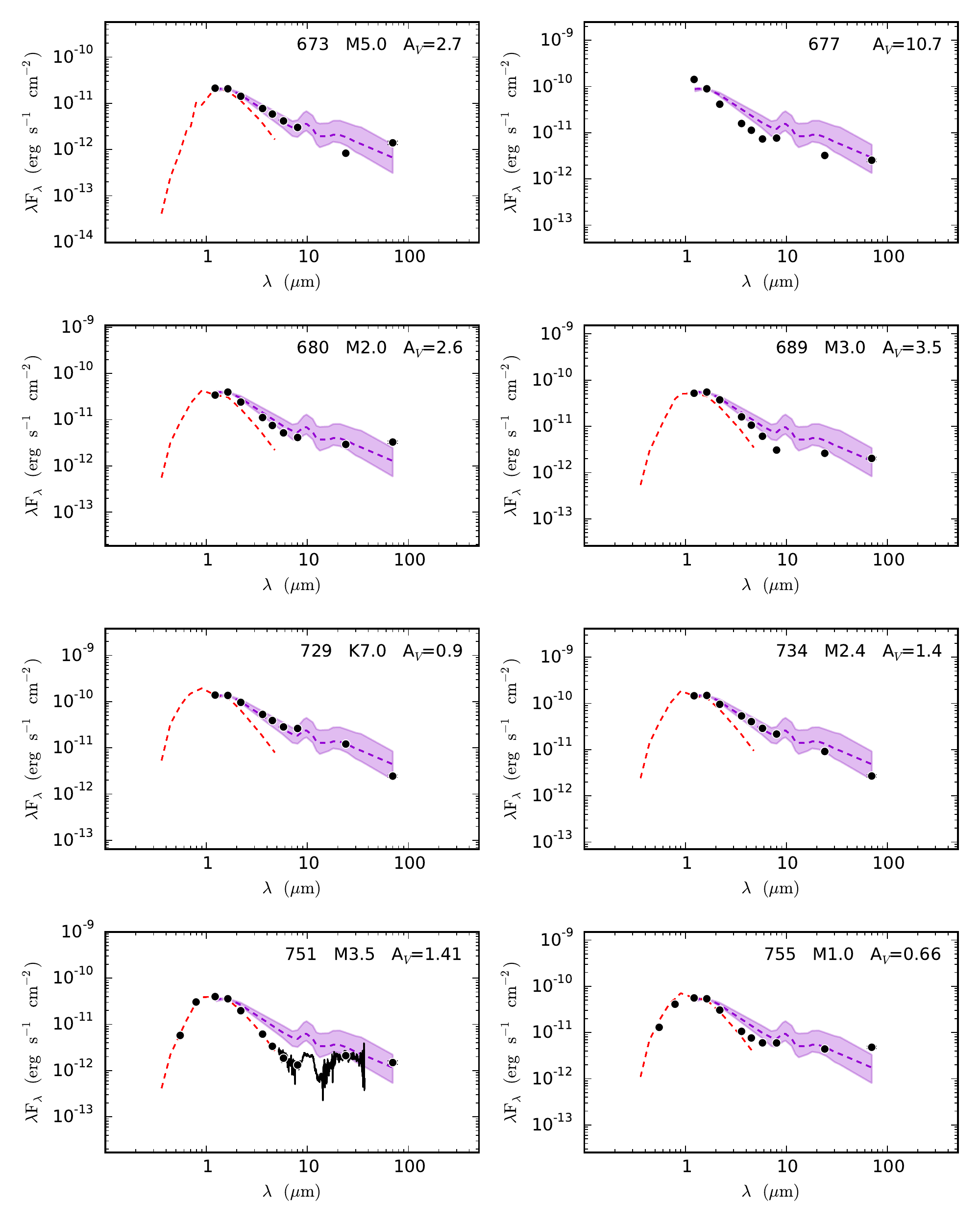}
\caption[]{Continued}
\end{figure*}
% \clearpage

% \centering
% \clearpage
\begin{figure*}
\ContinuedFloat
\includegraphics[angle=0,scale=0.7]{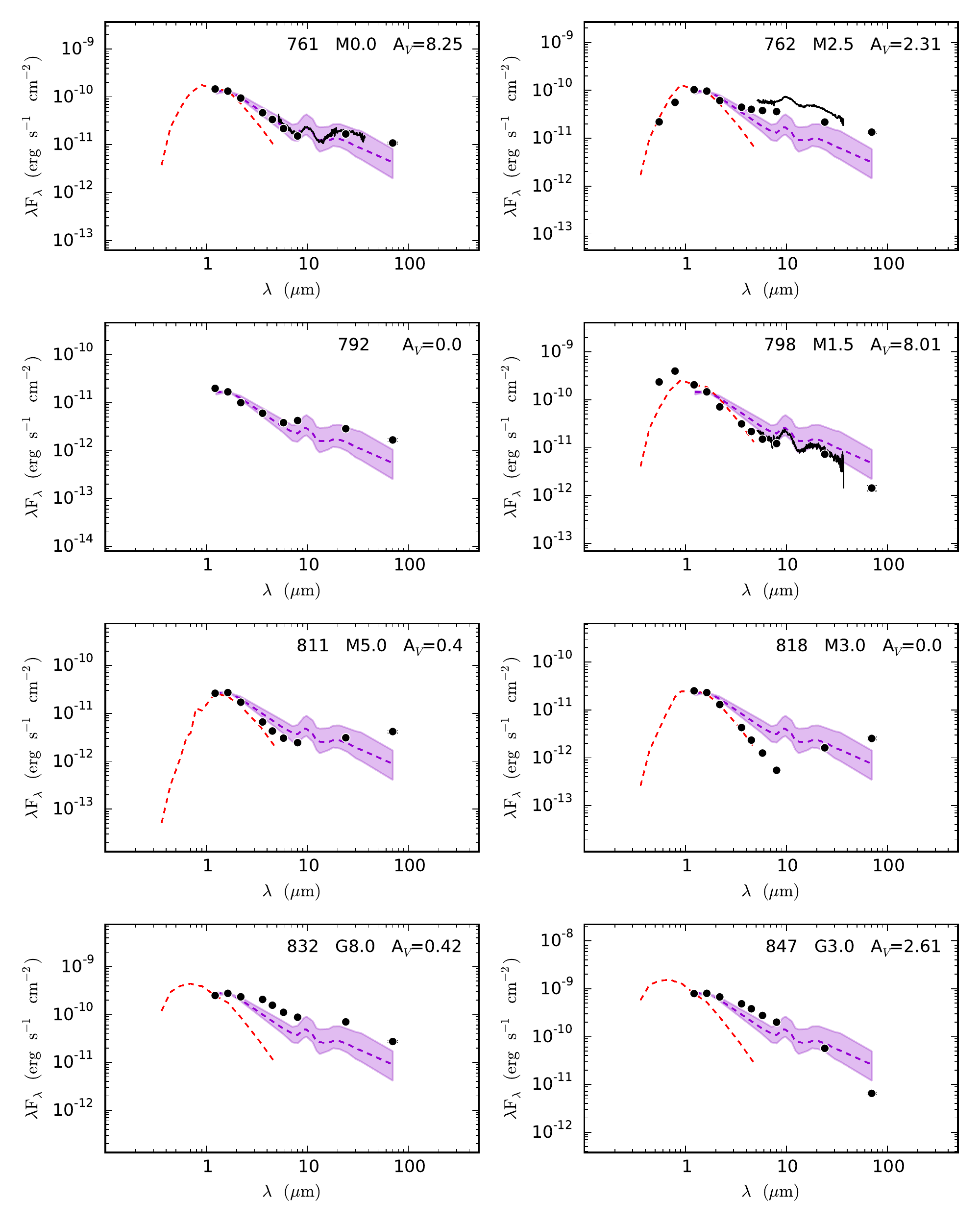}
\caption[]{Continued}
\end{figure*}
% \clearpage

% \centering
% \clearpage
\begin{figure*}
\ContinuedFloat
\includegraphics[angle=0,scale=0.7]{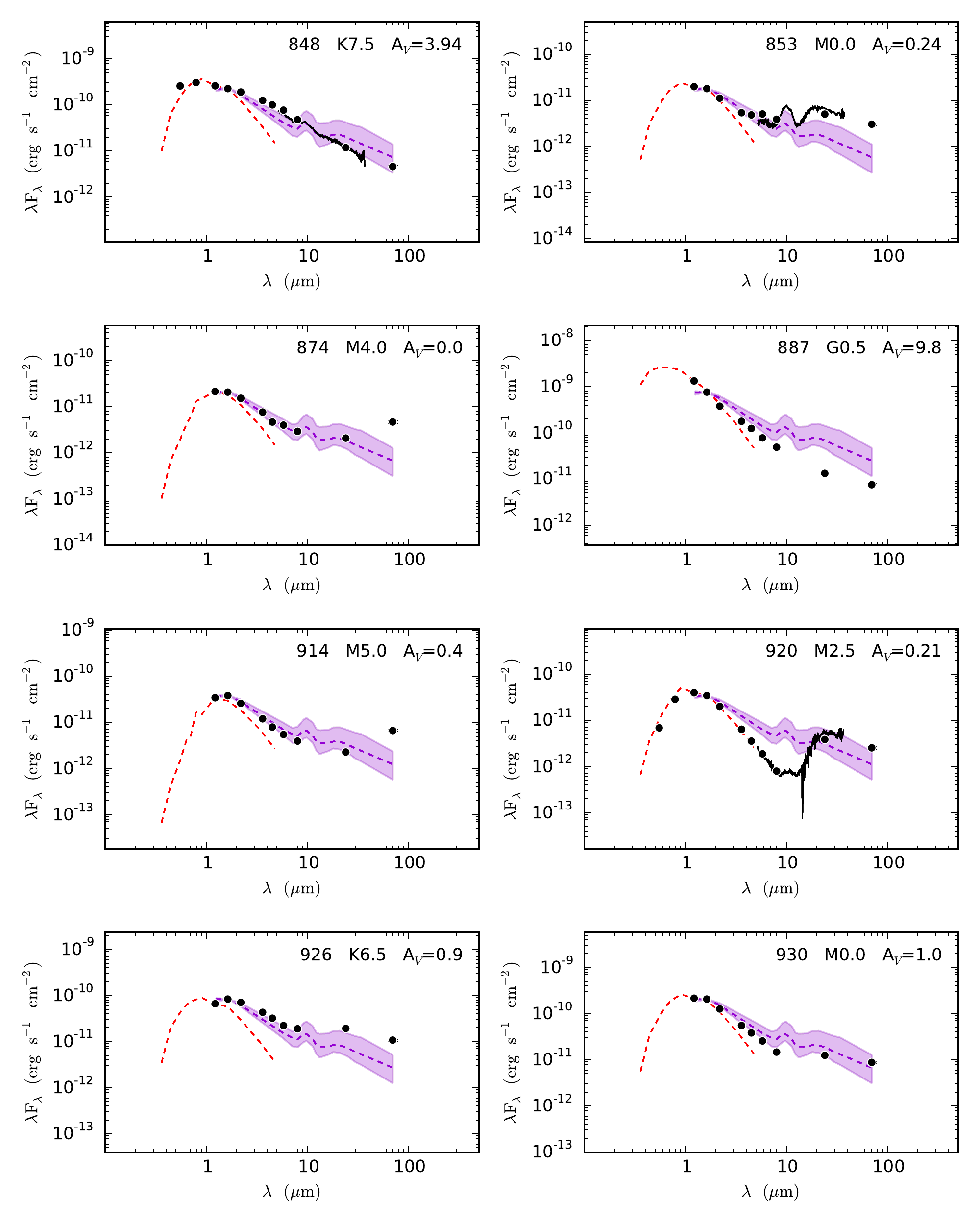}
\caption[]{Continued}
\end{figure*}
% \clearpage

% \centering
% \clearpage
\begin{figure*}
\ContinuedFloat
\includegraphics[angle=0,scale=0.7]{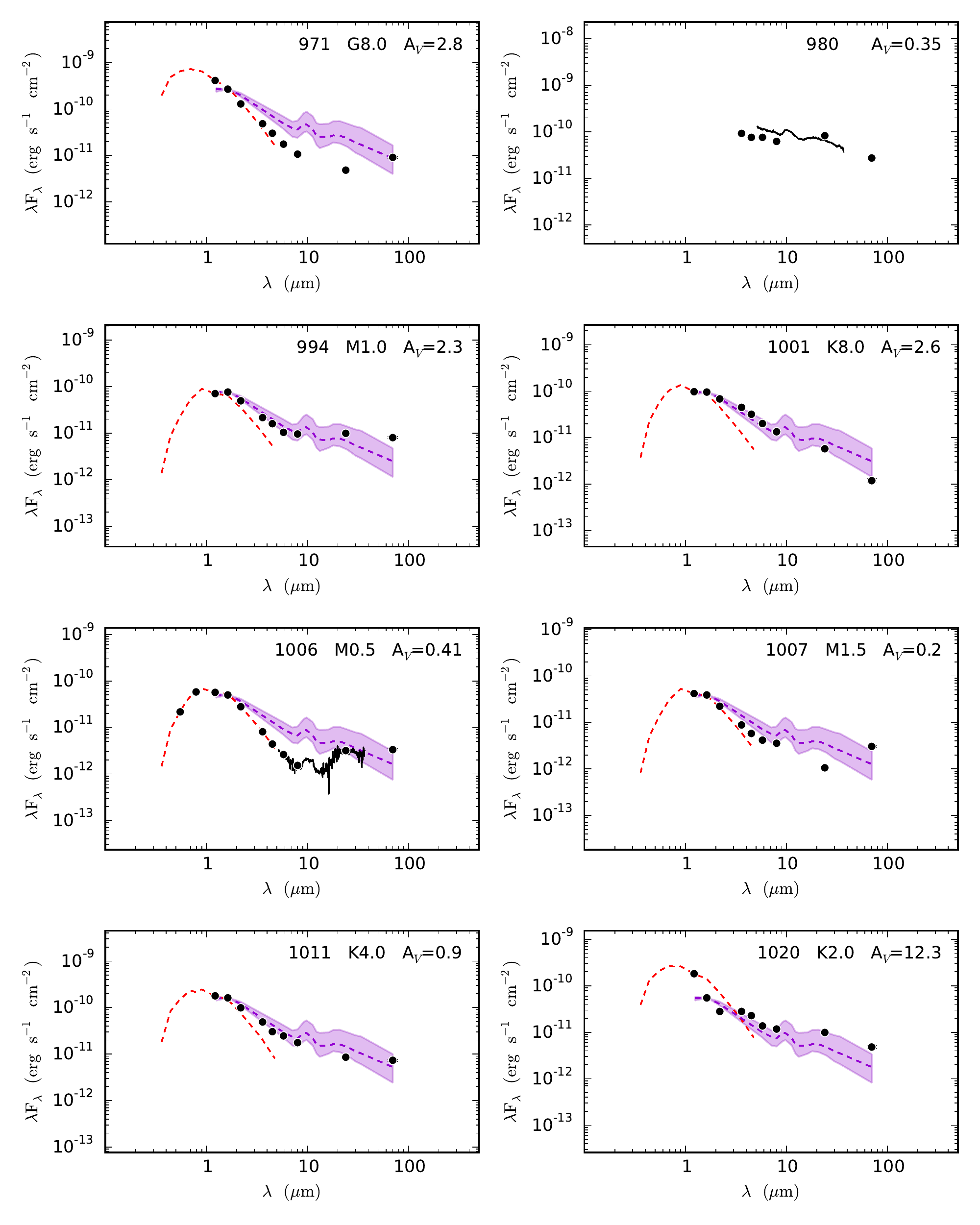}
\caption[]{Continued}
\end{figure*}
% \clearpage

% \centering
% \clearpage
\begin{figure*}
\ContinuedFloat
\includegraphics[angle=0,scale=0.7]{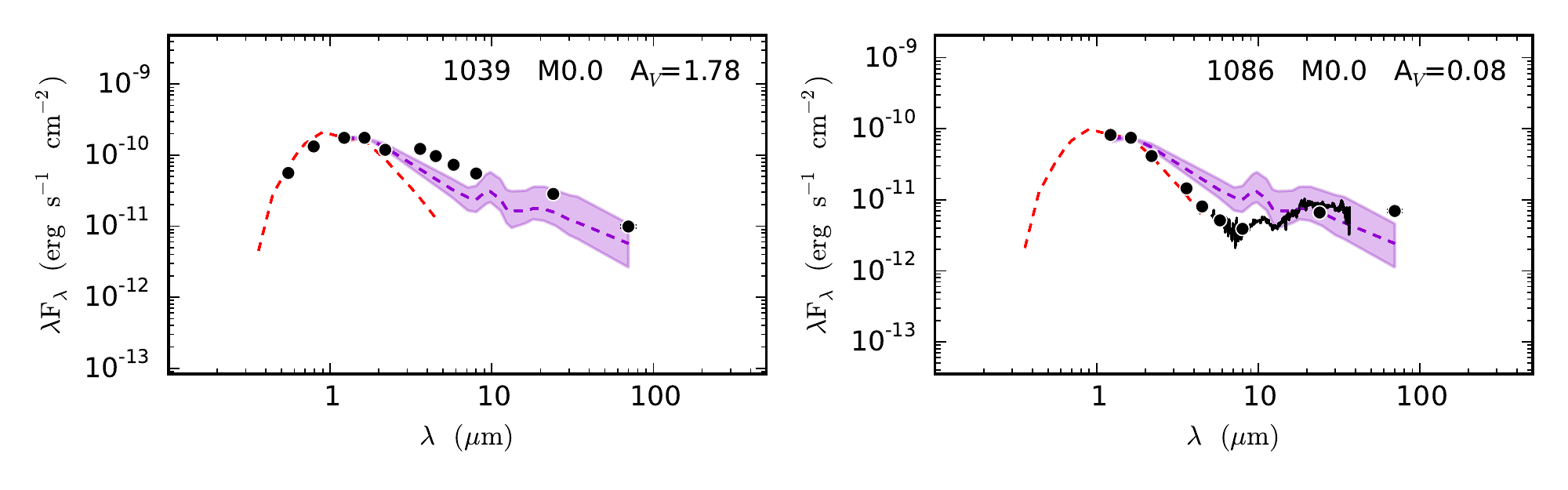}
\caption[]{Continued}
\end{figure*}
% \clearpage
% \FloatBarrier

% \phn = phantom number width; \phd = phantom decimal point width
%\startlongtable    
\FloatBarrier
\begin{deluxetable}{cccccc}
%\rotate
%\tabletypesize{\scriptsize}

\tablewidth{0pt}
\tablecaption{Transitional Disk Candidates}
\tablehead{  \colhead{M12 Num} & \colhead{Fang et al.} & \colhead{Mer\'{i}n et al.} & \colhead{Cieza et al.} & \colhead{Muzerolle et al.}  & \colhead{This work} \\  \colhead{} & \colhead{(2009, 2013)} & \colhead{(2010)} & \colhead{(2010)} & \colhead{(2010)} }
\startdata
198&      $\times$&      $\times$&      $\times$&      $\times$& $\times$\\ 
227&      $\times$&      $\times$&      $\times$&      $\times$& $\times$\\ 
231&      \ &      $\times$&      $\times$&      \ \\ 
232&      \ &      $\times$&      \ &      \ \\ 
250&      $\times$&      $\times$&      $\times$&      $\times$& $\times$\\ 
263&      $\times$&      $\times$&      $\times$&      $\times$& $\times$\\ 
269&      $\times$&      $\times$&      $\times$&      $\times$& $\times$\\ 
278&      $\times$&      $\times$&      $\times$&      $\times$& $\times$\\ 
282&      \ &      $\times$&      \ &      \ \\ 
291&      $\times$&      $\times$&      \ &      \ \\ 
296&      $\times$&      $\times$&      $\times$&      $\times$& $\times$\\ 
307&      $\times$&      $\times$&      $\times$&      $\times$& $\times$\\ 
313&      \ &      \ &      $\times$&      $\times$\\ 
342&      $\times$&      \ &      $\times$&      \ \\ 
387&      $\times$&      $\times$&      $\times$&      $\times$& $\times$\\ 
400&      $\times$&      $\times$&      \ &      \ \\ 
402&      \ &      $\times$&      $\times$&      \ \\ 
403&      $\times$&      \ &      $\times$&      $\times$& $\times$\\ 
463&      $\times$&      $\times$&      $\times$&      $\times$& $\times$\\ 
468&      \ &      \ &      $\times$&      \ \\ 
483&      \ &      \ &      \ &      $\times$\\ 
485&      $\times$&      \ &      $\times$&      $\times$& $\times$\\ 
487&      $\times$&      $\times$&      $\times$&      $\times$& $\times$\\ 
488&      \ &      \ &      \ &      $\times$\\ 
491&      $\times$&      \ &      \ &      $\times$\\ 
556&      $\times$&      $\times$&      $\times$&      $\times$& $\times$\\ 
561&      $\times$&      \ &      \ &      \ \\ 
597&      \ &      $\times$&      \ &      \ \\ 
598&      $\times$&      \ &      \ &      \ \\ 
619&      $\times$&      \ &      $\times$&      $\times$& $\times$\\ 
633&      $\times$&      $\times$&      $\times$&      $\times$& $\times$\\ 
644&      $\times$&      \ &      \ &      \ \\ 
666&      \ &      \ &      \ &      $\times$\\ 
689&      $\times$&      $\times$&      \ &      $\times$& $\times$\\ 
751&      $\times$&      $\times$&      $\times$&      $\times$& $\times$\\ 
761&      \ &      $\times$&      \ &      \ \\ 
811&      $\times$&      $\times$&      $\times$&      \ & $\times$\\ 
818&      $\times$&      $\times$&      $\times$&      $\times$& $\times$\\ 
832&      \ &      $\times$&      \ &      \ \\ 
874&      \ &      \ &      $\times$&      \ \\ 
920&      $\times$&      $\times$&      $\times$&      $\times$& $\times$\\ 
926&      \ &      $\times$&      \ &      \ \\ 
930&      \ &      $\times$&      \ &      \ \\ 
971&      $\times$&      \ &      $\times$&      $\times$& $\times$\\ 
994&      \ &      $\times$&      \ &      \ \\ 
1006&      $\times$&      $\times$&      $\times$&      $\times$& $\times$\\ 
1011&      \ &      \ &      $\times$&      \ \\ 
1020&      \ &      $\times$&      \ &      \ \\ 
1086&      $\times$&      $\times$&      $\times$&      $\times$& $\times$\\ 
\enddata
%From Indices_Megatable.py
\label{tab: TDs}
\end{deluxetable}
\FloatBarrier

% \phn = phantom number width; \phd = phantom decimal point width
%\startlongtable    
\begin{deluxetable}{cccc}
%\rotate
%\tabletypesize{\scriptsize}
\tablewidth{0pt}
\tablecaption{L1641 Median SED}
\tablehead{  \colhead{Wavelength} & \colhead{Median} & \colhead{Lower Quartile} & \colhead{Upper Quartile} \\  \colhead{(\mic)} &\colhead{($\log_{10}[\lambda F_\lambda]$)} &\colhead{($\log_{10}[\lambda F_\lambda]$)} &\colhead{($\log_{10}[\lambda F_\lambda]$)}}
\startdata
0.55&      -10.567&      -10.663&      -10.252   \\ 
0.79&      -10.006&      -10.121&      -9.838   \\ 
1.22&      -9.891&      -9.928&      -9.824   \\ 
1.63&      -9.935&      -9.935&      -9.935   \\ 
2.19&      -10.150&      -10.189&      -10.103   \\ 
3.6&      -10.487&      -10.641&      -10.378   \\ 
4.5&      -10.655&      -10.805&      -10.486   \\ 
5.8&      -10.817&      -10.974&      -10.616   \\ 
8.0&      -10.917&      -11.160&      -10.662   \\ 
24.0&      -11.054&      -11.239&      -10.836   \\ 
70.0&      -11.230&      -11.619&      -10.950   \\  
\enddata
%From SEDs.py & MedianSED_L1641_all.txt
\label{tab: median SED all}
\end{deluxetable}

% \phn = phantom number width; \phd = phantom decimal point width
%\startlongtable    
\begin{deluxetable}{cccccccc}
%\rotate
%\tabletypesize{\scriptsize}
\tablewidth{0pt}
\tablecaption{L1641 Median SED for Full Disks}
\tablehead{ 
\colhead{Wavelength}& \colhead{Median} & \colhead{Lower Quartile} & \colhead{Upper Quartile}  \\  \colhead{(\mic)} 
&\colhead{($\log_{10}[\lambda F_\lambda]$)} &\colhead{($\log_{10}[\lambda F_\lambda]$)} &\colhead{($\log_{10}[\lambda F_\lambda]$)}}
\startdata
0.55&      -10.554&      -10.691&      -9.939&         \\ 
0.79&      -10.053&      -10.175&      -9.728&         \\ 
1.22&      -9.892&      -9.939&      -9.821&         \\ 
1.63&      -9.935&      -9.935&      -9.935&         \\ 
2.19&      -10.145&      -10.179&      -10.069&         \\ 
3.6&      -10.453&      -10.537&      -10.346&         \\ 
4.5&      -10.586&      -10.682&      -10.438&         \\ 
5.8&      -10.750&      -10.855&      -10.555&         \\ 
8.0&      -10.817&      -10.978&      -10.606&         \\ 
24.0&      -11.022&      -11.239&      -10.743&         \\ 
70.0&      -11.246&      -11.679&      -10.950&         \\ 
\enddata
%From SEDs.py & MedianSED_L1641_TDandFD_top&bottom.txt
\label{tab: median SED FD}
\end{deluxetable}   

% \phn = phantom number width; \phd = phantom decimal point width
%\startlongtable
\begin{deluxetable}{cccccccc}
%\rotate
%\tabletypesize{\scriptsize}
\tablewidth{0pt}
\tablecaption{L1641 Median SED for Transitional Disks}
\tablehead{ 
\colhead{Wavelength}& \colhead{Median} & \colhead{Lower Quartile} & \colhead{Upper Quartile}  \\  \colhead{(\mic)} 
&\colhead{($\log_{10}[\lambda F_\lambda]$)} &\colhead{($\log_{10}[\lambda F_\lambda]$)} &\colhead{($\log_{10}[\lambda F_\lambda]$)}}
\startdata
0.55&      -10.582&      -10.656&      -10.474&         \\ 
0.79&      -9.990&      -10.015&      -9.876&         \\ 
1.22&      -9.887&      -9.915&      -9.862&         \\ 
1.63&      -9.935&      -9.935&      -9.935&         \\ 
2.19&      -10.186&      -10.194&      -10.147&         \\ 
3.6&      -10.666&      -10.712&      -10.605&         \\ 
4.5&      -10.907&      -10.966&      -10.834&         \\ 
5.8&      -11.154&      -11.215&      -11.062&         \\ 
8.0&      -11.351&      -11.460&      -11.210&         \\ 
24.0&      -11.092&      -11.221&      -10.886&         \\ 
70.0&      -11.140&      -11.377&      -10.960&         \\ 
\enddata
%From SEDs.py & MedianSED_L1641_TDandFD_top&bottom.txt
\label{tab: median SED TD}
\end{deluxetable}

\begin{deluxetable}{ccccc}
\centering
\tablewidth{0pt}
\tablecaption{\textit{Herschel} Fluxes For Flagged Objects}
\tablehead{
\colhead{M12 Num} & \colhead{RA(J2000)} & \colhead{Dec(J2000)} & \colhead{F$_{70}$} &\colhead{Flag} \\ \colhead{}& \colhead{}& \colhead{}&  \colhead{(Jy)} & \colhead{} 
}
\startdata
206&      05:43:00.9&      -08:44:18.4&      0.235$\pm$0.012&      2   \\ 
229&      05:41:22.6&      -08:39:16.0&      0.155$\pm$0.008&      1   \\ 
230&      05:41:22.9&      -08:39:11.0&      0.173$\pm$0.009&      1   \\ 
236&      05:42:48.7&      -08:38:30.8&      0.449$\pm$0.023&      1   \\ 
237&      05:42:50.5&      -08:38:29.1&      0.171$\pm$0.009&      1   \\ 
239&      05:42:49.7&      -08:38:25.7&      0.091$\pm$0.006&      1   \\ 
241&      05:42:47.4&      -08:38:08.6&      0.131$\pm$0.007&      1   \\ 
259&      05:42:50.3&      -08:34:37.6&      0.057$\pm$0.004&      2   \\ 
286&      05:42:49.4&      -08:17:07.2&      0.113$\pm$0.007&      1   \\ 
300&      05:40:20.9&      -08:14:06.4&      0.63$\pm$0.03&      1   \\ 
328&      05:42:44.0&      -08:09:26.7&      0.053$\pm$0.004&      1   \\ 
349&      05:40:44.2&      -08:07:34.8&      0.205$\pm$0.011&      1   \\ 
350&      05:40:25.0&      -08:07:33.0&      0.508$\pm$0.026&      1   \\ 
351&      05:40:46.6&      -08:07:12.9&      0.84$\pm$0.04&      1   \\ 
364&      05:40:48.0&      -08:05:58.7&      3.77$\pm$0.19&      1   \\ 
371&      05:40:46.2&      -08:05:24.5&      0.80$\pm$0.04&      1   \\ 
376&      05:40:46.8&      -08:04:54.6&      0.048$\pm$0.003&      1   \\ 
421&      05:40:20.5&      -07:56:39.6&      2.90$\pm$0.15&      1   \\ 
431&      05:41:20.1&      -07:55:23.8&      0.143$\pm$0.008&      1   \\ 
438&      05:41:30.0&      -07:54:21.1&      0.0431$\pm$0.0027&      2   \\ 
448&      05:41:21.7&      -07:53:16.3&      0.054$\pm$0.003&      2   \\ 
467&      05:41:22.7&      -07:49:16.1&      0.0413$\pm$0.0027&      2   \\ 
478&      05:40:58.9&      -07:48:01.5&      0.0378$\pm$0.0026&      1   \\ 
479&      05:40:38.5&      -07:47:46.8&      0.104$\pm$0.006&      2   \\ 
510&      05:40:31.3&      -07:37:01.2&      0.052$\pm$0.003&      2   \\ 
523&      05:39:53.5&      -07:30:09.5&      0.85$\pm$0.05&      1,2   \\ 
526&      05:39:55.1&      -07:29:37.0&      0.166$\pm$0.022&      1   \\ 
533&      05:39:54.7&      -07:27:44.1&      0.193$\pm$0.010&      2   \\ 
534&      05:40:08.0&      -07:27:41.2&      0.160$\pm$0.008&      1   \\ 
535&      05:40:10.3&      -07:27:38.2&      0.576$\pm$0.029&      1   \\ 
540&      05:39:22.3&      -07:26:44.5&      3.88$\pm$0.19&      1   \\ 
541&      05:39:58.1&      -07:26:41.2&      0.062$\pm$0.004&      2   \\ 
571&      05:39:48.4&      -07:24:14.9&      0.270$\pm$0.014&      2   \\ 
576&      05:39:42.5&      -07:23:16.5&      0.051$\pm$0.003&      2   \\ 
586&      05:38:52.4&      -07:21:09.4&      1.01$\pm$0.05&      1   \\ 
590&      05:39:28.6&      -07:20:31.1&      0.266$\pm$0.014&      2   \\ 
612&      05:38:58.6&      -07:16:45.7&      8.1$\pm$0.4&      1   \\ 
634&      05:39:09.5&      -07:09:17.7&      0.086$\pm$0.005&      2   \\ 
643&      05:39:00.6&      -07:06:30.0&      0.105$\pm$0.006&      2   \\ 
648&      05:39:05.2&      -07:05:42.0&      0.038$\pm$0.003&      2   \\ 
676&      05:38:45.7&      -07:01:58.5&      0.69$\pm$0.03&      1,2   \\ 
679&      05:38:47.2&      -07:01:53.4&      0.341$\pm$0.017&      1   \\ 
686&      05:38:21.2&      -07:01:20.5&      0.045$\pm$0.004&      1   \\ 
712&      05:38:40.1&      -06:59:14.6&      0.107$\pm$0.006&      1   \\ 
720&      05:38:43.8&      -06:58:22.3&      0.388$\pm$0.020&      1   \\ 
725&      05:38:44.9&      -06:58:14.6&      0.108$\pm$0.007&      1   \\ 
726&      05:38:43.2&      -06:58:08.9&      2.89$\pm$0.15&      1   \\ 
760&      05:38:09.3&      -06:49:16.8&      3.34$\pm$0.17&      1   \\ 
776&      05:36:21.4&      -06:45:36.8&      1.91$\pm$0.10&      1   \\ 
782&      05:36:05.0&      -06:44:42.9&      0.137$\pm$0.015&      1,2   \\ 
786&      05:36:33.0&      -06:44:29.3&      0.177$\pm$0.010&      1   \\ 
788&      05:36:32.9&      -06:44:21.0&      0.158$\pm$0.009&      1   \\ 
796&      05:36:25.4&      -06:42:57.4&      10.1$\pm$0.6&      1   \\ 
804&      05:37:47.0&      -06:42:30.2&      1.13$\pm$0.06&      3   \\ 
812&      05:36:32.4&      -06:40:42.9&      0.103$\pm$0.008&      1   \\ 
827&      05:35:58.2&      -06:36:42.9&      0.028$\pm$0.006&      1   \\ 
838&      05:37:13.2&      -06:35:00.6&      0.288$\pm$0.015&      1   \\ 
886&      05:36:21.8&      -06:26:01.9&      0.036$\pm$0.006&      1   \\ 
918&      05:36:12.9&      -06:23:30.6&      0.280$\pm$0.015&      1   \\ 
925&      05:36:23.8&      -06:23:11.1&      0.543$\pm$0.028&      1   \\ 
927&      05:36:21.5&      -06:22:52.4&      0.402$\pm$0.021&      1   \\ 
939&      05:36:11.4&      -06:22:22.1&      0.093$\pm$0.005&      1   \\ 
950&      05:36:21.0&      -06:21:53.1&      1.03$\pm$0.06&      1,2   \\ 
960&      05:36:18.5&      -06:20:38.7&      0.057$\pm$0.007&      1   \\ 
1065&      05:35:34.6&      -06:02:47.1&      0.131$\pm$0.008&      1   \\ 
\enddata
\tablecomments{For each object, we list the corresponding identification number from \citet{megeath12} and the objects associated flag. A flag of ``1'' denotes a protoplanetary system with a 70 \mic\ flux that is likely contaminated by close sources and/or nebulosity; ``2'' identifies systems with \Av $\geq$ 15; ``3'' labels systems that do not have colors characteristic of CTTSs. These sources are not included in our analysis}
\label{tab: flagged observations}
\end{deluxetable}

%%% TABLE 2:
%\startlongtable
% \startlongtable
\begin{deluxetable}{cccccccccc}
% \rotate
%\tabletypesize{\scriptsize}
\tablewidth{0pt}
\tablecaption{Stellar Properties For Flagged Objects}
\tablehead{
\colhead{M12 Num} & \colhead{SpT} & \colhead{SpT Ref.}  & \colhead{\Av} & \colhead{\Av \ Ref.} & \colhead{L} & \colhead{L Ref.} & \colhead{log\Mdot} & \colhead{\Mdot \ Ref.} & \colhead{Flag} \\
\colhead{} & \colhead{} & \colhead{} & \colhead{} & \colhead{} & \colhead{(\Lsun)} & \colhead{} & \colhead{(\msunyr)} & \colhead{}
}
\startdata
  206 & \ & \ & 22.87 & CTTS J-H & \ & \ & \ & \ & 2\\
  229 & \ & \ & 3.41 & CTTS J-H & \ & \ & \ & \ & 1\\
  230 & M1.8 & a & 2.9 & c & 0.773 & c & -7.8 & c & 1\\
  236 & K7.5 & a & 4 & c & 3.748 & c & -6.8 & c & 1\\
  237 & \ & \ & 14.91 & CTTS J-H & \ & \ & \ & \ & 1\\
  239 & \ & \ & 14.51 & CTTS J-H & \ & \ & \ & \ & 1\\
  241 & \ & \ & 10.55 & CTTS J-H & \ & \ & \ & \ & 1\\
  259 & \ & \ & 19.9 & b & 1.72 & b & \ & \ & 2\\
  %286 & \ & \ & \ & \ & \ & \ & \ & \ & 1\\
  %300 & \ & \ & \ & \ & \ & \ & \ & \ & 1\\
  328 & M0.0 & a & 5.2 & c & 0.79 & c & $<$ -8.7 & c & 1\\
  349 & \ & \ & 8.77 & CTTS J-H & \ & \ & \ & \ & 1\\
  350 & G4.0 & a & 2 & b & 5.5 & b & -7.13 & a & 1\\
  351 & K7.0 & c & 8.58 & b & 7.19 & b & $<$ -8.19 & b & 1\\
  364 & K5.0 & c & 2.4 & c & 7.394 & c & $<$ -8 & c & 1\\
  371 & K3.0 & a & 3.09 & b & 8.18 & b & -7.86 & b & 1\\
  376 & M2.5 & d & 5.2 & d & 1.231 & d & \ & \ & 1\\
  421 & \ & \ & 12.3 & b & 9.19 & b & \ & \ & 1\\
  431 & M4.0 & d & 10.2 & d & 0.581 & d & -8.23 & d & 1\\
  438 & \ & \ & 16.11 & CTTS J-H & \ & \ & \ & \ & 2\\
  448 & \ & \ & 27.27 & CTTS J-H & \ & \ & \ & \ & 2\\
  467 & \ & \ & 17.6 & b & 1.22 & b & \ & \ & 2\\
  478 & \ & \ & 2.75 & b & 0.11 & b & \ & \ & 1\\
  479 & \ & \ & 22 & b & 4.13 & b & \ & \ & 2\\
  510 & \ & \ & 15.47 & CTTS J-H & \ & \ & \ & \ & 2\\
  523 & \ & \ & 21.5 & b & 1.75 & b & \ & \ & 1,2\\
  %526 & \ & \ & \ & \ & \ & \ & \ & \ & 1\\
  533 & \ & \ & 34.5 & b & 53.61 & b & \ & \ & 2\\
  %534 & \ & \ & \ & \ & \ & \ & \ & \ & 1\\
  535 & K7.0 & b & 9.96 & b & 4.6 & b & \ & \ & 1\\
  540 & K7.0 & a & 6 & c & 19.795 & c & -5.6 & c & 1\\
  541 & \ & \ & 22 & b & 3.14 & b & \ & \ & 2\\
  571 & \ & \ & 19.3 & b & 1.99 & b & \ & \ & 2\\
  576 & \ & \ & 15.7 & b & 1.81 & b & \ & \ & 2\\
  586 & G4.5 & a & 4.6 & c & 10.504 & c & -6.8 & c & 1\\
  590 & \ & \ & 20.4 & b & 0.66 & b & \ & \ & 2\\
  612 & A9.0 & e & 0.0 & CTTS J-H & \ & \ & \ & \ & 1\\
  634 & \ & \ & 21.87 & CTTS J-H & \ & \ & \ & \ & 2\\
  643 & \ & \ & 21.5 & CTTS J-H & \ & \ & \ & \ & 2\\
  648 &  &  & 15.04 & CTTS J-H &  &  &  & &  2\\
  676 & \ & \ & 26.79 & CTTS J-H & \ & \ & \ & \ & 1,2\\
  %679 & \ & \ & \ & \ & \ & \ & \ & \ & 1\\
  686 & M4.5 & a & 1.4 & d & 0.102 & d & -9.78 & d & 1\\
  712 & M0.5 & a & 3.7 & b & 0.68 & b & -7.7 & c & 1\\
  720 & K7.5 & a & 4.5 & d & 0.359 & d & -8.83 & d & 1\\
  725 & K7.5 & a & 8.42 & b & 1.89 & b & -7.64 & d & 1\\
  726 & K7.5 & d & 11.3 & b & 14.57 & b & \ & \ & 1\\
  760 & A3.5 & b & 3.58 & b & 34.38 & b & -7.9 & b & 1\\
  776 & K5.0 & a & 5.92 & b & 9.09 & b & $<$ -7.46 & b & 1\\
  782 & \ & \ & 15.2 & CTTS J-H & \ & \ & \ & \ & 1,2\\
  786 & \ & \ & 8.35 & CTTS J-H & \ & \ & \ & \ & 1\\
  788 & M2.0 & a & 3.74 & b & 0.55 & b & -7.99 & d & 1\\
  796 & \ & \ & 0.56 & CTTS J-H & \ & \ & \ & \ & 1\\
  804 & A1.0 & e & \ & \ & \ & \ & \ & \ & 3\\
  812 & M4.5 & a & 0 & c & 0.166 & c & -9.45 & c & 1\\
  827 & K7.5 & a & 1.55 & b & 1.06 & b & -8.31 & b & 1\\
  838 & A7.0 & a & 0.83 & b & 27.41 & b & -7.15 & b & 1\\
  886 & K7.4 & a & 3.6 & c & 1.066 & c & -7.9 & c & 1\\
  918 & M1.5 & a & 1.37 & b & 1.19 & b & \ & \ & 1\\
  925 & \ & \ & 12.2 & b & 2.15 & b & \ & \ & 1\\
  927 & M0.5 & d & 9.5 & d & 0.636 & d & -7.75 & d & 1\\
  939 & M0.0 & a & 6.3 & d & 1.436 & d & -7.76 & d & 1\\
  950 & \ & \ & 15.89 & CTTS J-H & \ & \ & \ & \ & 1,2\\
  960 & M0.5 & a & 0.6 & d & 0.533 & d & -8.16 & d & 1\\
  1065 & M1.5 & a & 9.79 & CTTS J-H & \ & \ & \ & \ & 1\\
\enddata
\tablecomments{References: $^a$\citet{hsu12}, $^b$\citet{kim16}, $^c$\citet{fang13}, $^d$\citet{fang09}, $^e$\citet{hsu13}. For sources with an \Av\ reference of “CTTS J-H,” visual extinctions were measured in this work as described in Section~\ref{subsect: stellar sample}. Flags are the same as noted in Table~\ref{tab: flagged observations}. Flagged objects are not included in our analysis}  
%From StellarPropsTable_justflagged which is made with StellarProperties_full.csv and Topcat
\label{tab: stellar properties flagged}
\end{deluxetable}

\FloatBarrier
\begin{figure*}
\includegraphics[angle=0,scale=0.7]{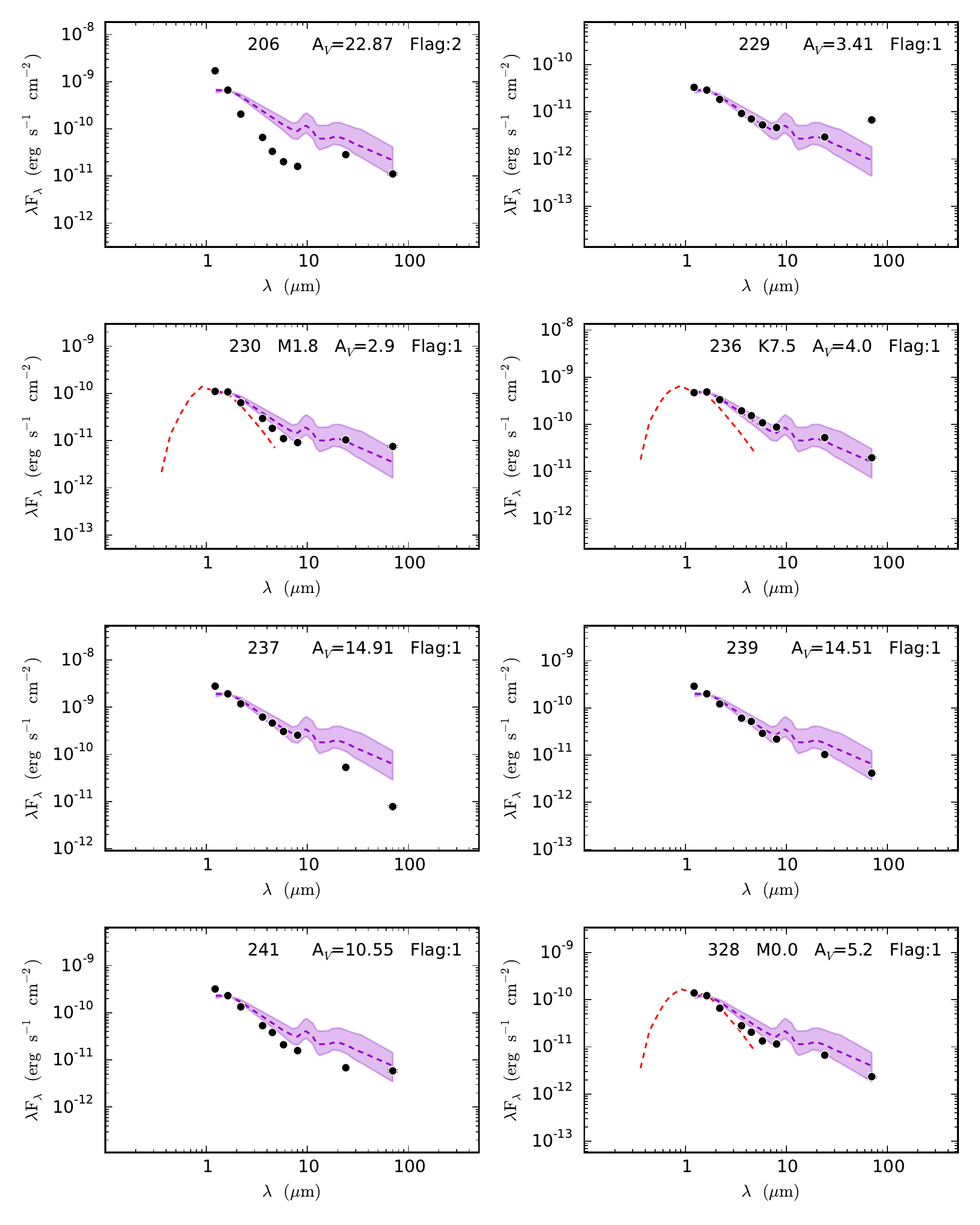}
\caption{SEDs for the L1641 sources that are flagged and not included in our analysis. Flags are the same as noted in Table~\ref{tab: flagged observations}.}\label{fig: flagged seds}
\end{figure*}
%\clearpage

%%%COMMENT
%\begin{comment}
\centering
\begin{figure*}
\ContinuedFloat
\includegraphics[angle=0,scale=0.75]{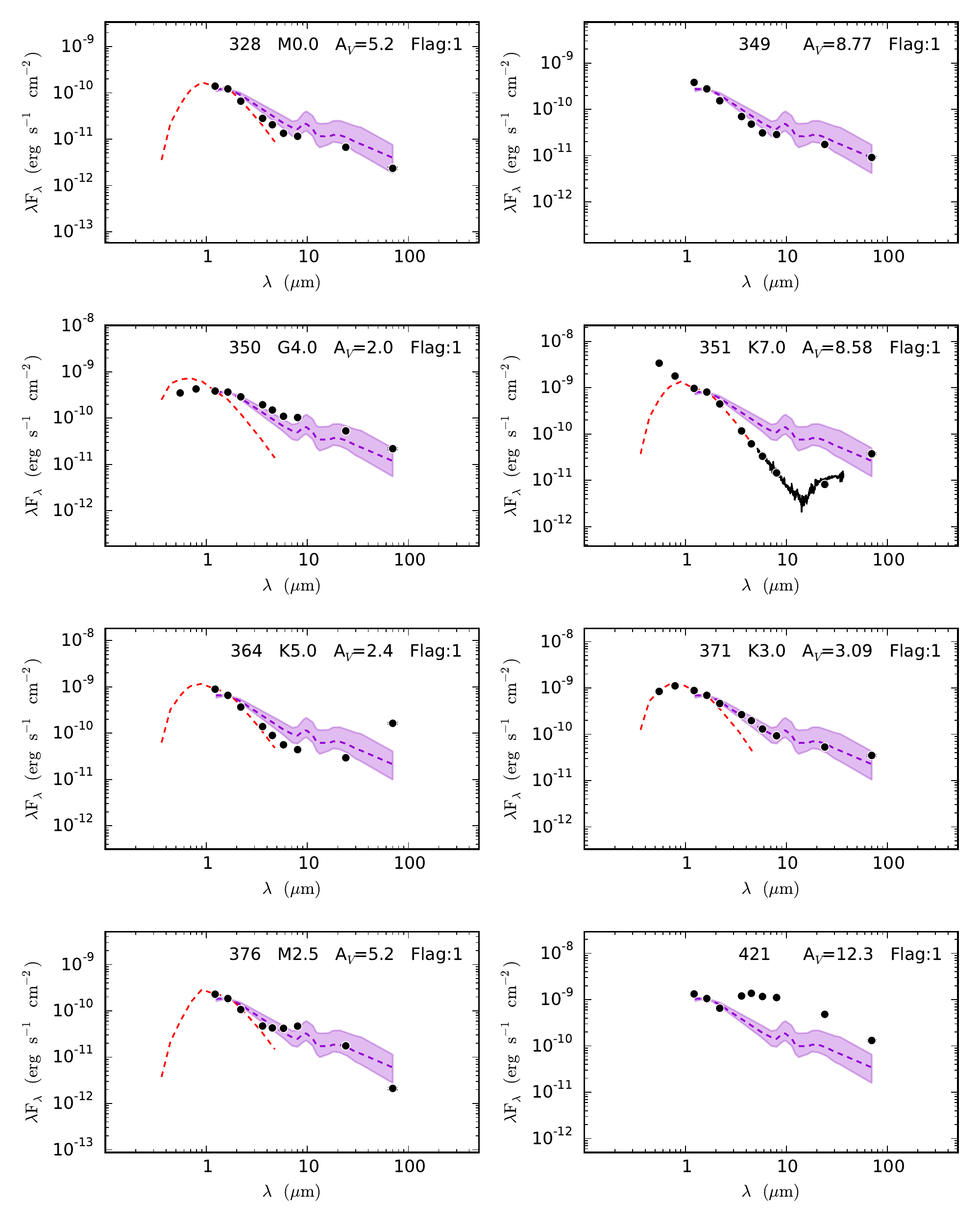}
\caption[]{Continued}
\end{figure*}
\clearpage

\clearpage
\begin{figure}
\ContinuedFloat
\includegraphics[angle=0,scale=0.75]{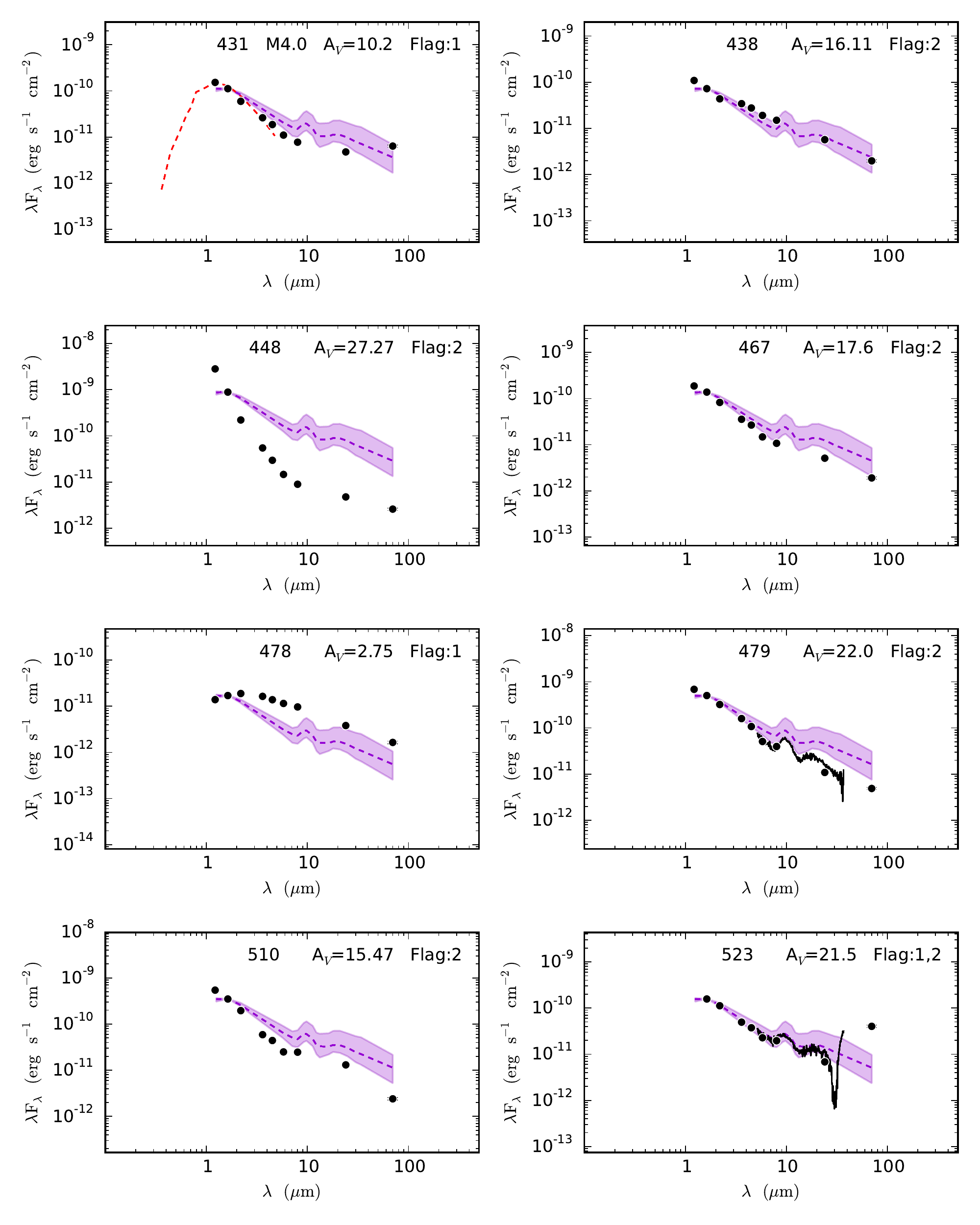}
\caption[]{Continued}
\end{figure}
\clearpage

\clearpage
\begin{figure}
\ContinuedFloat
\includegraphics[angle=0,scale=0.75]{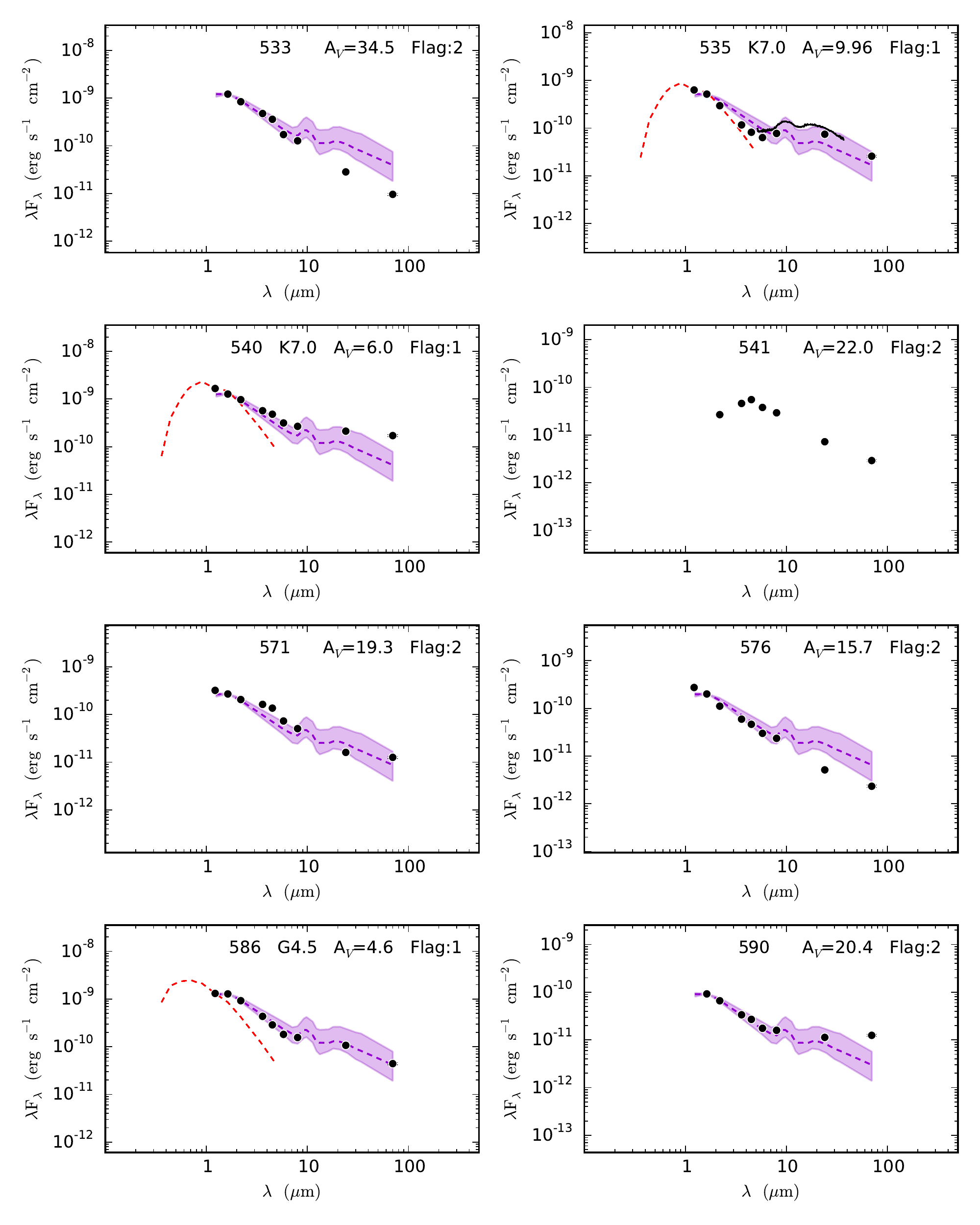}
\caption[]{Continued}
\end{figure}
\clearpage

\clearpage
\begin{figure}
\ContinuedFloat
\includegraphics[angle=0,scale=0.75]{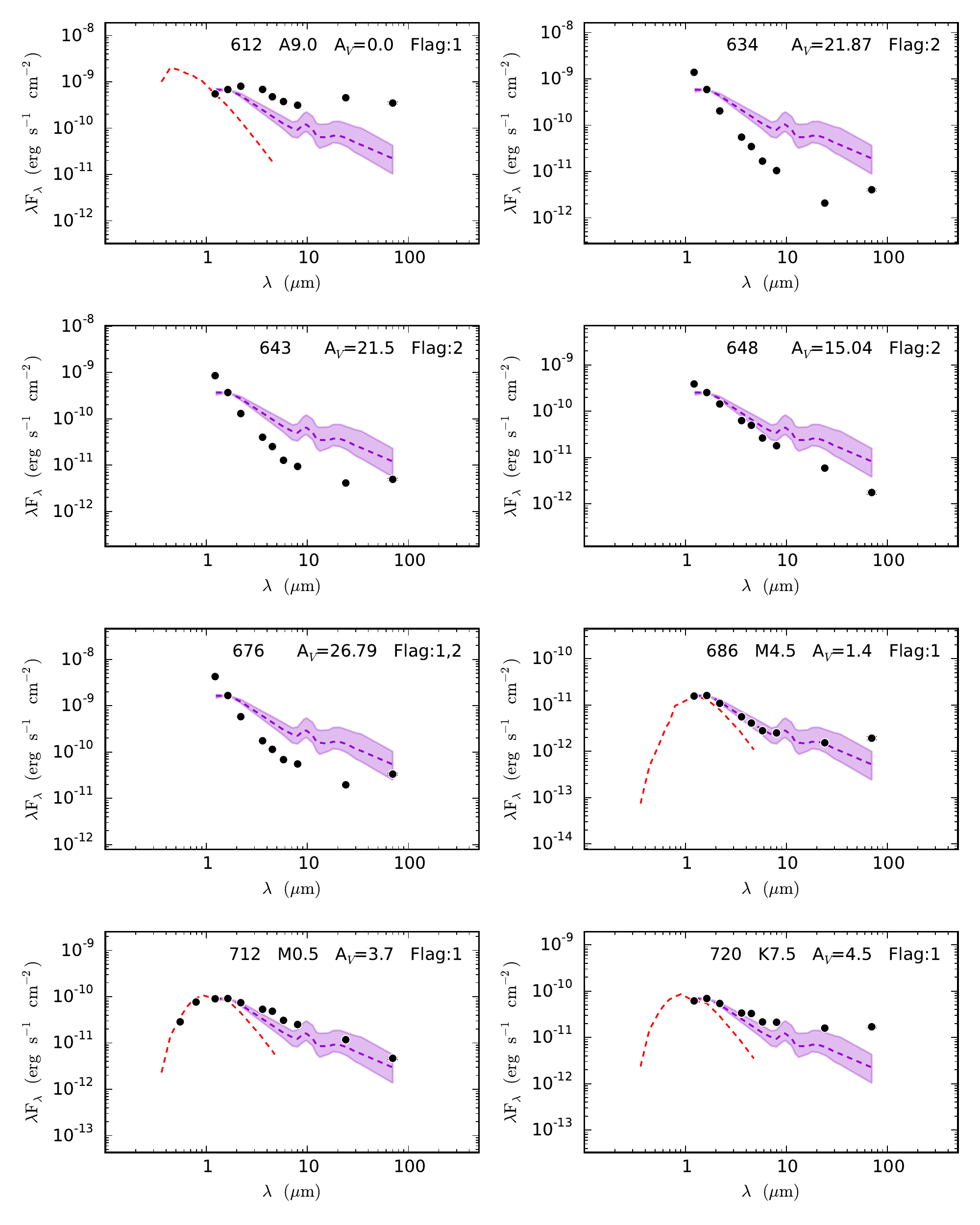}
\caption[]{Continued}
\end{figure}
\clearpage

\clearpage
\begin{figure}
\ContinuedFloat
\includegraphics[angle=0,scale=0.75]{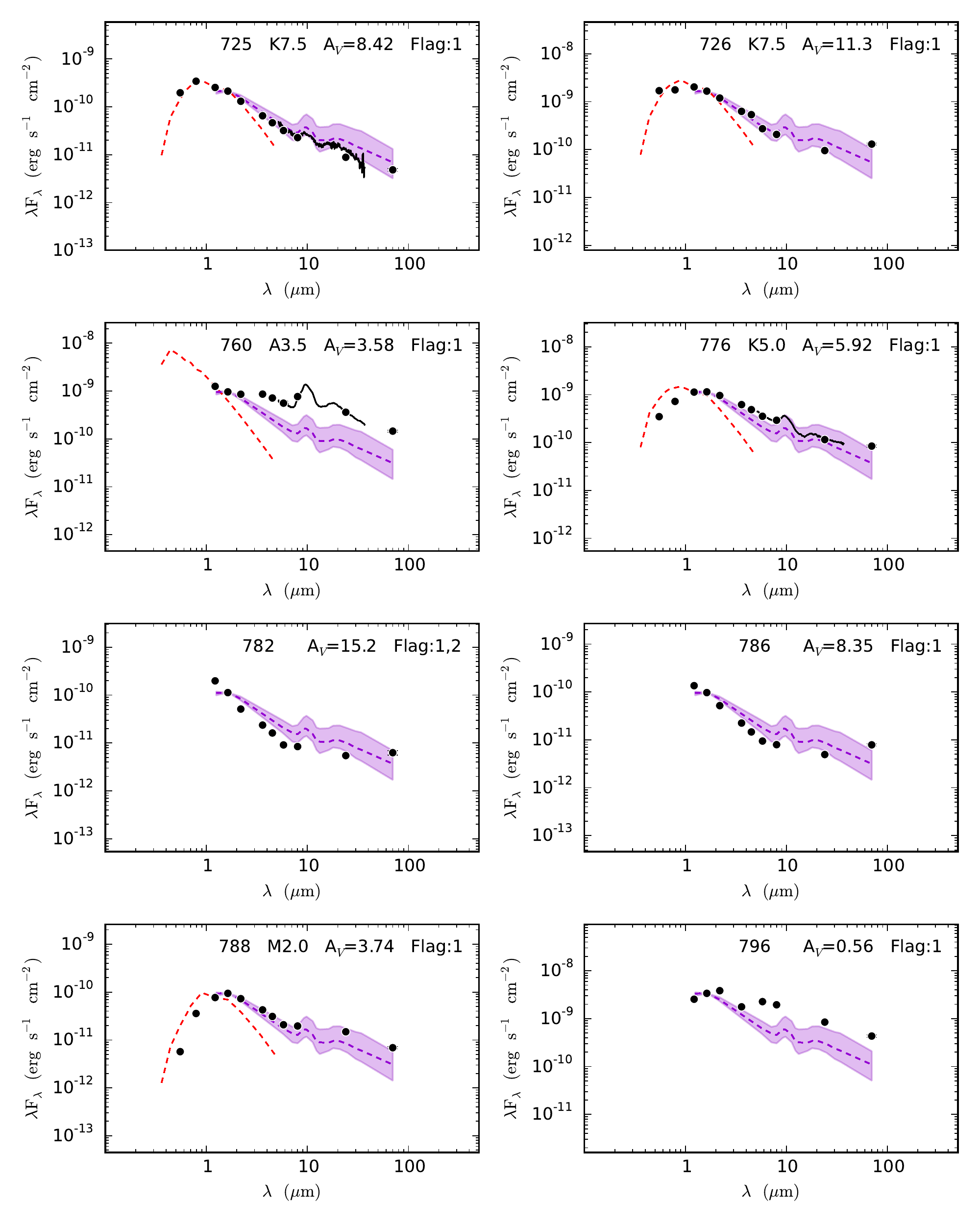}
\caption[]{Continued}
\end{figure}
\clearpage

\clearpage
\begin{figure}
\ContinuedFloat
\includegraphics[angle=0,scale=0.75]{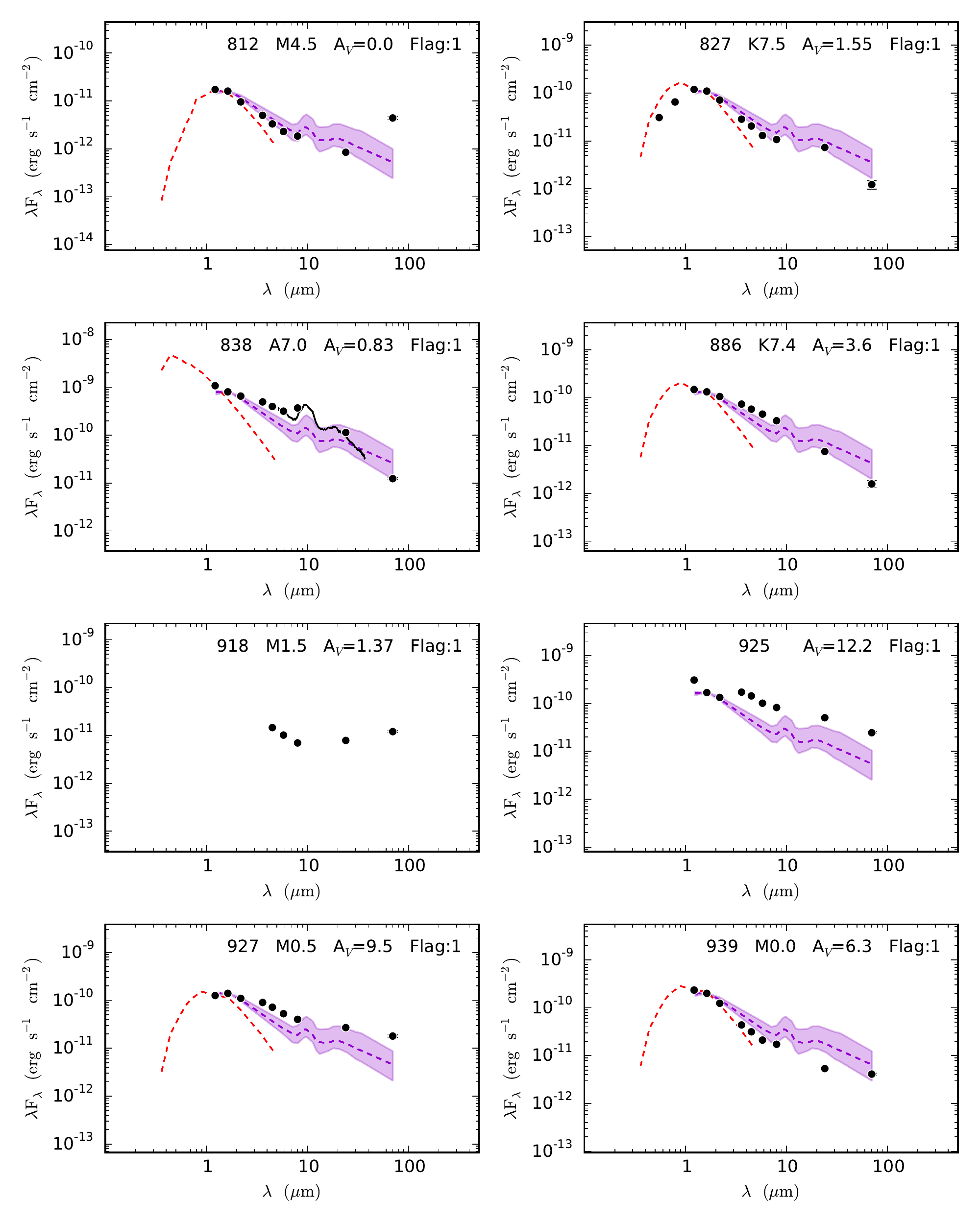}
\caption[]{Continued}
\end{figure}
\clearpage

\clearpage
\begin{figure}
\ContinuedFloat
\includegraphics[angle=0,scale=0.75]{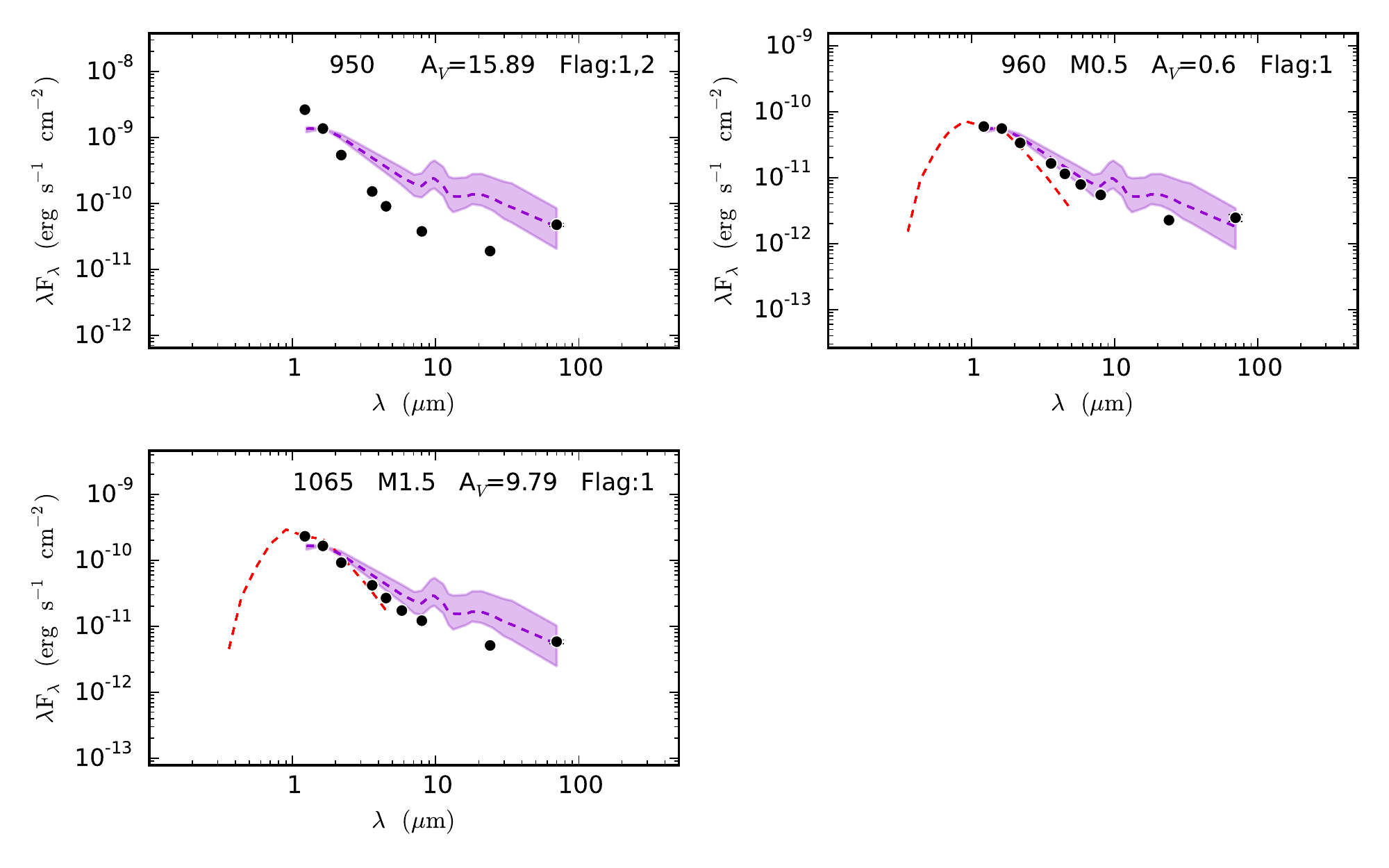}
\caption[]{Continued}
\end{figure}
\FloatBarrier

\bibliographystyle{aasjournal}
\bibliography{biblio}

\begin{thebibliography}{}
\expandafter\ifx\csname natexlab\endcsname\relax\def\natexlab#1{#1}\fi

\bibitem[{{Adams} {et~al.}(1987){Adams}, {Lada}, \& {Shu}}]{adams87}
{Adams}, F.~C., {Lada}, C.~J., \& {Shu}, F.~H. 1987, \apj, 312, 788

\bibitem[{{Alexander} {et~al.}(2014){Alexander}, {Pascucci}, {Andrews},
  {Armitage}, \& {Cieza}}]{alexander14}
{Alexander}, R., {Pascucci}, I., {Andrews}, S., {Armitage}, P., \& {Cieza}, L.
  2014, Protostars and Planets VI, 475

\bibitem[{{Andre} \& {Montmerle}(1994)}]{andre&montmerle94}
{Andre}, P., \& {Montmerle}, T. 1994, \apj, 420, 837

\bibitem[{{Andrews} {et~al.}(2013){Andrews}, {Rosenfeld}, {Kraus}, \&
  {Wilner}}]{andrews13}
{Andrews}, S.~M., {Rosenfeld}, K.~A., {Kraus}, A.~L., \& {Wilner}, D.~J. 2013,
  \apj, 771, 129

\bibitem[{{Armitage}(2018)}]{armitage18}
{Armitage}, P.~J. 2018, ArXiv e-prints, arXiv:1803.10526

\bibitem[{{Baraffe} {et~al.}(1998){Baraffe}, {Chabrier}, {Allard}, \&
  {Hauschildt}}]{baraffe98}
{Baraffe}, I., {Chabrier}, G., {Allard}, F., \& {Hauschildt}, P.~H. 1998, \aap,
  337, 403

\bibitem[{{Bessell} \& {Brett}(1988)}]{bessell&brett88}
{Bessell}, M.~S., \& {Brett}, J.~M. 1988, \pasp, 100, 1134

\bibitem[{{Bustamante} {et~al.}(2015){Bustamante}, {Mer{\'{\i}}n}, {Ribas},
  {Bouy}, {Prusti}, {Pilbratt}, \& {Andr{\'e}}}]{bustamante15}
{Bustamante}, I., {Mer{\'{\i}}n}, B., {Ribas}, {\'A}., {et~al.} 2015, \aap,
  578, A23

\bibitem[{{Calvet} \& {Gullbring}(1998)}]{calvet&gullbring98}
{Calvet}, N., \& {Gullbring}, E. 1998, \apj, 509, 802

\bibitem[{{Calvet} {et~al.}(1994){Calvet}, {Hartmann}, {Kenyon}, \&
  {Whitney}}]{calvet94}
{Calvet}, N., {Hartmann}, L., {Kenyon}, S.~J., \& {Whitney}, B.~A. 1994, \apj,
  434, 330

\bibitem[{{Calvet} {et~al.}(1992){Calvet}, {Magris}, {Patino}, \&
  {D'Alessio}}]{calvet92}
{Calvet}, N., {Magris}, G.~C., {Patino}, A., \& {D'Alessio}, P. 1992, Revista
  Mexicana de Astronomia y Astrofisica, 24

\bibitem[{{Caratti o Garatti} {et~al.}(2012){Caratti o Garatti}, {Garcia
  Lopez}, {Antoniucci}, {Nisini}, {Giannini}, {Eisl{\"o}ffel}, {Ray},
  {Lorenzetti}, \& {Cabrit}}]{carattiogaratti12}
{Caratti o Garatti}, A., {Garcia Lopez}, R., {Antoniucci}, S., {et~al.} 2012,
  \aap, 538, A64

\bibitem[{{Carpenter}(2001)}]{carpenter01}
{Carpenter}, J.~M. 2001, \aj, 121, 2851

\bibitem[{{Chiang} \& {Murray-Clay}(2007)}]{chiang&mc07}
{Chiang}, E., \& {Murray-Clay}, R. 2007, Nature Physics, 3, 604

\bibitem[{{Chiang} \& {Goldreich}(1997)}]{chiang&goldreich97}
{Chiang}, E.~I., \& {Goldreich}, P. 1997, \apj, 490, 368

\bibitem[{{Ciesla}(2007)}]{ciesla07}
{Ciesla}, F.~J. 2007, \apjl, 654, L159

\bibitem[{{Cieza} {et~al.}(2010){Cieza}, {Schreiber}, {Romero}, {Mora},
  {Merin}, {Swift}, {Orellana}, {Williams}, {Harvey}, \& {Evans}}]{cieza10}
{Cieza}, L.~A., {Schreiber}, M.~R., {Romero}, G.~A., {et~al.} 2010, \apj, 712,
  925

\bibitem[{{Clarke} {et~al.}(2001){Clarke}, {Gendrin}, \&
  {Sotomayor}}]{clarke01}
{Clarke}, C.~J., {Gendrin}, A., \& {Sotomayor}, M. 2001, \mnras, 328, 485

\bibitem[{{D'Alessio} {et~al.}(2001){D'Alessio}, {Calvet}, \&
  {Hartmann}}]{d'alessio01}
{D'Alessio}, P., {Calvet}, N., \& {Hartmann}, L. 2001, \apj, 553, 321

\bibitem[{{D'Alessio} {et~al.}(2006){D'Alessio}, {Calvet}, {Hartmann},
  {Franco-Hern{\'a}ndez}, \& {Serv{\'{\i}}n}}]{d'alessio06}
{D'Alessio}, P., {Calvet}, N., {Hartmann}, L., {Franco-Hern{\'a}ndez}, R., \&
  {Serv{\'{\i}}n}, H. 2006, \apj, 638, 314

\bibitem[{{D'Alessio} {et~al.}(1999){D'Alessio}, {Calvet}, {Hartmann},
  {Lizano}, \& {Cant{\'o}}}]{d'alessio99}
{D'Alessio}, P., {Calvet}, N., {Hartmann}, L., {Lizano}, S., \& {Cant{\'o}}, J.
  1999, \apj, 527, 893

\bibitem[{{D'Alessio} {et~al.}(1998){D'Alessio}, {Cant{\"o}}, {Calvet}, \&
  {Lizano}}]{d'alessio98}
{D'Alessio}, P., {Cant{\"o}}, J., {Calvet}, N., \& {Lizano}, S. 1998, \apj,
  500, 411

\bibitem[{{D'Alessio} {et~al.}(2005){D'Alessio}, {Hartmann}, {Calvet},
  {Franco-Hern{\'a}ndez}, {Forrest}, {Sargent}, {Furlan}, {Uchida}, {Green},
  {Watson}, {Chen}, {Kemper}, {Sloan}, \& {Najita}}]{d'alessio05}
{D'Alessio}, P., {Hartmann}, L., {Calvet}, N., {et~al.} 2005, \apj, 621, 461

\bibitem[{{Dorschner} {et~al.}(1995){Dorschner}, {Begemann}, {Henning},
  {Jaeger}, \& {Mutschke}}]{dorschner95}
{Dorschner}, J., {Begemann}, B., {Henning}, T., {Jaeger}, C., \& {Mutschke}, H.
  1995, \aap, 300, 503

\bibitem[{{Dotter} {et~al.}(2008){Dotter}, {Chaboyer}, {Jevremovi{\'c}},
  {Kostov}, {Baron}, \& {Ferguson}}]{dotter08}
{Dotter}, A., {Chaboyer}, B., {Jevremovi{\'c}}, D., {et~al.} 2008, \apjs, 178,
  89

\bibitem[{{Draine} \& {Lee}(1984)}]{draine&lee84}
{Draine}, B.~T., \& {Lee}, H.~M. 1984, \apj, 285, 89

\bibitem[{{Dullemond} \& {Dominik}(2004)}]{dullemond&dominik04b}
{Dullemond}, C.~P., \& {Dominik}, C. 2004, \aap, 421, 1075

\bibitem[{{Dullemond} \& {Dominik}(2005)}]{dullemond&dominik05}
---. 2005, \aap, 434, 971

\bibitem[{{Dunham} {et~al.}(2014){Dunham}, {Stutz}, {Allen}, {Evans},
  {Fischer}, {Megeath}, {Myers}, {Offner}, {Poteet}, {Tobin}, \&
  {Vorobyov}}]{dunham14}
{Dunham}, M.~M., {Stutz}, A.~M., {Allen}, L.~E., {et~al.} 2014, Protostars and
  Planets VI, 195

\bibitem[{{Espaillat} {et~al.}(2014){Espaillat}, {Muzerolle}, {Najita},
  {Andrews}, {Zhu}, {Calvet}, {Kraus}, {Hashimoto}, {Kraus}, \&
  {D'Alessio}}]{espaillat14}
{Espaillat}, C., {Muzerolle}, J., {Najita}, J., {et~al.} 2014, Protostars and
  Planets VI, 497

\bibitem[{{Evans} {et~al.}(2009){Evans}, {Dunham}, {J{\o}rgensen}, {Enoch},
  {Mer{\'{\i}}n}, {van Dishoeck}, {Alcal{\'a}}, {Myers}, {Stapelfeldt},
  {Huard}, {Allen}, {Harvey}, {van Kempen}, {Blake}, {Koerner}, {Mundy},
  {Padgett}, \& {Sargent}}]{evans09}
{Evans}, II, N.~J., {Dunham}, M.~M., {J{\o}rgensen}, J.~K., {et~al.} 2009,
  \apjs, 181, 321

\bibitem[{{Fang} {et~al.}(2013){Fang}, {Kim}, {van Boekel}, {Sicilia-Aguilar},
  {Henning}, \& {Flaherty}}]{fang13}
{Fang}, M., {Kim}, J.~S., {van Boekel}, R., {et~al.} 2013, \apjs, 207, 5

\bibitem[{{Fang} {et~al.}(2009){Fang}, {van Boekel}, {Wang}, {Carmona},
  {Sicilia-Aguilar}, \& {Henning}}]{fang09}
{Fang}, M., {van Boekel}, R., {Wang}, W., {et~al.} 2009, \aap, 504, 461

\bibitem[{{Fazio} {et~al.}(2004){Fazio}, {Hora}, {Allen}, {Ashby}, {Barmby},
  {Deutsch}, {Huang}, {Kleiner}, {Marengo}, {Megeath}, {Melnick}, {Pahre},
  {Patten}, {Polizotti}, {Smith}, {Taylor}, {Wang}, {Willner}, {Hoffmann},
  {Pipher}, {Forrest}, {McMurty}, {McCreight}, {McKelvey}, {McMurray}, {Koch},
  {Moseley}, {Arendt}, {Mentzell}, {Marx}, {Losch}, {Mayman}, {Eichhorn},
  {Krebs}, {Jhabvala}, {Gezari}, {Fixsen}, {Flores}, {Shakoorzadeh}, {Jungo},
  {Hakun}, {Workman}, {Karpati}, {Kichak}, {Whitley}, {Mann}, {Tollestrup},
  {Eisenhardt}, {Stern}, {Gorjian}, {Bhattacharya}, {Carey}, {Nelson},
  {Glaccum}, {Lacy}, {Lowrance}, {Laine}, {Reach}, {Stauffer}, {Surace},
  {Wilson}, {Wright}, {Hoffman}, {Domingo}, \& {Cohen}}]{fazio04}
{Fazio}, G.~G., {Hora}, J.~L., {Allen}, L.~E., {et~al.} 2004, \apjs, 154, 10

\bibitem[{{Fischer} {et~al.}(2011){Fischer}, {Edwards}, {Hillenbrand}, \&
  {Kwan}}]{fischer11}
{Fischer}, W., {Edwards}, S., {Hillenbrand}, L., \& {Kwan}, J. 2011, \apj, 730,
  73

\bibitem[{{Fischer} {et~al.}(2017){Fischer}, {Megeath}, {Furlan}, {Ali},
  {Stutz}, {Tobin}, {Osorio}, {Stanke}, {Manoj}, {Poteet}, {Booker},
  {Hartmann}, {Wilson}, {Myers}, \& {Watson}}]{fischer17}
{Fischer}, W.~J., {Megeath}, S.~T., {Furlan}, E., {et~al.} 2017, \apj, 840, 69

\bibitem[{{Flaherty} {et~al.}(2015){Flaherty}, {Hughes}, {Rosenfeld},
  {Andrews}, {Chiang}, {Simon}, {Kerzner}, \& {Wilner}}]{flahery15}
{Flaherty}, K.~M., {Hughes}, A.~M., {Rosenfeld}, K.~A., {et~al.} 2015, \apj,
  813, 99

\bibitem[{{Flaherty} {et~al.}(2017){Flaherty}, {Hughes}, {Rose}, {Simon}, {Qi},
  {Andrews}, {K{\'o}sp{\'a}l}, {Wilner}, {Chiang}, {Armitage}, \&
  {Bai}}]{flaherty17}
{Flaherty}, K.~M., {Hughes}, A.~M., {Rose}, S.~C., {et~al.} 2017, \apj, 843,
  150

\bibitem[{{Furlan} {et~al.}(2006){Furlan}, {Hartmann}, {Calvet}, {D'Alessio},
  {Franco-Hern{\'a}ndez}, {Forrest}, {Watson}, {Uchida}, {Sargent}, {Green},
  {Keller}, \& {Herter}}]{furlan06}
{Furlan}, E., {Hartmann}, L., {Calvet}, N., {et~al.} 2006, \apjs, 165, 568

\bibitem[{{Furlan} {et~al.}(2009){Furlan}, {Watson}, {McClure}, {Manoj},
  {Espaillat}, {D'Alessio}, {Calvet}, {Kim}, {Sargent}, {Forrest}, \&
  {Hartmann}}]{furlan09a}
{Furlan}, E., {Watson}, D.~M., {McClure}, M.~K., {et~al.} 2009, \apj, 703, 1964

\bibitem[{{Furlan} {et~al.}(2011){Furlan}, {Luhman}, {Espaillat}, {D'Alessio},
  {Adame}, {Manoj}, {Kim}, {Watson}, {Forrest}, {McClure}, {Calvet}, {Sargent},
  {Green}, \& {Fischer}}]{furlan11}
{Furlan}, E., {Luhman}, K.~L., {Espaillat}, C., {et~al.} 2011, \apjs, 195, 3

\bibitem[{{Furlan} {et~al.}(2016){Furlan}, {Fischer}, {Ali}, {Stutz}, {Stanke},
  {Tobin}, {Megeath}, {Osorio}, {Hartmann}, {Calvet}, {Poteet}, {Booker},
  {Manoj}, {Watson}, \& {Allen}}]{furlan16}
{Furlan}, E., {Fischer}, W.~J., {Ali}, B., {et~al.} 2016, \apjs, 224, 5

\bibitem[{{G{\^a}lfalk} \& {Olofsson}(2008)}]{galfalk&olofsson08}
{G{\^a}lfalk}, M., \& {Olofsson}, G. 2008, \aap, 489, 1409

\bibitem[{{Goldreich} \& {Ward}(1973)}]{goldreich&ward73}
{Goldreich}, P., \& {Ward}, W.~R. 1973, \apj, 183, 1051

\bibitem[{{Greene} {et~al.}(1994){Greene}, {Wilking}, {Andre}, {Young}, \&
  {Lada}}]{greene94}
{Greene}, T.~P., {Wilking}, B.~A., {Andre}, P., {Young}, E.~T., \& {Lada},
  C.~J. 1994, \apj, 434, 614

\bibitem[{{Gutermuth} {et~al.}(2008){Gutermuth}, {Myers}, {Megeath}, {Allen},
  {Pipher}, {Muzerolle}, {Porras}, {Winston}, \& {Fazio}}]{gutermuth08}
{Gutermuth}, R.~A., {Myers}, P.~C., {Megeath}, S.~T., {et~al.} 2008, \apj, 674,
  336

\bibitem[{{Hartmann} {et~al.}(2001){Hartmann}, {Ballesteros-Paredes}, \&
  {Bergin}}]{hartmann01}
{Hartmann}, L., {Ballesteros-Paredes}, J., \& {Bergin}, E.~A. 2001, \apj, 562,
  852

\bibitem[{{Hartmann} {et~al.}(1998){Hartmann}, {Calvet}, {Gullbring}, \&
  {D'Alessio}}]{hartmann98}
{Hartmann}, L., {Calvet}, N., {Gullbring}, E., \& {D'Alessio}, P. 1998, \apj,
  495, 385

\bibitem[{{Hendler} {et~al.}(2018){Hendler}, {Pinilla}, {Pascucci}, {Pohl},
  {Mulders}, {Henning}, {Dong}, {Clarke}, {Owen}, \& {Hollenbach}}]{hendler18}
{Hendler}, N.~P., {Pinilla}, P., {Pascucci}, I., {et~al.} 2018, \mnras, 475,
  L62

\bibitem[{{Hern{\'a}ndez} {et~al.}(2004){Hern{\'a}ndez}, {Calvet},
  {Brice{\~n}o}, {Hartmann}, \& {Berlind}}]{hernandez04}
{Hern{\'a}ndez}, J., {Calvet}, N., {Brice{\~n}o}, C., {Hartmann}, L., \&
  {Berlind}, P. 2004, \aj, 127, 1682

\bibitem[{{Houck} {et~al.}(2004){Houck}, {Roellig}, {van Cleve}, {Forrest},
  {Herter}, {Lawrence}, {Matthews}, {Reitsema}, {Soifer}, {Watson}, {Weedman},
  {Huisjen}, {Troeltzsch}, {Barry}, {Bernard-Salas}, {Blacken}, {Brandl},
  {Charmandaris}, {Devost}, {Gull}, {Hall}, {Henderson}, {Higdon}, {Pirger},
  {Schoenwald}, {Sloan}, {Uchida}, {Appleton}, {Armus}, {Burgdorf},
  {Fajardo-Acosta}, {Grillmair}, {Ingalls}, {Morris}, \& {Teplitz}}]{houck04}
{Houck}, J.~R., {Roellig}, T.~L., {van Cleve}, J., {et~al.} 2004, \apjs, 154,
  18

\bibitem[{{Howard} {et~al.}(2013){Howard}, {Sandell}, {Vacca}, {Duch{\^e}ne},
  {Mathews}, {Augereau}, {Barrado}, {Dent}, {Eiroa}, {Grady}, {Kamp}, {Meeus},
  {M{\'e}nard}, {Pinte}, {Podio}, {Riviere-Marichalar}, {Roberge}, {Thi},
  {Vicente}, \& {Williams}}]{howard13}
{Howard}, C.~D., {Sandell}, G., {Vacca}, W.~D., {et~al.} 2013, \apj, 776, 21

\bibitem[{{Hsu} {et~al.}(2012){Hsu}, {Hartmann}, {Allen}, {Hern{\'a}ndez},
  {Megeath}, {Mosby}, {Tobin}, \& {Espaillat}}]{hsu12}
{Hsu}, W.-H., {Hartmann}, L., {Allen}, L., {et~al.} 2012, \apj, 752, 59

\bibitem[{{Hsu} {et~al.}(2013){Hsu}, {Hartmann}, {Allen}, {Hern{\'a}ndez},
  {Megeath}, {Tobin}, \& {Ingleby}}]{hsu13}
---. 2013, \apj, 764, 114

\bibitem[{{Ingleby} {et~al.}(2011){Ingleby}, {Calvet}, {Bergin}, {Herczeg},
  {Brown}, {Alexander}, {Edwards}, {Espaillat}, {France}, {Gregory},
  {Hillenbrand}, {Roueff}, {Valenti}, {Walter}, {Johns-Krull}, {Brown},
  {Linsky}, {McClure}, {Ardila}, {Abgrall}, {Bethell}, {Hussain}, \&
  {Yang}}]{ingleby11}
{Ingleby}, L., {Calvet}, N., {Bergin}, E., {et~al.} 2011, \apj, 743, 105

\bibitem[{{Ingleby} {et~al.}(2013){Ingleby}, {Calvet}, {Herczeg}, {Blaty},
  {Walter}, {Ardila}, {Alexander}, {Edwards}, {Espaillat}, {Gregory},
  {Hillenbrand}, \& {Brown}}]{ingleby13}
{Ingleby}, L., {Calvet}, N., {Herczeg}, G., {et~al.} 2013, \apj, 767, 112

\bibitem[{{Kenyon} \& {Hartmann}(1987)}]{k&h87}
{Kenyon}, S.~J., \& {Hartmann}, L. 1987, \apj, 323, 714

\bibitem[{{Kenyon} \& {Hartmann}(1995)}]{k&h95}
---. 1995, \apjs, 101, 117

\bibitem[{{Kim} {et~al.}(2013){Kim}, {Watson}, {Manoj}, {Forrest}, {Najita},
  {Furlan}, {Sargent}, {Espaillat}, {Muzerolle}, {Megeath}, {Calvet}, {Green},
  \& {Arnold}}]{kim13}
{Kim}, K.~H., {Watson}, D.~M., {Manoj}, P., {et~al.} 2013, \apj, 769, 149

\bibitem[{{Kim} {et~al.}(2016){Kim}, {Watson}, {Manoj}, {Forrest}, {Furlan},
  {Najita}, {Sargent}, {Hern{\'a}ndez}, {Calvet}, {Adame}, {Espaillat},
  {Megeath}, {Muzerolle}, \& {McClure}}]{kim16}
---. 2016, \apjs, 226, 8

\bibitem[{{Koenig} {et~al.}(2012){Koenig}, {Leisawitz}, {Benford}, {Rebull},
  {Padgett}, \& {Assef}}]{koenig12}
{Koenig}, X.~P., {Leisawitz}, D.~T., {Benford}, D.~J., {et~al.} 2012, \apj,
  744, 130

\bibitem[{{Kounkel} {et~al.}(2016){Kounkel}, {Megeath}, {Poteet}, {Fischer}, \&
  {Hartmann}}]{kounkel16}
{Kounkel}, M., {Megeath}, S.~T., {Poteet}, C.~A., {Fischer}, W.~J., \&
  {Hartmann}, L. 2016, \apj, 821, 52

\bibitem[{{Kounkel} {et~al.}(2017){Kounkel}, {Hartmann}, {Loinard},
  {Ortiz-Le{\'o}n}, {Mioduszewski}, {Rodr{\'{\i}}guez}, {Dzib}, {Torres},
  {Pech}, {Galli}, {Rivera}, {Boden}, {Evans}, {Brice{\~n}o}, \&
  {Tobin}}]{kounkel17}
{Kounkel}, M., {Hartmann}, L., {Loinard}, L., {et~al.} 2017, \apj, 834, 142

\bibitem[{{Kryukova} {et~al.}(2014){Kryukova}, {Megeath}, {Hora}, {Gutermuth},
  {Bontemps}, {Kraemer}, {Hennemann}, {Schneider}, {Smith}, \&
  {Motte}}]{kryukova14}
{Kryukova}, E., {Megeath}, S.~T., {Hora}, J.~L., {et~al.} 2014, \aj, 148, 11

\bibitem[{{Lada}(1987)}]{lada87}
{Lada}, C.~J. 1987, in IAU Symposium, Vol. 115, Star Forming Regions, ed.
  M.~{Peimbert} \& J.~{Jugaku}, 1--17

\bibitem[{{Lebouteiller} {et~al.}(2011){Lebouteiller}, {Barry}, {Spoon},
  {Bernard-Salas}, {Sloan}, {Houck}, \& {Weedman}}]{lebouteiller11}
{Lebouteiller}, V., {Barry}, D.~J., {Spoon}, H.~W.~W., {et~al.} 2011, \apjs,
  196, 8

\bibitem[{{Lewis} \& {Lada}(2016)}]{lewis&lada16}
{Lewis}, J.~A., \& {Lada}, C.~J. 2016, \apj, 825, 91

\bibitem[{{Lubow} \& {D'Angelo}(2006)}]{lubow&d'angelo06}
{Lubow}, S.~H., \& {D'Angelo}, G. 2006, \apj, 641, 526

\bibitem[{{Luhman}(2004)}]{luhman04}
{Luhman}, K.~L. 2004, \apj, 602, 816

\bibitem[{{Luhman} {et~al.}(2010){Luhman}, {Allen}, {Espaillat}, {Hartmann}, \&
  {Calvet}}]{luhman10}
{Luhman}, K.~L., {Allen}, P.~R., {Espaillat}, C., {Hartmann}, L., \& {Calvet},
  N. 2010, \apjs, 186, 111

\bibitem[{{Luhman} \& {Rieke}(1999)}]{luhman&rieke99}
{Luhman}, K.~L., \& {Rieke}, G.~H. 1999, \apj, 525, 440

\bibitem[{{Manoj} {et~al.}(2011){Manoj}, {Kim}, {Furlan}, {McClure}, {Luhman},
  {Watson}, {Espaillat}, {Calvet}, {Najita}, {D'Alessio}, {Adame}, {Sargent},
  {Forrest}, {Bohac}, {Green}, \& {Arnold}}]{manoj11}
{Manoj}, P., {Kim}, K.~H., {Furlan}, E., {et~al.} 2011, \apjs, 193, 11

\bibitem[{{Manoj} {et~al.}(2013){Manoj}, {Watson}, {Neufeld}, {Megeath},
  {Vavrek}, {Yu}, {Visser}, {Bergin}, {Fischer}, {Tobin}, {Stutz}, {Ali},
  {Wilson}, {Di Francesco}, {Osorio}, {Maret}, \& {Poteet}}]{manoj13}
{Manoj}, P., {Watson}, D.~M., {Neufeld}, D.~A., {et~al.} 2013, \apj, 763, 83

\bibitem[{{Marsh} \& {Mahoney}(1992)}]{marsh&mahoney92}
{Marsh}, K.~A., \& {Mahoney}, M.~J. 1992, \apjl, 395, L115

\bibitem[{{Mathis}(1990)}]{mathis90}
{Mathis}, J.~S. 1990, \araa, 28, 37

\bibitem[{{Mauc{\'o}} {et~al.}(2016){Mauc{\'o}}, {Hern{\'a}ndez}, {Calvet},
  {Ballesteros-Paredes}, {Brice{\~n}o}, {McClure}, {D'Alessio}, {Anderson}, \&
  {Ali}}]{mauco16}
{Mauc{\'o}}, K., {Hern{\'a}ndez}, J., {Calvet}, N., {et~al.} 2016, \apj, 829,
  38

\bibitem[{{McClure}(2009)}]{mcclure09}
{McClure}, M. 2009, \apjl, 693, L81

\bibitem[{{McClure} {et~al.}(2010){McClure}, {Furlan}, {Manoj}, {Luhman},
  {Watson}, {Forrest}, {Espaillat}, {Calvet}, {D'Alessio}, {Sargent}, {Tobin},
  \& {Chiang}}]{mcclure10}
{McClure}, M.~K., {Furlan}, E., {Manoj}, P., {et~al.} 2010, \apjs, 188, 75

\bibitem[{{McClure} {et~al.}(2013{\natexlab{a}}){McClure}, {Calvet},
  {Espaillat}, {Hartmann}, {Hern{\'a}ndez}, {Ingleby}, {Luhman}, {D'Alessio},
  \& {Sargent}}]{mcclure13a}
{McClure}, M.~K., {Calvet}, N., {Espaillat}, C., {et~al.} 2013{\natexlab{a}},
  \apj, 769, 73

\bibitem[{{McClure} {et~al.}(2013{\natexlab{b}}){McClure}, {D'Alessio},
  {Calvet}, {Espaillat}, {Hartmann}, {Sargent}, {Watson}, {Ingleby}, \&
  {Hern{\'a}ndez}}]{mcclure13b}
{McClure}, M.~K., {D'Alessio}, P., {Calvet}, N., {et~al.} 2013{\natexlab{b}},
  \apj, 775, 114

\bibitem[{{Megeath} {et~al.}(2012){Megeath}, {Gutermuth}, {Muzerolle},
  {Kryukova}, {Flaherty}, {Hora}, {Allen}, {Hartmann}, {Myers}, {Pipher},
  {Stauffer}, {Young}, \& {Fazio}}]{megeath12}
{Megeath}, S.~T., {Gutermuth}, R., {Muzerolle}, J., {et~al.} 2012, \aj, 144,
  192

\bibitem[{{Megeath} {et~al.}(2016){Megeath}, {Gutermuth}, {Muzerolle},
  {Kryukova}, {Hora}, {Allen}, {Flaherty}, {Hartmann}, {Myers}, {Pipher},
  {Stauffer}, {Young}, \& {Fazio}}]{megeath16}
---. 2016, \aj, 151, 5

\bibitem[{{Mer{\'{\i}}n} {et~al.}(2010){Mer{\'{\i}}n}, {Brown}, {Oliveira},
  {Herczeg}, {van Dishoeck}, {Bottinelli}, {Evans}, {Cieza}, {Spezzi},
  {Alcal{\'a}}, {Harvey}, {Blake}, {Bayo}, {Geers}, {Lahuis}, {Prusti},
  {Augereau}, {Olofsson}, {Walter}, \& {Chiu}}]{merin10}
{Mer{\'{\i}}n}, B., {Brown}, J.~M., {Oliveira}, I., {et~al.} 2010, \apj, 718,
  1200

\bibitem[{{Meyer} {et~al.}(1997){Meyer}, {Calvet}, \& {Hillenbrand}}]{meyer97}
{Meyer}, M.~R., {Calvet}, N., \& {Hillenbrand}, L.~A. 1997, \aj, 114, 288

\bibitem[{{Muzerolle} {et~al.}(2010){Muzerolle}, {Allen}, {Megeath},
  {Hern{\'a}ndez}, \& {Gutermuth}}]{muzerolle10}
{Muzerolle}, J., {Allen}, L.~E., {Megeath}, S.~T., {Hern{\'a}ndez}, J., \&
  {Gutermuth}, R.~A. 2010, \apj, 708, 1107

\bibitem[{{Pilbratt} {et~al.}(2010){Pilbratt}, {Riedinger}, {Passvogel},
  {Crone}, {Doyle}, {Gageur}, {Heras}, {Jewell}, {Metcalfe}, {Ott}, \&
  {Schmidt}}]{pilbratt10}
{Pilbratt}, G.~L., {Riedinger}, J.~R., {Passvogel}, T., {et~al.} 2010, \aap,
  518, L1

\bibitem[{{Pillitteri} {et~al.}(2013){Pillitteri}, {Wolk}, {Megeath}, {Allen},
  {Bally}, {Gagn{\'e}}, {Gutermuth}, {Hartman}, {Micela}, {Myers}, {Oliveira},
  {Sciortino}, {Walter}, {Rebull}, \& {Stauffer}}]{pillitteri13}
{Pillitteri}, I., {Wolk}, S.~J., {Megeath}, S.~T., {et~al.} 2013, \apj, 768, 99

\bibitem[{{Pinilla} {et~al.}(2017){Pinilla}, {P{\'e}rez}, {Andrews}, {van der
  Marel}, {van Dishoeck}, {Ataiee}, {Benisty}, {Birnstiel}, {Juh{\'a}sz},
  {Natta}, {Ricci}, \& {Testi}}]{pinilla17}
{Pinilla}, P., {P{\'e}rez}, L.~M., {Andrews}, S., {et~al.} 2017, \apj, 839, 99

\bibitem[{{Poglitsch} {et~al.}(2010){Poglitsch}, {Waelkens}, {Geis},
  {Feuchtgruber}, {Vandenbussche}, {Rodriguez}, {Krause}, {Renotte}, {van
  Hoof}, {Saraceno}, {Cepa}, {Kerschbaum}, {Agn{\`e}se}, {Ali}, {Altieri},
  {Andreani}, {Augueres}, {Balog}, {Barl}, {Bauer}, {Belbachir}, {Benedettini},
  {Billot}, {Boulade}, {Bischof}, {Blommaert}, {Callut}, {Cara}, {Cerulli},
  {Cesarsky}, {Contursi}, {Creten}, {De Meester}, {Doublier}, {Doumayrou},
  {Duband}, {Exter}, {Genzel}, {Gillis}, {Gr{\"o}zinger}, {Henning},
  {Herreros}, {Huygen}, {Inguscio}, {Jakob}, {Jamar}, {Jean}, {de Jong},
  {Katterloher}, {Kiss}, {Klaas}, {Lemke}, {Lutz}, {Madden}, {Marquet},
  {Martignac}, {Mazy}, {Merken}, {Montfort}, {Morbidelli}, {M{\"u}ller},
  {Nielbock}, {Okumura}, {Orfei}, {Ottensamer}, {Pezzuto}, {Popesso},
  {Putzeys}, {Regibo}, {Reveret}, {Royer}, {Sauvage}, {Schreiber}, {Stegmaier},
  {Schmitt}, {Schubert}, {Sturm}, {Thiel}, {Tofani}, {Vavrek}, {Wetzstein},
  {Wieprecht}, \& {Wiezorrek}}]{poglitsch10}
{Poglitsch}, A., {Waelkens}, C., {Geis}, N., {et~al.} 2010, \aap, 518, L2

\bibitem[{{Rayner} {et~al.}(2003){Rayner}, {Toomey}, {Onaka}, {Denault},
  {Stahlberger}, {Vacca}, {Cushing}, \& {Wang}}]{rayner03}
{Rayner}, J.~T., {Toomey}, D.~W., {Onaka}, P.~M., {et~al.} 2003, \pasp, 115,
  362

\bibitem[{{Rebollido} {et~al.}(2015){Rebollido}, {Mer{\'{\i}}n}, {Ribas},
  {Bustamante}, {Bouy}, {Riviere-Marichalar}, {Prusti}, {Pilbratt},
  {Andr{\'e}}, \& {{\'A}brah{\'a}m}}]{rebollido15}
{Rebollido}, I., {Mer{\'{\i}}n}, B., {Ribas}, {\'A}., {et~al.} 2015, \aap, 581,
  A30

\bibitem[{{Ribas} {et~al.}(2013){Ribas}, {Mer{\'{\i}}n}, {Bouy}, {Alves de
  Oliveira}, {Ardila}, {Puga}, {K{\'o}sp{\'a}l}, {Spezzi}, {Cox}, {Prusti},
  {Pilbratt}, {Andr{\'e}}, {Matr{\`a}}, \& {Vavrek}}]{ribas13}
{Ribas}, {\'A}., {Mer{\'{\i}}n}, B., {Bouy}, H., {et~al.} 2013, \aap, 552, A115

\bibitem[{{Rieke} {et~al.}(2004){Rieke}, {Young}, {Engelbracht}, {Kelly},
  {Low}, {Haller}, {Beeman}, {Gordon}, {Stansberry}, {Misselt}, {Cadien},
  {Morrison}, {Rivlis}, {Latter}, {Noriega-Crespo}, {Padgett}, {Stapelfeldt},
  {Hines}, {Egami}, {Muzerolle}, {Alonso-Herrero}, {Blaylock}, {Dole}, {Hinz},
  {Le Floc'h}, {Papovich}, {P{\'e}rez-Gonz{\'a}lez}, {Smith}, {Su}, {Bennett},
  {Frayer}, {Henderson}, {Lu}, {Masci}, {Pesenson}, {Rebull}, {Rho}, {Keene},
  {Stolovy}, {Wachter}, {Wheaton}, {Werner}, \& {Richards}}]{rieke04}
{Rieke}, G.~H., {Young}, E.~T., {Engelbracht}, C.~W., {et~al.} 2004, \apjs,
  154, 25

\bibitem[{{Rosotti} {et~al.}(2016){Rosotti}, {Juhasz}, {Booth}, \&
  {Clarke}}]{rosotti16}
{Rosotti}, G.~P., {Juhasz}, A., {Booth}, R.~A., \& {Clarke}, C.~J. 2016,
  \mnras, 459, 2790

\bibitem[{{Shakura} \& {Sunyaev}(1973)}]{shakura&sunyaev73}
{Shakura}, N.~I., \& {Sunyaev}, R.~A. 1973, \aap, 24, 337

\bibitem[{{Siess} {et~al.}(2000){Siess}, {Dufour}, \& {Forestini}}]{siess00}
{Siess}, L., {Dufour}, E., \& {Forestini}, M. 2000, \aap, 358, 593

\bibitem[{{Skrutskie} {et~al.}(2006){Skrutskie}, {Cutri}, {Stiening},
  {Weinberg}, {Schneider}, {Carpenter}, {Beichman}, {Capps}, {Chester},
  {Elias}, {Huchra}, {Liebert}, {Lonsdale}, {Monet}, {Price}, {Seitzer},
  {Jarrett}, {Kirkpatrick}, {Gizis}, {Howard}, {Evans}, {Fowler}, {Fullmer},
  {Hurt}, {Light}, {Kopan}, {Marsh}, {McCallon}, {Tam}, {Van Dyk}, \&
  {Wheelock}}]{skrutskie06}
{Skrutskie}, M.~F., {Cutri}, R.~M., {Stiening}, R., {et~al.} 2006, \aj, 131,
  1163

\bibitem[{{Stutz} \& {Gould}(2016)}]{stutz&gould16}
{Stutz}, A.~M., \& {Gould}, A. 2016, \aap, 590, A2

\bibitem[{{Stutz} \& {Kainulainen}(2015)}]{stutz&kainulainen15}
{Stutz}, A.~M., \& {Kainulainen}, J. 2015, \aap, 577, L6

\bibitem[{{Stutz} {et~al.}(2013){Stutz}, {Tobin}, {Stanke}, {Megeath},
  {Fischer}, {Robitaille}, {Henning}, {Ali}, {di Francesco}, {Furlan},
  {Hartmann}, {Osorio}, {Wilson}, {Allen}, {Krause}, \& {Manoj}}]{stutz13}
{Stutz}, A.~M., {Tobin}, J.~J., {Stanke}, T., {et~al.} 2013, \apj, 767, 36

\bibitem[{{Takeuchi} \& {Lin}(2002)}]{takeuchi&lin02}
{Takeuchi}, T., \& {Lin}, D.~N.~C. 2002, \apj, 581, 1344

\bibitem[{{Terebey} {et~al.}(1984){Terebey}, {Shu}, \& {Cassen}}]{terebey84}
{Terebey}, S., {Shu}, F.~H., \& {Cassen}, P. 1984, \apj, 286, 529

\bibitem[{{Testi} {et~al.}(2014){Testi}, {Birnstiel}, {Ricci}, {Andrews},
  {Blum}, {Carpenter}, {Dominik}, {Isella}, {Natta}, {Williams}, \&
  {Wilner}}]{testi14}
{Testi}, L., {Birnstiel}, T., {Ricci}, L., {et~al.} 2014, Protostars and
  Planets VI, 339

\bibitem[{{van der Marel} {et~al.}(2018){van der Marel}, {Williams}, {Ansdell},
  {Manara}, {Miotello}, {Tazzari}, {Testi}, {Hogerheijde}, {Bruderer}, {van
  Terwisga}, \& {van Dishoeck}}]{vanderMarel18}
{van der Marel}, N., {Williams}, J.~P., {Ansdell}, M., {et~al.} 2018, \apj,
  854, 177

\bibitem[{{Weidenschilling}(1980)}]{weidenschilling80}
{Weidenschilling}, S.~J. 1980, Icarus, 44, 172

\bibitem[{{White} \& {Ghez}(2001)}]{white&ghez01}
{White}, R.~J., \& {Ghez}, A.~M. 2001, \apj, 556, 265

\bibitem[{{Wright} {et~al.}(2010){Wright}, {Eisenhardt}, {Mainzer}, {Ressler},
  {Cutri}, {Jarrett}, {Kirkpatrick}, {Padgett}, {McMillan}, {Skrutskie},
  {Stanford}, {Cohen}, {Walker}, {Mather}, {Leisawitz}, {Gautier}, {McLean},
  {Benford}, {Lonsdale}, {Blain}, {Mendez}, {Irace}, {Duval}, {Liu}, {Royer},
  {Heinrichsen}, {Howard}, {Shannon}, {Kendall}, {Walsh}, {Larsen}, {Cardon},
  {Schick}, {Schwalm}, {Abid}, {Fabinsky}, {Naes}, \& {Tsai}}]{wright10}
{Wright}, E.~L., {Eisenhardt}, P.~R.~M., {Mainzer}, A.~K., {et~al.} 2010, \aj,
  140, 1868

\end{thebibliography}

\end{document}